\documentclass[11pt]{article}

\usepackage{amsmath} 
\usepackage{amssymb}
\usepackage{amsthm} 
\usepackage{bbm}
\usepackage{bm} 
\usepackage{graphicx}
\usepackage{subcaption}
\usepackage{cite}
\usepackage{pifont}
\usepackage[makeroom]{cancel}
\usepackage[usenames,dvipsnames]{color}
\usepackage{blindtext}

\usepackage{jheppub}

\usepackage{todonotes}

\hypersetup{colorlinks=true,citecolor=blue,linkcolor=black,urlcolor=blue,linktocpage,linktoc=all}

\numberwithin{equation}{section}

\newcommand{\de}{\partial}
\newcommand{\lr}[1]{\left(#1\right)}
\newcommand{\pder}[2]{\frac{\partial #1}{\partial #2}}
\newcommand{\pderr}[2]{\frac{\partial^2 #1}{\partial #2^2}}

\newcommand{\oser}[1]{{\cal O}\lr{#1}}

\def\b{\mathfrak{b}}
\def\c{\mathfrak{c}}

\def\Re{{\rm Re}}
\def\Im{{\rm Im}}
\def\k{{\bf k}}
\def\x{{\bf x}}
\def\z{{\bf z}}
\def\vo{{\bf v}_0}

\def\B{{\bf B}}

\def\q{{\mathfrak{q}}}
\def\w{{\mathfrak{w}}}
\def\qv{\vec{{\mathfrak q}}}

\begin{document}

\title{Causality constraints and anisotropic states}
\author{Raphael E. Hoult}
\affiliation{
{\it\small Department of Physics \& Astronomy, University of Victoria}\\
{\it\small PO Box 1700 STN CSC, Victoria, BC, V8W 2Y2, Canada}\\
}
\emailAdd{rhoult@uvic.ca}
\abstract{We investigate the effects of anisotropy on dispersion relations and convergence in relativistic hydrodynamics. In particular, we show that for dispersion relations with a branch point at the origin (such as sound modes), there exists a continuum of collisions between hydrodynamic modes at complex wavevector. These collisions are then explicitly demonstrated to be present in a holographic plasma. We lay out a criterion for when the continuum of collisions affects the convergence of the hydrodynamic derivative expansion. Finally, the radius of convergence of hydrodynamic dispersion relations in anisotropic systems is bounded from above on the basis of compatibility with microscopic causality.}

\maketitle

\section{Introduction}
\label{sec:intro}

Relativistic hydrodynamics is an effective theory that describes the long-distance, long-time behaviour of many systems~\cite{LL6,Rezzolla-Zanotti,Romatschke:2017ejr}. Both the idealized and the viscous versions of the theory have found applications in neutron star mergers~\cite{Faber:2012rw}, in black hole accretion disks~\cite{Abramowicz:2011xu}, and in the description of the quark-gluon plasma formed in heavy-ion collisions~\cite{Romatschke:2017ejr,Heinz:2024jwu}.

By its nature as an effective theory, relativistic hydrodynamics has a finite regime of applicability. This regime of validity is often characterized by the existence of a large separation between the microscopic scale of the underlying physics, and the macroscopic\footnote{``Macroscopic" is a relative term. The macroscopic scale in the quark-gluon plasma is on the order of 10 fermi ($10^{-14}$ m)~\cite{Bernhard:2018hnz}.} scale on which the hydrodynamic description is employed. The exact bounds of the regime of validity of relativistic hydrodynamics is an open question. Hydrodynamic theories are typically written in terms of a derivative expansion, and the fate of that expansion \textit{vis-a-vis} convergence or divergence can vary. In the context of highly symmetric flows (Bjorken flow~\cite{Bjorken:1982qr}), it has been demonstrated that the hydrodynamic derivative expansion in real space is divergent~\cite{Heller:2013fn,Denicol:2016bjh,Florkowski:2017olj}, showing factorial growth at high orders. Borel-resummation of divergent hydrodynamic series in these contexts reveals the presence of a late-time attractor solution~\cite{Heller:2015dha,Basar:2015ava,Heller:2016rtz}. Even when the requirements of Bjorken flow are relaxed, divergent expansions have been found~\cite{Heller:2021oxl,Heller:2021yjh} for longitudinal flows. On the other hand, some holographic fluids such as the ${\cal N}=4$ supersymmetric Yang-Mills (SYM) plasma~\cite{Withers:2018srf,Grozdanov:2019kge,Grozdanov:2019uhi} at strong coupling have been shown to have finite, non-zero radii of convergence for the momentum space hydrodynamic frequency modes $\omega_i(\k)$ in terms of a wavevector $\k$. The radius of convergence of the small-$|\k|$ expansion in these holographic fluids is set by collisions between quasinormal modes of black branes at (complex) finite wavevector. The convergence or divergence of the hydrodynamic derivative expansion can also depend on support in momentum space; the introduction of cutoffs can render a divergent expansion convergent~\cite{Heller:2020uuy,Heller:2021oxl,Gavassino:2024pgl}. The radius of convergence can also be set by the collisions of hydrodynamic modes with logarithmic branch points, as in~\cite{Heller:2020hnq}\footnote{While the analysis to follow is not inherently incompatible with the presence of logarithmic branch cuts in the retarded two-point function (so long as they stay far enough away from the origin, unlike in the weak-coupling limit of kinetic theory~\cite{Romatschke:2017ejr,Romatschke:2015gic,Kurkela:2017xis}), we will not consider their effects in this paper.}.

When considering convergence in momentum space, there has for the most part been an underlying assumption of isotropy for the (potentially fictitious) equilibrium state. In the context of dispersion relations, this amounts to utilizing a spatial $SO(d)$ rotation symmetry to rotate the wavevector $\k$ to align with a definite coordinate axis.  This allows the frequency $\omega$ to become a function of only one component of $\k$. 

The goals of this paper are two-fold. Firstly, we wish to investigate how dependence on more than one component of the wavevector $\k$ can affect the details of the radii of convergence in hydrodynamic theories when the vector $\k$ is complexified. Utilizing the analysis of several complex variables (see e.g. Appendix~\ref{appendix:scv}), we will find that the frequency $\omega(\k)$ can contain within it a continuum of branch points at complex $\k$. This continuum of branch points, when present, allows the radius of convergence to depend on the direction of $\k$, i.e. $R(\hat{\k})$. In particular, in at least one toy\footnote{i.e. not containing the full derivative expansion to all orders.} example to follow (first order relativistic magnetohydrodynamics), this dependence on direction means that the expansion of $\omega(\k)$ in terms of small $|\k|$ can converge in one direction, and diverge in another.

The existence of continua of branch points within $\omega(\k)$ will imply that for hydrodynamic modes which collide at the origin $\k=0$ (such as the hydrodynamic sound modes in an uncharged fluid~\cite{LL6,Kovtun:2012rj}, or the Alfv\'en modes in MHD~\cite{Krall-Trivelpiece-book,Anile}), the origin is part of a surface (or surfaces) of branch points. This immediately leads to the fact that at $|\k| \neq 0$, including when $|\k|$ is very small, one can find a continuum of points at complex $\k$ at which collisions between hydrodynamic modes occur. We will demonstrate the existence of this continuum of collisions between hydrodynamic modes by investigating the behaviour of gapless quasinormal sound modes in the dual to an ${\cal N} = 4$ SYM plasma. 

Secondly, given the assumption of convergence, we will obtain constraints on hydrodynamic transport coefficients and radii of convergence in terms of one another. Specifically, compatibility with microcausality\footnote{i.e. demanding that dispersion relations come from singularities of retarded two-point functions in causal microscopic theories, which decay rapidly enough at infinity. This condition can be related to the demand of covariant stability~\cite{Gavassino:2023myj}.} yields bounds on the magnitudes of transport coefficients from above in terms of radii of convergence, and vice-versa. This second goal generalizes the work of~\cite{Heller:2022ejw} to systems with a single anisotropy.

The remainder of the paper is organized as follows. In Section~\ref{sec:examples}, we build up an intuition for the effects of the dependence of $\omega$ on several components of $\k$ with some simple examples. These examples showcase three scenarios: dispersion relations with a) no branch point at the origin, b) a branch point at the origin which does not affect the radius of convergence of the small-$|\k|$ expansion, and c) a branch point at the origin which \textit{does} affect the radius of convergence of the small-$|\k|$ expansion, sending it to zero. In Section~\ref{sec:holography}, we investigate radii of convergence for boosted shear and boosted sound modes of the asymptotically AdS$_{5}$ black brane. We find that for $|\k|/(2 \pi T) \neq 0$, where $T$ is the Hawking temperature of the black brane, there are a continuum of collisions between the hydrodynamic sound modes at complex $\k$ -- however, these collisions do not obstruct the hydrodynamic derivative expansion. In Section~\ref{sec:caus_con}, we take the lessons of the previous two sections, and use them to obtain causality constraints on anisotropic transport coefficients, bounding coefficients from above. We also obtain an infinite set of upper bounds on radii of convergence in a causal theory. Finally, in Section~\ref{sec:conc}, we conclude and look to future directions. Appendix~\ref{appendix:scv} contains a (very) brief introduction to the analysis of several complex variables.

\paragraph{Notation:} We denote spacetime indices with greek letters, and spatial indices with lowercase latin letters. In Section~\ref{sec:holography}, bulk spacetime indices are denoted with uppercase latin letters. We use the mostly positive convention for the metric, $\eta_{\mu\nu} = \text{diag}(-1,+1,...,+1)$. Bolded variables such as $\k$ denote spatial vectors, while gothic variables such as $\w=\frac{\omega}{2\pi T}$ denote variables normalized by the temperature; the exception is $\qv = \frac{\k}{2\pi T}$, which falls into both categories. The number of spatial dimensions is denoted by $d$. We work in units where $c=k_B=\hbar=1$.

\section{Examples}
\label{sec:examples}
In this paper, we are interested in dispersion relations $\omega = \omega(\k)$ describing perturbations about a homogeneous equilibrium state. When the background equilibrium state possesses a spatial $SO(d)$ symmetry, then $\omega$ can depend only on the rotation-invariant $\k^2$, and the wavevector can be rotated to align with a coordinate axis. Without loss of generality let us choose the $x$-axis, such that $\k = (k_x,0,0,...,0)$; then $\omega$ is a function of only $k_x^2$, i.e. $\omega = \omega(k_x^2)$. If one likes, one may then proceed to complexify the variable $k_x$, an endeavour that has given rise to many fruitful results in diverse areas such as pole-skipping and quantum chaos~\cite{Grozdanov:2017ajz,Blake:2017ris,Blake:2018leo,Grozdanov:2018kkt,Blake:2019otz}, bounds on transport coefficients from causality~\cite{Heller:2022ejw,Gavassino:2023myj,Heller:2023jtd,Hoult:2023clg,Wang:2023csj}, and convergence properties of the hydrodynamic derivative expansion~\cite{Grozdanov:2019uhi,Abbasi:2020ykq,Jansen:2020hfd,Grozdanov:2021jfw,Heller:2020hnq}. 

Suppose, however, that one does \textit{not} make use of the $SO(d)$ symmetry to write $\omega$ solely as a function of $k_x$. Then $\omega$ will depend on all of the components of $\k$ by way of $\k^2$. One then has the freedom to complexify all of the components of $\k$ individually, such that $\k \in \mathbb{C}^d$. This gives a larger complex space than the complexification of a single component, and there is accordingly a richer structure. 

\paragraph{The wave equation.} As a nearly trivial example to compare the two situations, let us consider the wave equation in $2+1$D, $\de_t^2 \phi(t,{\mathbf x}) - v^2 (\de_x^2+\de_y^2)\phi(t,{\mathbf x}) = 0$, where $v$ is the (constant) propagation speed, and $\phi(t,{\mathbf x})$ is a scalar field. The wave equation is, of course, isotropic. Fourier transforming yields the controlling equation
\begin{equation}
\label{eq:spectral_curve_waveq}
    F(\omega,k_x,k_y)=\omega^2 - v^2(k_x^2+k_y^2) = 0\,.
\end{equation}
Solving for $\omega$ yields the dispersion relations 
\begin{equation}
\label{eq:solution_wave_eq}
\omega_{\pm} = \pm |v| \sqrt{k_x^2+k_y^2} = \pm |v| \sqrt{\k^2}\,,
\end{equation}
the standard dispersion relations for sound without attenuation. The dispersion relations $\omega_{\pm}$ have a branch point when $\k=0$, which means that the two solutions collide there.

If we now consider complexifying $k_x$ and $k_y$ independently, then in addition to the branch point at the origin, $\omega_{\pm}(\k)$ \textit{also} have a continuum of branch points in $\mathbb{C}^2$ that sit on the complex lines $k_x = \pm i k_y$. Therefore, in addition to the two sound modes colliding at the origin, they also undergo an infinity of collisions in the space of complex $k_x$, $k_y$.

These surfaces\footnote{Since ${\mathbb C}^2 \simeq \mathbb{R}^4$, we can equivalently consider the complex lines $k_x = \pm i k_y$ as dimension-$2$ surfaces in the real and imaginary components of $k_x$, $k_y$.} of branch points are in fact not particularly esoteric in origin. The branch points for the solutions of a polynomial equation are given by the zeroes of the discriminant. The discriminant $\Delta_\omega$ in $\omega$ of equation~\eqref{eq:spectral_curve_waveq} is a function of $k_x$ and $k_y$. The equation $\Delta_\omega(k_x,k_y) = 0$ is one equation in two variables; naturally, the solutions are no longer isolated points.

Despite these seemingly pedestrian origins, the idea of the sound mode undergoing a continuum of collisions in the space of complex $\k$ at any fixed value of $|\k|$, especially for very small $|\k|$, is striking due to the role collisions between modes have played in limiting the radius of convergence of the small-$|\k|$ expansion. In this section, we will consider various examples which demonstrate when these collisions are (and are not) relevant in terms of the convergence of the hydrodynamic derivative expansion.

In the preceding example of the wave equation, we could have taken advantage of the $SO(2)$ symmetry and chosen to simply work with the single coordinate $k_x$. In the following, we will instead consider simple setups which do not have a spatial $SO(d)$ symmetry, so that this reduction in the number of variables is not an option. We will investigate three examples. The first example is shear diffusion of a normal uncharged fluid towards a boosted equilibrium state, while the second example is sound waves in the same system. Finally, the third example is that of Alfv\'en waves in a plasma~\cite{ALFVN1942,Krall-Trivelpiece-book,Anile}. We have elected to describe the Alfv\'en waves using the so-called ``one-form formulation" of magnetohydrodynamics~\cite{Schubring:2014iwa,Grozdanov:2016tdf,Hernandez:2017mch,Armas:2018atq,Hoult:2024qph,Lier:2025wfw} for simplicity. 

In all examples given, we work with the hydrodynamic derivative expansion truncated at first order; we therefore do not expect the small-$|\k|$ expansions written down here to be accurate beyond ${\cal O}(\k^2)$. Nevertheless, we do expect the continuum of collisions to be present to all orders due to their presence for $|\k|$ very small.

\subsection{Uncharged hydrodynamics}
\label{subsec:hydro}
An uncharged normal fluid is the long-wavelength, low-frequency limit of many quantum field theories which do not admit a global $U(1)$ symmetry; one example is a theory of a real scalar $\varphi$ with a $\varphi^4$ interaction~\cite{Jeon:1994if,Jeon:1995zm}. We will here consider first perturbations of the fluid about an isotropic, homogeneous equilibrium state, and then about a homogeneous equilibrium state which has been boosted.

Up to first order in the hydrodynamic derivative expansion, the constitutive relations for the stress-energy tensor $T^{\mu\nu}$ in flat space are given\footnote{In the Landau frame.} by~\cite{LL6,Kovtun:2012rj}
\begin{equation}
    T^{\mu\nu} = \epsilon u^\mu u^\nu + \lr{p - \zeta \de_\mu u^\mu} \Delta^{\mu\nu} - \eta \sigma^{\mu\nu}\,,
\end{equation}
where $\epsilon$ is the energy density, $p$ is the isotropic pressure, $u^\mu$ is the fluid velocity which is normalized to $u^2 = -1$, $\Delta^{\mu\nu} = u^\mu u^\nu + \eta^{\mu\nu}$ is the projector onto the space orthogonal to $u^\mu$, $\eta_{\mu\nu}$ is the Minkowski metric, $\zeta$ is the bulk viscosity, $\eta$ is the shear viscosity, and $\sigma^{\mu\nu}$ is the shear tensor, which is given by
\[
\sigma^{\mu\nu} = \lr{\Delta^{\mu\alpha} \Delta^{\nu\beta} + \Delta^{\mu\beta} \Delta^{\nu\alpha} - \frac{2}{d} \Delta^{\mu\nu} \Delta^{\alpha\beta}} \de_\alpha u_\beta\,.
\]
All of the coefficients $p$, $\epsilon$, $\zeta$, $\eta$ are functions of $T$, the temperature of the fluid. The functions $p$ and $\epsilon$ are related to one another, $\epsilon(T) = -p(T) + \pder{p}{T} T$. Once the equation of state\footnote{Note that the parametrization of $p$ in terms of $T$ is a choice; one could just as easily work with $p = p(\epsilon)$, in which case one can write $T=T(\epsilon)$.} $p=p(T)$ is given, $\epsilon(T)$ is fixed. By the demands of positivity of entropy production, it must be the case that $\eta, \,\zeta \geq 0$.

Now, one may consider the behaviour of this system upon applying a small perturbation away from a homogeneous equilibrium state. More specifically, let us consider plane-wave perturbations of the form
\begin{equation}
\begin{split}
    T(x^\nu) &= T_0 + \delta T(\omega,\k) \exp\lr{-i \omega t  + i \k{\cdot}\x},\\
    u^\mu(x^\nu) &= u_0^\mu + \delta u^\mu(\omega,\k) \exp\lr{-i \omega t + i \k{\cdot}\x}.
\end{split}
\end{equation}
We have introduced here the frequency $\omega$ and the wave-vector $\k$. We demand that $u_0^\mu \delta u_\mu = 0$ so that (up to quadratic order in $\delta u^\mu$) the condition $u^2 = -1$ is maintained. 

\subsubsection{Isotropic state}
Let us begin by considering a perturbation about a homogeneous solution as viewed by a comoving observer, such that $u_0^\mu = \delta^\mu_t$. This background state is isotropic (i.e. it preserves a spatial SO($d$) symmetry), and so without loss of generality let us align $\k$ with the $x$-axis, such that
\begin{equation}
\label{eq:plane_wave_perts}
    T(x^\mu) = T_0 + \delta T(\omega,k_x) e^{-i\omega t + i k_x x}, \quad u^\mu(x^\nu) = \delta^\mu_t + \delta u^\mu(\omega,k_x) e^{-i \omega t + i k_x x}\,.
\end{equation} 
The equations of motion for $T^{\mu\nu}$ are simply the conservation equations, $\de_\mu T^{\mu\nu} = 0$. Linearizing the conservation equations in the perturbations~\eqref{eq:plane_wave_perts} yields a system of equations which only admit non-trivial solutions $\delta T$, $\delta u^\mu$ if the frequency $\omega$ and the wavevector $\k$ are non-trivially related to one another by dispersion relations $\omega = \omega(k_x)$. 

There are $d+1$ dispersion relations. The first $d-1$ are the shear diffusion modes, which are identical and each given by the solution of
\begin{equation}\label{eq:isotropic_shear_speccurve}
    F_{\rm shear}(\omega,k_x^2) = \omega + i D k_x^2 = 0\,,
\end{equation}
where $D = \eta/(p_0 + \epsilon_0)$ is the shear diffusion constant, and $p_0$, $\epsilon_0$ denote the values of the pressure and energy in the homogeneous equilibrium state. The remaining two modes are the sound modes, which are given by the roots of the following controlling equation:
\begin{equation}\label{eq:isotropic_sound_speccurve}
    F_{\rm sound}(\omega,k_x^2) = \omega^2 + i \Gamma k_x^2 \omega - v_s^2 k_x^2 = 0\,,
\end{equation}
where $v_s^2 = \pder{p}{\epsilon}$ is the squared speed of sound, and $\Gamma = \lr{\zeta + (2(d-1)/d) \eta}/(p_0+\epsilon_0)$ is the sound attenuation constant. Note that, because of the isotropy, the dependence on the wavevector is only through $k_x^2$. We will in general refer to functions $F(\omega,\k) =0$ which control dispersion relations as spectral curves~\cite{Grozdanov:2019uhi,Hoult:2023clg}. 

The solution to equation~\eqref{eq:isotropic_shear_speccurve} is given by
$
\omega = - i D k_x^2\,,
$
while the roots of equation~\eqref{eq:isotropic_sound_speccurve} in $\omega$ are given by
\begin{equation}\label{eq:isotropic_sound_mode_full}
    \omega(k_x) =  - i \frac{\Gamma}{2} k_x^2 \pm \frac{1}{2} \sqrt{4 k_x^2 v_s^2 - \Gamma^2 k_x^4}\,.
\end{equation}
We would like to understand the expansion of these solutions about $k_x=0$. The radius of convergence of any such expansion is given by the distance to the nearest non-removable singularity of $\omega(k_x)$ in the plane of complex $k_x$. The shear mode $\omega(k_x) = - i D k_x^2$ is an entire function, and so it formally has an infinite radius of convergence\footnote{Of course, at large enough $k_x$ the first-order theory itself breaks down.}. For the sound modes~\eqref{eq:isotropic_sound_mode_full}, the radius of convergence is finite, and is controlled by the branch points of the square root; these branch points occur when the discriminant in $\omega$ is zero, i.e. when
\begin{equation}
\label{eq:discriminant-isotropic}
    \Delta_\omega \equiv k_x^2 \lr{4 v_s^2  - \Gamma^2 k_x^2} = 0\,.
\end{equation}

\begin{figure}[t]
    \centering
    \includegraphics[width=0.5\linewidth]{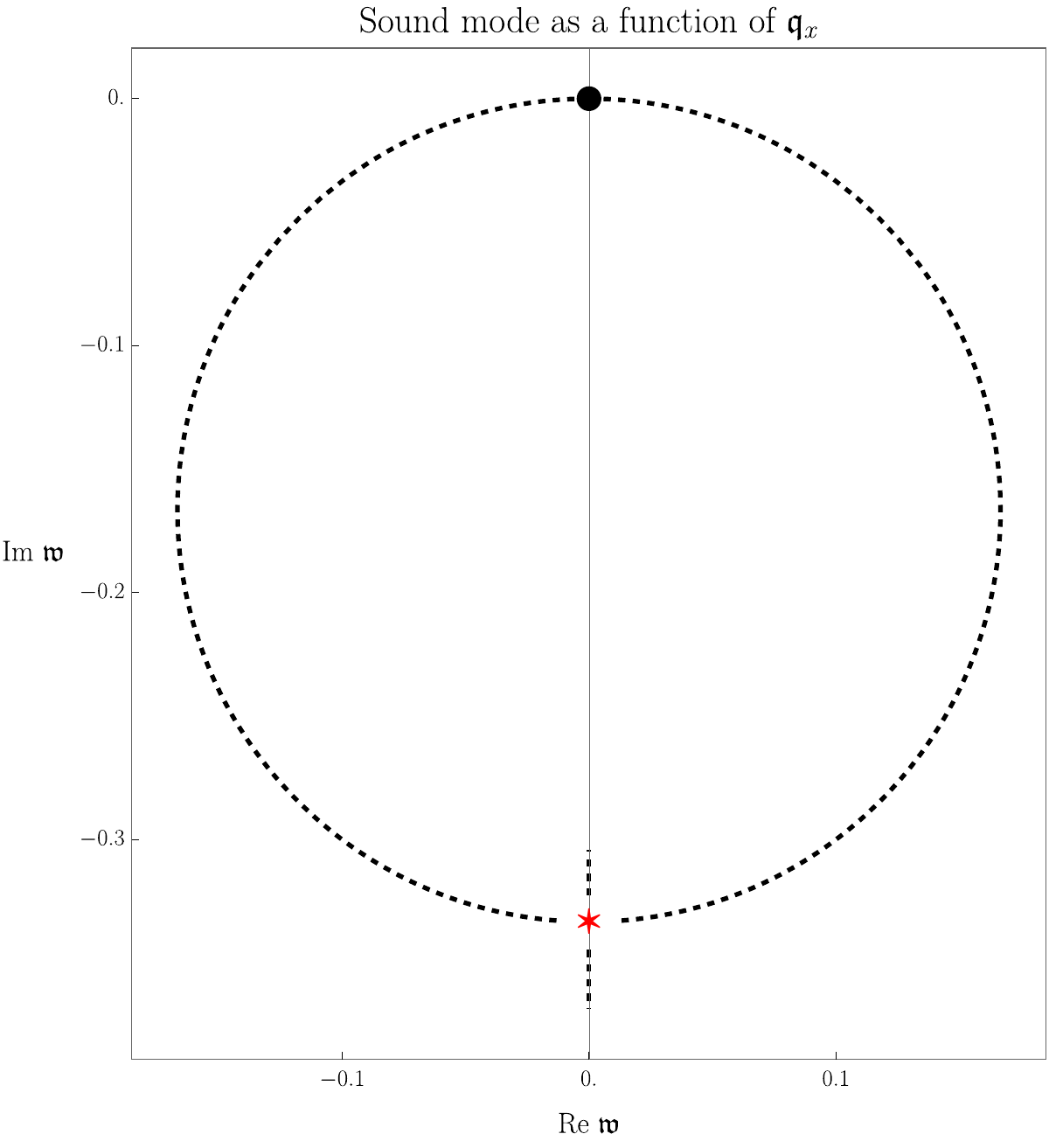}
    \caption{A plot of the two first-order sound modes as they increase with $k_x$, written in terms of unitless quantities $\mathfrak{w} = \omega/(2 \pi T_0)$, $\mathfrak{q}_x = k_x/(2 \pi T_0)$, and with values $v_s = 1/\sqrt{3}$ and $\Gamma = 2/(2 \pi T_0)$. The branch point at the origin ($\q_x=0, \w=0$) is marked with a black dot, while the branch point that sets the radius of convergence of the Puiseux series ($\q_x = \frac{1}{\sqrt{3}}, \w=-\frac{i}{3}$) is marked with a red star.}
    \label{fig:sound_mode_iso}
\end{figure}

More generally, branch points of solutions to polynomials occur when the discriminant $\Delta$ vanishes. We will interchangeably refer to these points where the discriminant $\Delta_\omega$ in $\omega$ vanishes as branch points or as ``critical points"\footnote{Note that for non-polynomial $F(\omega,\k)$, critical points can also arise due to singularities of $F(\omega,\k)$ itself. We do not consider such situations here.}, so-called because they are also the points where $\de_\omega F(\omega,\k) = 0$, i.e. the points where the implicit function theorem~\cite{Krantz2012-bk} no longer holds.

The equation~\eqref{eq:discriminant-isotropic} has solutions when $k_x^2 = 0$ and $k_x^2 = 4 v_s^2/\Gamma^2$. The branch point at $k_x^2 = 0$ means that $\omega = \omega(k_x)$ cannot be expanded in a power series about $k_x^2 = 0$. Rather, it must be expanded in a Puiseux series~\cite{Wall-singular-points} in $\xi \equiv k_x^2$. In terms of $\xi$, the sound mode is given by
\begin{equation}
\label{eq:iso_sound_Puiseux}
    \omega(\xi) = \pm v_s \xi^{1/2} - i \frac{\Gamma}{2} \xi + \,\oser{\xi^{3/2}}\,,
\end{equation}
The radius of convergence of the Puiseux series~\eqref{eq:iso_sound_Puiseux} is set by the branch point at $\xi = k_x^2 = 4 v_s^2/\Gamma^2$, which is where the two sound modes collide in the first-order theory (refer to Figure~\ref{fig:sound_mode_iso}). We now investigate how the situation changes when the equilibrium state is boosted.

\subsubsection{Boosted equilibrium state}
Let us now consider perturbations about a homogeneous equilibrium state with a non-zero spatial velocity, i.e. a state with $u_0^\mu =\gamma(1, \vo)$, where $\gamma = (1-\vo^2)^{-1/2}$ is the Lorentz factor, and $\vo \neq 0$ is the spatial velocity. The presence of this background velocity $\vo$ means that the equilibrium state no longer enjoys a spatial $SO(d)$ symmetry\footnote{There is still an unbroken $SO(d)$ symmetry, which is the little group of $SO(d,1)$ with respect to $u^\mu_0$.}; there does, however, remain a spatial $SO(d{-}1)$ symmetry, which corresponds to rotations in the directions perpendicular to the spatial fluid velocity. 

Without loss of generality, we can fix $\vo$ to point in the $z$-direction, and denote the $z$-component of $\vo$ by $v_0$. Let us therefore define $k_z$ as the component of $\k$ aligned with $\vo$ and $\k_{\perp}$ as the set of components of $\k$ perpendicular to $\vo$. By the $SO(d{-}1)$ symmetry, we can align $\k_{\perp}$ with the $x$-axis, and denote $\k_{\perp}^2 = k_x^2$. We will now consider the shear modes and the sound modes separately.
\paragraph{Boosted shear modes.}
In the presence of a background spatial velocity, the controlling equation for the shear mode~\eqref{eq:isotropic_shear_speccurve} becomes
\begin{equation}\label{eq:anisotropic_shear_speccurve}
    F_{\rm shear}(\omega,k_z,k_x) = \gamma (\omega - v_0 k_z) + i D \lr{k_x^2 + \gamma^2 (k_z - v_0 \omega)^2} = 0\,.
\end{equation}
This spectral curve gives rise to two modes $\omega = \omega(k_x,k_z)$, one of which is famously gapped and unstable~\cite{Hiscock:1983zz}. We will ignore this issue, and press on. The discriminant of~\eqref{eq:anisotropic_shear_speccurve} in $\omega$ is given by
\begin{equation}
\label{eq:discriminant_boosted_shear}
    \Delta_\omega = \gamma (\gamma + 4 D^2 k_x^2 v_0^2 \gamma - 4 i D k_z v_0)\,,
\end{equation}
which is non-vanishing in the limit $\k \to 0$; therefore, $\k = 0$ is not a branch point, and the solutions $\omega = \omega(k_z,k_x)$ of equation~\eqref{eq:anisotropic_shear_speccurve} admit\footnote{By the implicit function theorem.} a Taylor series expansion in a neighbourhood of the origin:
\begin{subequations}
\label{eq:boosted_shear_all_modes}
\begin{align}
    \omega_1(k_z,k_x) &= v_0 k_z - \frac{i D}{\gamma} k_x^2 - \frac{i D}{\gamma^3} k_z^2 + ...\,,\label{eq:boosted_shear_hydro_mode}\\
    \omega_2(k_z,k_x)&= \frac{i}{D v_0^2 \gamma} +\lr{\frac{1+ \gamma^2}{v_0 \gamma^2}} k_z + \frac{i D}{\gamma} k_x^2 + \frac{i D}{\gamma^3} k_z^2 + ...\,,
\end{align}
\end{subequations}
where the $...$ refer to higher-order terms in $k_x$ and $k_z$. We can now try to determine the radius of convergence of the expansion~\eqref{eq:boosted_shear_hydro_mode} about $\k=0$. As one might expect, the radius of convergence depends on the direction of the wavevector relative to the background velocity -- it is no longer simply a number. 

The convergence radius will be set by a collision between the shear mode $\omega_1$ and the gapped mode $\omega_2$. Collisions occur whenever $\Delta_\omega(k_x,k_z)=0$ and so, analogously to the case of the wave equation, there exists a surface of branch points in the space of complex $k_x,k_z$ at which $\omega_{1,2}$ will collide. The surface of critical points, or the ``critical surface" as we will interchangeably refer to it, sets the radius of convergence $R$ for each particular angle $\theta = \arccos(\hat{\k}{\cdot}\hat{{\mathbf v}}_0)$. Strictly, it would be more accurate to say that the expansion~\eqref{eq:boosted_shear_hydro_mode} has radii of convergence, rather than a radius of convergence.

Taylor series in several complex variables, such as the expansions~\eqref{eq:boosted_shear_all_modes}, have natural domains on which they can be defined known as Reinhardt domains\footnote{Technically, a logarithmically convex complete Reinhardt domain; refer to Appendix~\ref{appendix:scv}.}~\cite{HormanderBook,Shabat:1992,Lebl:2025}. Reinhardt domains are defined by the property that if a vector $\k\in \mathbb{C}^n$ is in the Reinhardt domain, then the vector $\k'\in\mathbb{C}^n$ defined by applying a phase rotation to any of the components of $\k$ (i.e. $k'_j= k_j e^{i \varphi_j}$ for any component $k_j$ of $\k$, with $0 \leq \varphi_j < 2 \pi$) is also in the Reinhardt domain. This gives the boundaries of Reinhardt domains the convenient property of being entirely characterized by a projection from the space $\mathbb{C}^n$ down to the positive orthant $\mathbb{R}_+^n$ spanned by the magnitudes of each of the complex components $|k_j|$. In other words, since the phases of the components $k_j$ don't affect whether $\k$ is in the Reinhardt domain, knowledge of only the magnitudes is sufficient to describe the boundary of the domain. For $\mathbb{C}^2$ and $\mathbb{C}^3$, this implies that the boundaries of Reinhardt domains may be straightforwardly visualized.

\begin{figure}[t]
    \centering
    \begin{subfigure}[t]{0.5\linewidth}
    \includegraphics[width=\linewidth]{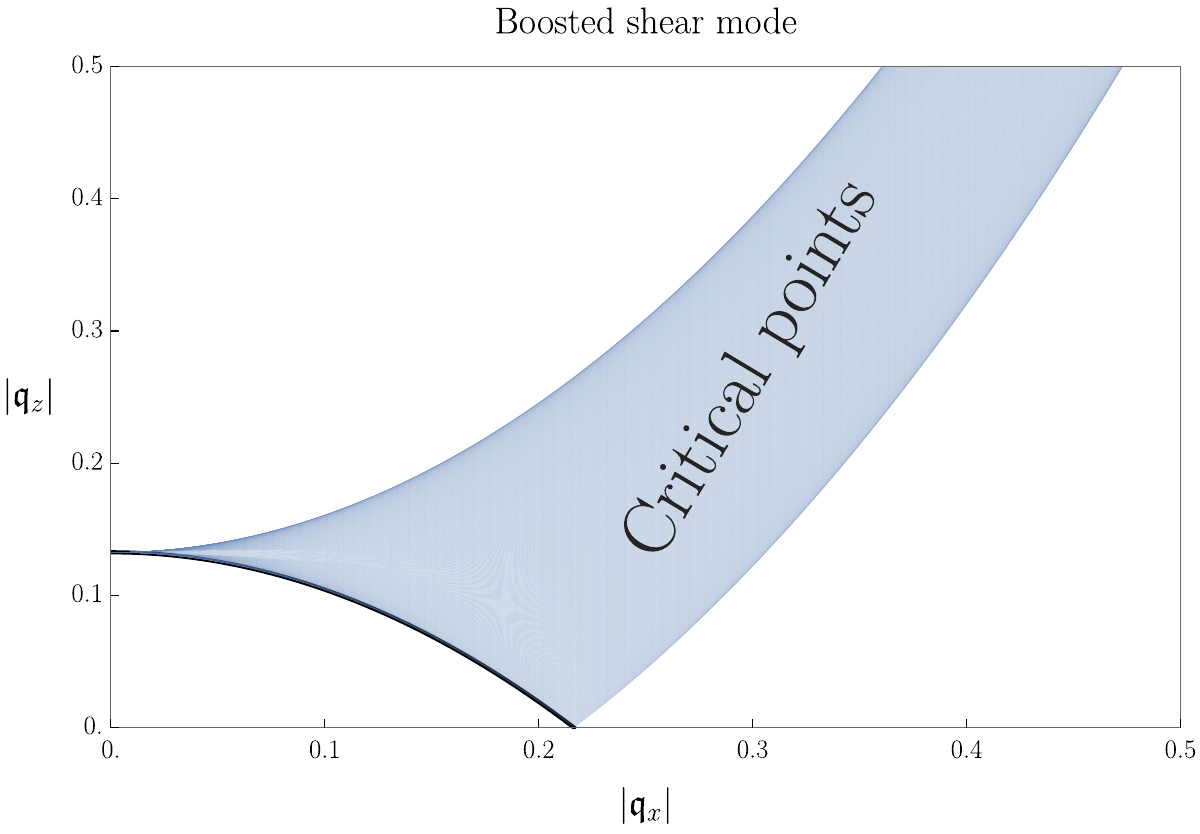}
    \end{subfigure}%
    ~
    \begin{subfigure}[t]{0.5\linewidth}
    \includegraphics[width=\linewidth]{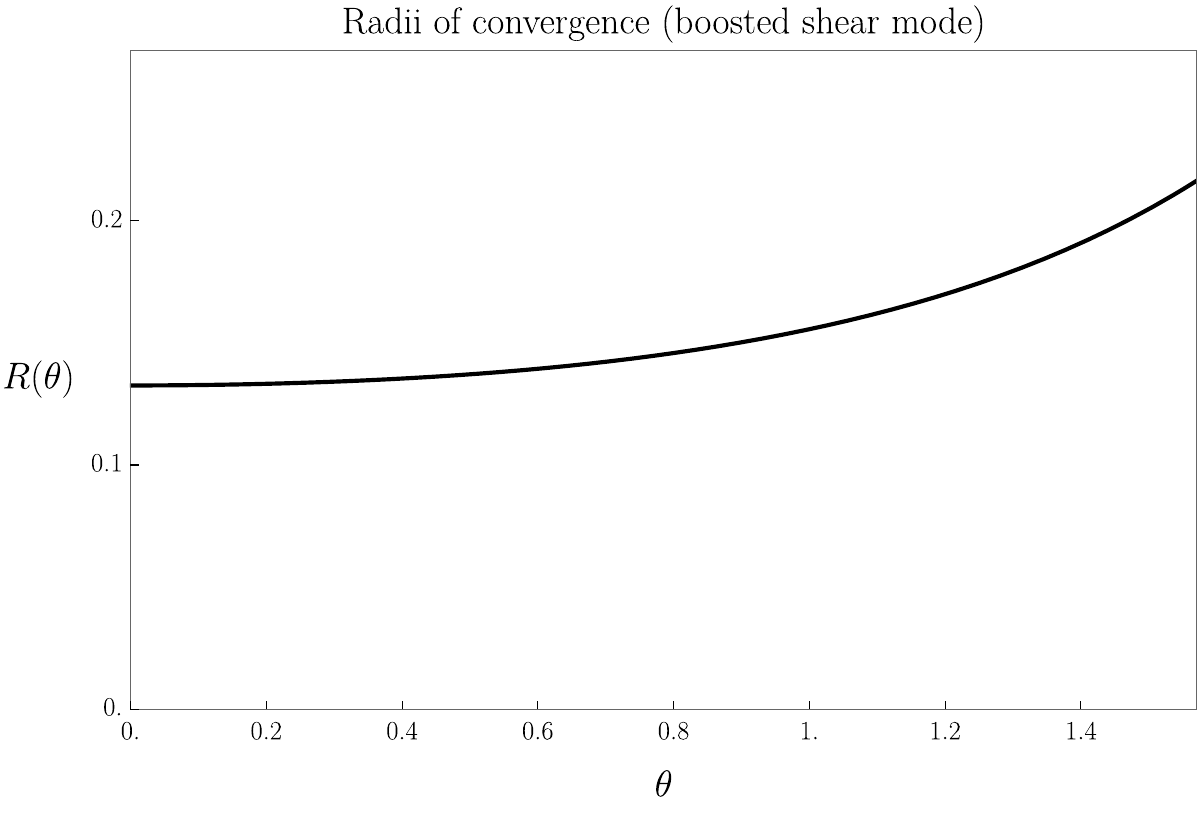}
    \end{subfigure}
    \caption{On the left is a plot of the critical surface for the boosted shear mode with $v_0 = \frac{1}{\sqrt{3}}$, and $D = 4/(2 \pi T_0)$. We work with dimensionless parameters $\mathfrak{w} = \omega/(2 \pi T_0)$, $\mathfrak{q}_x = k_x/(2 \pi T_0)$, and $\mathfrak{q}_z = k_z/(2 \pi T_0)$. The closest set of critical points to the origin, marked in black and with distance from the origin given analytically by equation~\eqref{eq:boosted_shear_radii_analytical}, set the radius of convergence and are given as a curve by equation~\eqref{eq:boundary_curve_boosted_shear}. The radii depend on the relative size of $|\mathfrak{q}_x|$ and $|\mathfrak{q}_z|$. On the right is a plot of the radius of convergence of the boosted shear mode as a function of $\theta$ when $v_0 = \frac{1}{\sqrt{3}}$, and $D = 4/(2 \pi T_0)$.}
    \label{fig:boosted_shear}
\end{figure}

The radius of convergence of the expansion~\eqref{eq:boosted_shear_hydro_mode} at a given $\theta$ is given by the distance in the space of $|k_z|$, $|k_x|$ from the origin to the nearest critical surface in that direction. This distance also characterizes the Reinhardt domain. The surface of critical points in the space of $|k_z|$, $|k_x|$, and the radius of convergence as a function of $\theta$ for the boosted shear mode are plotted in Figure~\ref{fig:boosted_shear} with sample values. In the simple case presented here, the closest curve to the origin may be obtained analytically\footnote{Specifically, solving for the roots of equation~\eqref{eq:discriminant_boosted_shear} in $k_z$ yields an expression for $k_z$ in terms of $k_x$. One can then in turn write $|k_z|$ as a function of $k_x = |k_x| \exp(i \xi_x)$. The surface $|k_z|(k_x)$ is minimized when $\xi_x = \pm \pi/2$, yielding equation~\eqref{eq:boundary_curve_boosted_shear}.}, and is given by
\begin{equation}
\label{eq:boundary_curve_boosted_shear}
    |k_z| = \frac{1}{4} \sqrt{ \frac{(1 - 4 D^2 |k_x|^2 v_0^2)^2}{D^2 v_0^2 (1-v_0^2)}}\,.
\end{equation}
Comparing to Figure~\ref{fig:boosted_shear}, which was obtained numerically, this is indeed what we find. The radius of convergence as a function of $|k_x|$, $R(|k_x|)$, is given by
\begin{equation}
\label{eq:boosted_shear_radii_analytical}
    R^2(|k_x|) = |k_x|^2 + \frac{1}{16} \lr{\frac{(1 - 4 D^2 |k_x|^2 v_0^2)^2}{D^2 v_0^2 (1-v_0^2)}} \,.
\end{equation}
 Casting $k_x$ and $k_z$ into polar coordinates, we can find the radius of convergence for the first-order solution $\omega_1(k_x,k_z)$ about the origin as a function of the angle $\theta$ between $\k$ and $\vo$, which is given by
\begin{equation}
\label{eq:Rtheta_boosted_shear}
    R(\theta) = \frac{\sqrt{\gamma^2 \sin^2(\theta) + 2 \cos^2(\theta) - \sqrt{2 (1 + \gamma^2 + (1- \gamma^2) \cos(2\theta))} \cos(\theta)}}{2  \gamma D |v_0| \sin^2(\theta)} \,,
\end{equation}
and is plotted in Figure~\ref{fig:boosted_shear}.

\paragraph{Boosted sound modes.}
After boosting, the controlling equation~\eqref{eq:isotropic_sound_speccurve} for the sound modes becomes
\begin{equation}
\begin{split}\label{eq:boosted_sound}
        F_{\rm sound}(\omega,k_z,k_x) = \,&i v_0^2 \gamma^3 \Gamma \omega^3 + \gamma^2 \lr{1 - v_0^2 v_s^2 - i k_z v_0 \lr{2 + v_0^2} \gamma \Gamma} \omega^2 \\
        &+ \gamma \lr{i k^2_x \Gamma + k_z \gamma \lr{i k_z \lr{1 + 2 v_0^2} \gamma \Gamma - 2 v_0 \lr{1 - v_s^2}}}\omega \\
        &+ k_z^2 v_0^2 \gamma^2 - \lr{k_x^2 + \gamma^2 k_z^2}\lr{v_s^2 + i k_z v_0 \gamma \Gamma} = 0.
\end{split}
\end{equation}
This, of course, yields an unstable gapped mode. We are once again interested in discussing when the solutions $\omega=\omega(k_x,k_z)$ of equation~\eqref{eq:boosted_sound} can be expanded about $\k=0$. In the isotropic case, $\omega$ could not be expanded in a Taylor series due to the critical point at the origin. This remains the case after boosting.

For simplicity and ease of presentation, let us restrict to the specific case where $v_s = 1/\sqrt{3}$, $v_0 = 1/\sqrt{2}$, and $\Gamma = 4/(2 \pi T_0)$, where $T_0$ is the background temperature. This choice will not affect the qualitative picture to follow. We then work with unitless variables $\mathfrak{w} = \omega/(2 \pi T_0)$, $\mathfrak{q}_x = k_x/(2 \pi T_0)$, and $\mathfrak{q}_z = k_z/(2 \pi T_0)$. The discriminant of equation~\eqref{eq:boosted_sound} in $\w$ is given by
\begin{equation}\label{eq:boosted_sound_disc}
\begin{split}
    \Delta_\w(\q_z,\q_x) = & - 4 \q_x^6 + \q_x^4 \lr{\frac{23}{72} + \frac{i}{3} \q_z - 4 \q_z^2} + \q_x^2 \lr{ \frac{125}{20736} - \frac{85 i}{864} \q_z + \frac{2i}{3} \q_z^3 - \q_z^4}\\
    &+ \frac{25}{6912} \q_z^2 - \frac{11 i}{192} \q_z^3 - \frac{3}{32} \q_z^4 + \frac{i}{4} \q_z^5\,.
\end{split}
\end{equation}
We can immediately see that there remains a critical point at the origin, as $\Delta_{\w}(0,0) = 0$. Let us begin by considering the expansion of $\mathfrak{w}(\mathfrak{q}_x,\mathfrak{q}_z)$ when first taking $\q_x$ to be small, followed by taking $\q_z$ to be small. Doing so yields the following three expansions:
\begin{subequations}\label{eq:kxkz-expansions}
    \begin{align}
        \w_{\pm}(\q_x,\q_z) &= \pm \frac{\q_x^2}{2 \sqrt{3} \q_z} + \frac{2 \sqrt{2}\pm\sqrt{3}}{5} \q_z \nonumber\\
        &- \frac{i}{125} \lr{5 \lr{36 \sqrt{2} \mp 17 \sqrt{3}} \q_x^2 + 6 \lr{27 \sqrt{2} \mp 19 \sqrt{3}} \q_z^2} + ...\,,\\
        \w_g(\q_x,\q_z) &=\frac{5 i}{12 \sqrt{2}} +\frac{17}{5 \sqrt{2}} \q_z + \frac{36}{125} i \sqrt{2} (10 \q_x^2 + 9 \q_z^2) + ...\,,
    \end{align}
\end{subequations}
where the $...$ denote higher-order\footnote{Specifically, higher order giving $\q_x$ and $\q_z$ the same scaling.} terms in $\q_x$, $\q_z$. The $\w_{\pm}$ modes are the boosted sound modes, while $\w_g$ is the (unphysical) unstable gapped mode. Taking the limit of $\q_z \to 0$ in the expansions~\eqref{eq:kxkz-expansions} while keeping $\q_x$ finite leads to a blowup in the boosted sound modes. Similarly, taking $\q_z$ small first and then $\q_x$ yields
\begin{subequations}\label{eq:kzkx-expansions}
    \begin{align}
      \w_{\pm}(\q_x,\q_z) &= \pm \frac{3 \q_z^2}{10 \sqrt{5} \q_x} \pm \frac{1}{\sqrt{5}} \q_x + \frac{2 \sqrt{2}}{5} \q_z \nonumber\\
      &- \frac{18 i}{125}   \lr{ 10 \sqrt{2} \q_x^2 \mp 6 \sqrt{5} \q_x \q_z + 9 \sqrt{2} \q_z^2}+...\,,\\
      \w_g(\q_x,\q_z) &= \frac{5 i}{12 \sqrt{2}} +\frac{17}{5 \sqrt{2}} \q_z + \frac{36}{125} i \sqrt{2} (10 \q_x^2 + 9 \q_z^2) + ...\,.
    \end{align}
\end{subequations}
We can see that taking $\q_x\to0$ while keeping $\q_z$ finite now leads to a blowup, while the gapped mode $\w_g$ is the same in both cases. The ($\q_x\to0$, $\q_z\to0$) and the ($\q_z\to 0$, $\q_x\to0$) limits do not commute in the boosted sound mode\footnote{Nor, indeed, in the sound mode for an equilibrium state at rest. The logic in the paragraph to follow carries through for the isotropic case written in terms of $\q_z$ and $\q_x$.}.

\begin{figure}[t]
    \centering
    \includegraphics[width=0.5\linewidth]{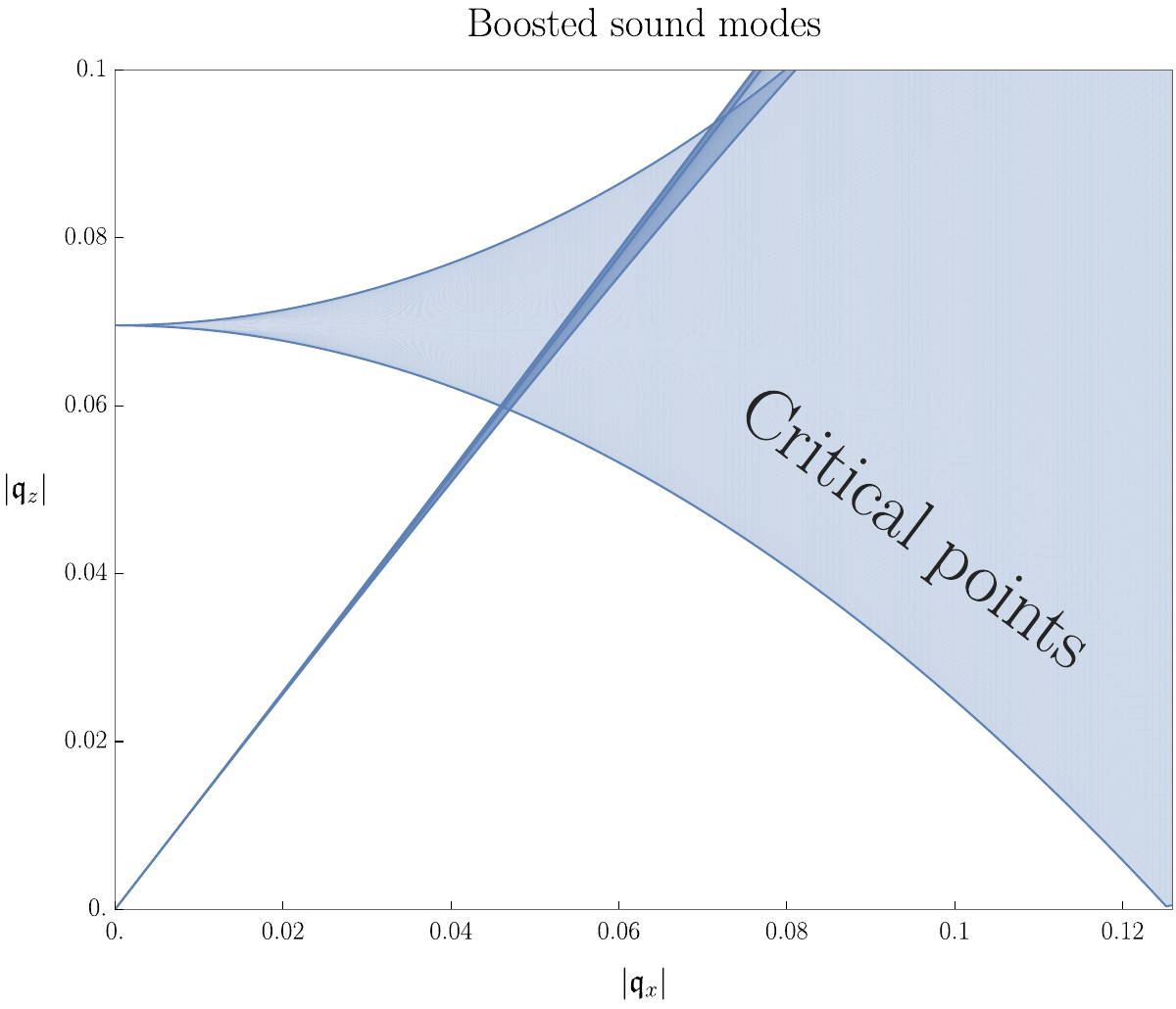}
    \caption{Surfaces of critical points for the boosted sound mode. There are surfaces, $\Sigma_0$, that cut in sharply to reach the origin, leading to a non-commutativity of limits.}
    \label{fig:boosted_sound}
\end{figure}

The source of this behaviour is the critical point at the origin, where the two boosted sound modes collide. We saw in the example of the wave equation that the branch point at the origin was part of a larger surface of critical points. This feature is generic. A theorem in the analysis of several complex variables known as Hartogs' extension theorem~\cite{HormanderBook,Shabat:1992,Lebl:2025} says (roughly, see Appendix~\ref{appendix:scv} for the precise statement) that in $\mathbb{C}^n$ for $n>1$,  singularities cannot be compact\footnote{This also provides another justification for the existence of a \textit{surface} of critical points in the case of the boosted shear mode.}. This means that the branch point at the origin must necessarily be connected to infinity via a critical surface $\Sigma_0$, as may be seen directly in Figure~\ref{fig:boosted_sound}. Put differently, after complexifying the wavevector $\vec{\q} \equiv \k/(2\pi T_0)$ there exists for \textit{any} fixed value of $|\vec{\q}|$ at least one compatible set of complex values for $\q_x$, $\q_z$ such that the boosted sound modes collide with one another. There is, therefore, no absolute notion in which a small $\q_z$, $\q_x$ expansion of $\w(\q_z,\q_x)$ converges when there is a branch point at the origin.

Nevertheless, one can attempt to proceed anyways. In the hydrodynamic derivative expansion, one typically wants to demand small $|\k|$, not necessarily small $|k_z|$, $|k_x|$ individually. Let us therefore write $\q_z = r \cos(\theta)$ and $\q_x = r \sin(\theta)$, where $r^2 =\q_z^2+\q_x^2$. The expansion of $\w(\q_z,\q_x)$ about $r=0$ is given (to quadratic order in $r$) by
\begin{subequations}\label{eq:boosted_sound_r_expansion}
    \begin{align}
        \w_{\pm} &= \lr{\frac{2 \sqrt{2} \cos(\theta) \pm \sqrt{3\cos^2(\theta) + 5 \sin^2(\theta)}}{5}}r \nonumber\\
        &- \frac{18i}{125}\biggl[\sqrt{2} \lr{9 \cos^2(\theta) + 10 \sin^2(\theta)}\mp  \frac{\cos(\theta) (19 \cos^2(\theta) + 30 \sin^2(\theta))}{\sqrt{3\cos^2(\theta) + 5 \sin^2(\theta)}}\biggr]r^2\nonumber \\
        &+ {\cal O}(r^3)\,,\label{eq:boosted_sound_r_specific}\\
        \w_g &= \frac{5 i}{12 \sqrt{2}} + \frac{17}{5 \sqrt{2}} r \cos(\theta) +\frac{36\sqrt{2} i}{125}(9 \cos^2(\theta) + 10 \sin^2(\theta)) r^2 + {\cal O}(r^3)\,.
    \end{align}
\end{subequations}
The expansion~\eqref{eq:boosted_sound_r_specific} still blows up if $\theta$ is complexified and set to $\theta = i\,\text{arctanh}\lr{\sqrt{\frac{3}{5}}} + \pi\,n$ for any value of $r>0$, where $n \in \mathbb{Z}$. However, since the expansions~\eqref{eq:boosted_sound_r_expansion} are expansions only in small $r$, there is no reason in particular to complexify $\theta$. If $\theta$ is restricted to be real, then the expansions~\eqref{eq:boosted_sound_r_expansion} do quite well at approximating the exact solutions to~\eqref{eq:boosted_sound} for small enough $r$. The breakdown of the small-$r$ expansion~\eqref{eq:boosted_sound_r_specific} with real $\theta$ occurs when $r$ grows large enough for the sound modes $\w_{\pm}$ to collide with the non-hydrodynamic mode $\w_g$ at complex $r$; this sets the radius of convergence $R(\theta)$. In Figure~\ref{fig:boosted_sound}, the breakdown corresponds to $r$ being large enough to reach the critical surface that touches the axes for $r\neq 0$. 

In the expansions~\eqref{eq:kxkz-expansions} and~\eqref{eq:kzkx-expansions}, the critical point at the origin was a fatal obstruction to convergence. On the other hand, in the expansion~\eqref{eq:boosted_sound_r_specific} it was completely irrelevant to convergence. Why the difference? The answer is that the expansions~\eqref{eq:kxkz-expansions},~\eqref{eq:kzkx-expansions} are expansions in two complex variables $\q_x$, $\q_z$, while the expansion~\eqref{eq:boosted_sound_r_specific} is in just one complex variable $r$. 

The expansion~\eqref{eq:boosted_sound_r_specific} proceeds on a disc ${\cal D}_r$ in the space of complex $r \in \mathbb{C}$. However, by $r^2 = \q_z^2 + \q_x^2$, this disc ${\cal D}_r$ corresponds to a set of embedded surfaces $\Sigma_r$ in the space of complex $\q_x$, $\q_z$. These surfaces $\Sigma_r$ are determined by the constraint that $\theta = \arctan(\q_x/\q_z)$ is real. This condition can be used to show that the surfaces $\Sigma_r$ upon which the small-$r$ expansions proceed are given by the relations $|r|^2 = |\q_z|^2 + |\q_x|^2$, $\arg(\q_z) = \arg(r) + n_1 \pi$, and $\arg(\q_x) = \arg(r) + n_2 \pi$, where $n_1, n_2 \in \mathbb{Z}$. We will refer to the surfaces $\Sigma_r$ in $\mathbb{C}^2$ as ``expansion surfaces", as they are the surfaces upon which the hydrodynamic derivative expansion actually proceeds in practice. 

The expansion surfaces $\Sigma_r$ may be straightforwardly described by writing $\qv$ in an extension of the so-called ``Hopf coordinates"~\cite{Dzhunushaliev:2020pyk}. We may parametrize $\qv$ as
\begin{equation}
\label{eq:Hopf}
    \q_z = |r| \cos(\theta) e^{i \xi_z}, \qquad \q_x = |r| \sin(\theta) e^{i \xi_x}\,,
\end{equation}
where $\theta \in [0,\frac{\pi}{2}]$, $\xi_z, \xi_x \in[0,2\pi)$, and $|r| \geq 0$. The expansion surfaces may then be captured by setting $\xi_x=\xi_z=\xi$, and extending the domain of $\theta$ such that $\theta \in [0, \pi)$:
\begin{equation}
\label{eq:Hopf_extended}
   \Sigma_r: \qquad  \q_z = |r| \cos(\theta) e^{i \xi}, \qquad \q_x = |r| \sin(\theta) e^{i \xi}\,.
\end{equation}
We will make extensive use of the parametrizations~\eqref{eq:Hopf},~\eqref{eq:Hopf_extended}. 

The critical surfaces connected to the origin, $\Sigma_0$, do not obstruct the small-$r$ expansion because they do not intersect the expansion surfaces $\Sigma_r$ except for at the origin. Therefore, the small-$r$ expansion does not encounter any non-removable singularities until $r$ becomes large enough to reach a different critical surface which does not connect to the origin. This feature is not generic; as we shall see in the final example, the critical surfaces $\Sigma_0$ can intersect with the expansion surfaces $\Sigma_r$, and can cause the radius of convergence of the small-$|\k|$ expansion to go to zero as $\theta$ varies.

\subsection{Magnetohydrodynamics}
\label{subsec:mhd}
Magnetohydrodynamics is the effective theory used to describe plasmas in the presence of electromagnetic fields~\cite{Krall-Trivelpiece-book,Anile}. On macroscopic length scales, the electric field is Debye-screened, while the magnetic field remains at full strength. The equations of motion are the conservation of the stress-energy tensor, and Maxwell's equations in matter. In the perfect fluid case, the electric field is taken to vanish due to the Debye screening, and the only contribution from Maxwell's equations are those which govern the magnetic field:
\begin{equation}
\label{eq:MaxwellConEq}
    \nabla_\mu T^{\mu\nu} = 0, \qquad \nabla_\mu J^{\mu\nu} = 0\,,
\end{equation}
where $J^{\mu\nu} = \frac{1}{2} \epsilon^{\mu\nu\alpha\beta} F_{\alpha \beta}$ is the dual field strength tensor. One can now proceed to parametrize the stress-energy tensor and dual field strength tensor in terms of equilibrium fields such as the temperature $T$, a fluid velocity $u^\mu$ (with $u^2 = -1$), and a magnetic field $B^\mu$ (with $B^\mu u_\mu = 0$), so that~\cite{Anile,Hernandez:2017mch}
\begin{subequations}
\label{eq:ideal-MHD}
\begin{align}
    T^{\mu\nu} &= \lr{\epsilon_m(T,B^2) + P_m(T,B^2) + \frac{B^2}{\mu_B}} u^\mu u^\nu\nonumber \\
    &+ \lr{P_m(T,B^2) - \frac{1}{2} B^2+ \frac{B^2}{\mu_B}} g^{\mu\nu} - \frac{B^\mu B^\nu}{\mu_B}\,,\\
    J^{\mu\nu} &= u^\mu B^\nu - u^\nu B^\mu\,,
\end{align}
\end{subequations}
where $\epsilon_m(T,B^2)$ is the energy density of the matter component, and $P_m(T,B^2)$ is the pressure associated with the matter component. If the matter is polarizable, then the magnetic permeability is given by $\mu_B = 1/(1-2 (\pder{P_m}{B^2})_T)$. In terms of the temperature $T$ and the magnetic field $B^2$, the energy density is defined by
$
    \epsilon_m(T,B^2) = -P_m(T,B^2) + \lr{\pder{P_m}{T}}_{B^2} T
$. These constitutive relations can be obtained from an equilibrium generating functional approach~\cite{Jensen:2012jh,Kovtun:2016lfw,Hernandez:2017mch}. The subscripts on brackets indicate which independent variable is being held constant when taking the partial derivative.

Let us note that the dual field strength tensor is conserved in equation~\eqref{eq:MaxwellConEq}. One interpretation of this statement is simply that the field strength tensor obeys the Bianchi identity. Another, equivalent interpretation, is that there is a type of generalized global symmetry that the system enjoys known as a one-form symmetry~\cite{Gaiotto:2014kfa,Schubring:2014iwa,Grozdanov:2016tdf,Hernandez:2017mch,Grozdanov:2017kyl,Hofman:2017vwr,Armas:2018atq,Armas:2018zbe}. To each continuous global symmetry, there is a conserved quantity -- and in the equations~\eqref{eq:MaxwellConEq}, the conserved quantity may be thought of as the number of magnetic field lines. Using the global magnetic one-form symmetry as a guiding basis, one may write down a dual formulation of magnetohydrodynamics which is, in some senses, easier to use.

We will not provide a complete introduction to the one-form formulation of magnetohydrodynamics here; the interested reader may instead refer to~\cite{Grozdanov:2016tdf,Armas:2018atq,Armas:2018zbe}. The spectrum of this dual formulation of MHD (``dMHD"~\cite{Hoult:2024qph}) has been shown to be the same as the standard formulation of MHD~\cite{Hernandez:2017mch}. The one-form formulation has been interpreted in terms of spontaneous symmetry breaking and superfluidity~\cite{Armas:2018atq,Armas:2018zbe}. The equilibrium state is described in terms of a temperature $T$, a fluid velocity $u^\mu$ (with $u^2=-1$), a spacelike vector related to the direction of the magnetic field $h^\mu$ (with $h^2 = 1$ and $h^\mu u_\mu = 0$), and a ``chemical potential" $\mu$ which is the conjugate variable to the magnetic flux density. The equilibrium constitutive relations (at zeroth order) are given by
\begin{subequations}
\label{eq:ideal-dMHD}
    \begin{align}
        T^{\mu\nu} &= \lr{\epsilon(T,\mu) + p(T,\mu)} u^\mu u^\nu + p(T,\mu) g^{\mu\nu} - \mu\, \rho(T,\mu) h^\mu h^\nu\,,\\
        J^{\mu\nu} &=\rho(T,\mu) (u^\mu h^\nu - u^\nu h^\mu)\,,
    \end{align}
\end{subequations}
where $p(T,\mu)$ is the total pressure which satisfies the Gibbs-Duhem-type relation
\begin{equation}
    dp = s\, dT + \, \rho \, d\mu\,,
\end{equation}
where $s \equiv (\de p/\de T)_\mu$ is the entropy density, and $\rho \equiv (\de p/ \de \mu)_T$ is the magnetic flux density. The total energy density is related to the pressure by $\epsilon = - p + s T + \rho  \mu$. Once an equation of state $p = p(T,\mu)$ has been provided, all other thermodynamic parameters $\epsilon$, $s$, $\rho$ are known in terms of $T$ and $\mu$.

The hydrodynamic variables $T,\mu$ and thermodynamic coefficients $\epsilon,\,p,\rho$ appearing in the dual formulation~\eqref{eq:ideal-dMHD} can be related\footnote{At ideal order; at higher-order in the derivative expansion, these relations can gain corrections.} to those appearing in the usual formulation~\eqref{eq:ideal-MHD} by
\begin{subequations}
\begin{alignat}{2}
    & \rho = |B|, \qquad \qquad \qquad \,\,\,h^\mu = \frac{B^\mu}{|B|}, && \qquad \qquad \qquad \qquad \qquad\,\mu = \frac{|B|}{\mu_B}\,,\\
    & p(T,\mu) =P_m(T,B^2) - \frac{B^2}{2} + \frac{B^2}{\mu_B}, && \qquad \epsilon(T,\mu) = \epsilon_m(T,B^2) + \frac{1}{2} B^2,
\end{alignat}
\end{subequations}
while $T$ and $s$ are the same in both formulations. With ideal-order dMHD written down~\eqref{eq:ideal-dMHD}, we now move on to the first-order formulation of the theory. For convenience, let us define the projector orthogonal to both $u^\mu$ and $h^\mu$ as $\Delta_{\perp}^{\mu\nu} \equiv u^\mu u^\nu + g^{\mu\nu} - h^\mu h^\nu.$ For the sake of brevity in writing the constitutive relations, we also define the following shorthand~\cite{Armas:2022wvb,Hoult:2024qph,Hoult:2025exx}
\begin{subequations}
\begin{alignat}{4}
&s_1 && \equiv \nabla_\mu u^\mu\,,\qquad &&s_2 && \equiv h^\mu h^\nu \nabla_\mu u_\nu\,,\\
& Y^\mu && \equiv 2T \Delta_{\perp}^{\mu\alpha} h^{\beta} \nabla_{[\alpha} \lr{\frac{\mu h_{\beta]}}{T}}\,, \qquad && \Sigma^{\mu}&& \equiv 2 \Delta_{\perp}^{\mu\alpha} h^\beta \nabla_{(\alpha} u_{\beta)}\,,
\end{alignat}
\begin{alignat}{2}
& \sigma_\perp^{\mu\nu} &&\equiv \lr{\frac{\Delta_{\perp}^{\mu\alpha} \Delta_{\perp}^{\nu\beta} + \Delta_{\perp}^{\mu\beta} \Delta_{\perp}^{\nu\alpha}}{2} - \frac{1}{d-1} \Delta_{\perp}^{\mu\nu} \Delta_{\perp}^{\alpha\beta}}\nabla_\alpha u_\beta\,,\\
& Z^{\mu\nu} &&\equiv 2\mu \Delta_{\perp}^{\mu\alpha} \Delta_{\perp}^{\nu\beta} \nabla_{[\alpha} h_{\beta]}\,,
\end{alignat}
\end{subequations}
where we have employed the notation $A_{(\mu} B_{\nu)} = (1/2) (A_\mu B_\nu + A_\nu B_\mu)$ and $A_{[\mu} B_{\nu]} = (1/2) (A_\mu B_\nu - A_\nu B_\mu)$. With these shorthands, the first-order constitutive relations in a parity-even theory\footnote{In the analogue of the Landau frame.} take on the form~\cite{Grozdanov:2016tdf,Hernandez:2017mch}
\begin{subequations}
\label{eq:dMHD-conrel}
    \begin{align}
        T^{\mu\nu} &= \epsilon u^\mu u^\nu + \lr{p - \mu \rho - \zeta_{\parallel} s_2 - \zeta_{\times} (s_1-s_2)} h^\mu h^\nu + \lr{p - \zeta_\times s_2 - \zeta_{\perp} (s_1 - s_2)} \Delta_{\perp}^{\mu\nu}\nonumber\\
        &- 2 \eta_{\parallel} \Sigma^{(\mu} h^{\nu)} - \eta_{\perp} \sigma^{\mu\nu}_\perp\,,\\
        J^{\mu\nu} &= 2 \rho u^{[\mu} h^{\nu]} - 2 r_{\perp} Y^{[\mu} h^{\nu]} - r_{\parallel} Z^{\mu\nu}\,,
    \end{align}
\end{subequations}
where $\zeta_{\parallel},$ $\zeta_{\times},$ $\zeta_{\perp}$ are bulk viscosities, $\eta_{\parallel},$ $\eta_{\perp}$ are shear viscosities, and $r_{\perp}$, $r_{\parallel}$ are resistivities. All of these transport coefficients are functions of $T$ and $\mu$. For entropy bounds on these coefficients, refer to~\cite{Grozdanov:2016tdf,Hernandez:2017mch}. 

The equations of motion at first order are the conservation equations~\eqref{eq:MaxwellConEq} with the constitutive relations~\eqref{eq:dMHD-conrel}. Let us now consider plane-wave perturbations about a homogeneous background state at rest, i.e. we consider perturbations $\delta T, \delta \mu, \delta u^\mu, \delta h^\mu$ about background equilibrium values $(T_0, \mu_0, u^\mu_0, h^\mu_0)$ with $u^\mu_0 = \delta^\mu_t$ such that each perturbation is proportional to $\exp(-i \omega t + i \k {\cdot} \x)$. We also demand that, to linear order, $u_\mu^{0} \delta u^\mu = h_\mu^{0} \delta h^\mu = 0$, and $u_\mu^{0} \delta h^\mu = - h_\mu^{0} \delta u^\mu$, so as to maintain $u^2=-1$, $h^2 = 1$, and $u_\mu h^\mu =0$. The spectral curve $F(\omega, \k)$ describing the hydrodynamic modes of magnetohydrodynamics may then be found in the usual way\footnote{Some care should be taken due to the fact that $\nabla_\mu J^{\mu 0} = 0$ is not an evolution equation. Rather, it is a constraint equation, the analogue of $\nabla{\cdot}\B = 0$. }.

 Without loss of generality, let us set the background magnetic field of the homogeneous equilibrium state to align with the $z$-axis, i.e. we set $h^\mu_0 = \delta^\mu_z$.  We can then use the $SO(d-1)$ symmetry for rotations orthogonal to the magnetic field to align $\k$ with the $xz$-plane. This means that $k_z = \k{\cdot}{\mathbf h}_0$ (where ${\mathbf h}_0$ is the spatial part of $h^\mu_0$), and $k_x^2 =\k_\perp^2$. Upon obtaining the spectral curve $F(\omega,k_z,k_x^2)$, one finds that there are four longitudinal modes (``slow and fast magnetosonic waves") and $2(d-2)$ transverse modes, referred to as ``Alfv\'en waves"~\cite{ALFVN1942,Anile}. The remainder of this section will discuss the transverse modes.
 
In dMHD, Alfv\'en waves are described by the following controlling equation~\cite{Grozdanov:2016tdf,Hernandez:2017mch}:
\begin{equation}
\label{eq:Alfven_spec_curve}
\begin{split}
   F_{\rm Alfv\acute{e}n}(\omega,k_x,k_z)\equiv \omega^2 + i \lr{ k_{x}^2 \lr{\frac{\eta_{\perp}}{p_0+\epsilon_0} + \frac{ \mu_0 r_{\parallel}}{\rho_0}} + k_z^2 \lr{ \frac{\eta_{\parallel}}{p_0 + \epsilon_0} + \frac{\mu_0 r_{\perp}}{\rho_0}}}\omega&\\
    - \frac{\mu_0}{\lr{p_0 + \epsilon_0}\rho_0}\biggl[r_{\perp} \eta_{\parallel} k_z^4 + r_{\parallel} \eta_{\perp} k_x^4 + \lr{r_{\parallel} \eta_{\parallel} + \eta_{\perp} r_{\perp}} k_x^2 k_z^2 + \rho_0^2 k_z^2\biggr] &= 0\,,
\end{split}
\end{equation}
where the subscript $0$ indicates the equilibrium value of the attached thermodynamic parameter. Equation~\eqref{eq:Alfven_spec_curve} is a quadratic equation; the solutions are given by
\begin{equation}
\label{eq:dMHD_Alfven_sol}
    \omega_{\pm} = -\frac{i}{2} \lr{ k_{x}^2 \lr{\frac{\eta_{\perp}}{p_0+\epsilon_0} + \frac{ \mu_0 r_{\parallel}}{\rho_0}} + k_z^2 \lr{ \frac{\eta_{\parallel}}{p_0 + \epsilon_0} + \frac{\mu_0 r_{\perp}}{\rho_0}}} \pm \frac{1}{2} \sqrt{\Delta_{\omega}}\,,
\end{equation}
where the discriminant $\Delta_\omega$ in $\omega$ is given by
\begin{equation}
\label{eq:dMHD_disc_kzkx}
    \Delta_\omega = 4 \lr{\frac{\mu_0 \rho_0}{\epsilon_0 + p_0}} k_z^2 - \biggl[ \lr{\frac{r_{\parallel} \mu_0}{\rho_0} - \frac{\eta_{\perp}}{\epsilon_0 + p_0}}k_x^2 + \lr{\frac{r_{\perp} \mu_0}{\rho_0} - \frac{\eta_{\parallel}}{\epsilon_0+p_0}} k_z^2 \biggr]^2\,.
\end{equation}
An immediate point to note in light of the preceding examples is that $\Delta_\omega(k_x=0,k_z=0)=0$, and so $(k_x,k_z)=0$ is a critical point, as in the case of the boosted sound mode. There will therefore exist critical surfaces $\Sigma_0$ extending from the origin, and any attempt at an expansion in both $k_x$ and $k_z$ will have zero radius of convergence. We will proceed by looking at expansions in just one variable, $r = |\k|$. Let us write $k_z = r\cos(\theta)$ and $k_x = r\sin(\theta)$, where $\theta$ is the angle between $\k$ and ${\mathbf h}$. Then an expansion in the limit of small-$r$ yields
\begin{equation}\label{eq:Alfven_expansion}
    \omega_{\pm} = \pm{\cal V}_A \cos(\theta) r - i \Gamma(\theta) r^2 + {\cal O}(|\k|^3)\,,
\end{equation}
where
\begin{subequations}
    \begin{align}
        {\cal V}_A  &= \sqrt{\frac{\mu_0 \rho_0}{\epsilon_0 + p_0}}\,,\\
        \Gamma(\theta) &= \frac{1}{2} \biggl[\lr{ \frac{r_{\perp} \mu_0}{\rho_0} + \frac{\eta_{\parallel}}{p_0 + \epsilon_0}} \cos^2(\theta) + \lr{\frac{r_{\parallel} \mu_0}{\rho_0} + \frac{\eta_{\perp}}{p_0 + \epsilon_0}}\sin^2(\theta)\biggr]\,.
    \end{align}
\end{subequations}
In the previous example of the boosted sound mode, the critical surfaces $\Sigma_0$ did not intersect (except at the origin) the expansion surfaces $\Sigma_r$ upon which the small-$|\k|$ expansions proceeded, and so they did not impede the small-$|\k|$ expansion. This is no longer the case. The Alfv\'en channel has a non-commutativity between the $r\to0$ limit and the $\theta \to \frac{\pi}{2}$ limit (see e.g.~\cite{Hernandez:2017mch,Grozdanov:2017kyl,Fang:2024skm,Fang:2024hxa,Hoult:2025exx}), except for in the fine-tuned case $r_{\parallel} \mu_0/\rho_0 = \eta_{\perp}/\lr{\epsilon_0+p_0}$. 

This may be directly seen by truncating the expansion~\eqref{eq:Alfven_expansion} at quadratic order, and comparing it in the limit $\theta \to \frac{\pi}{2}$ to the solutions~\eqref{eq:dMHD_Alfven_sol} in the limit $\theta \to \frac{\pi}{2}$ (without necessarily taking $r$ small):
\begin{subequations}
\label{eq:Alfven_noncommutativity}
    \begin{align}
        \text{\eqref{eq:Alfven_expansion}}:\quad \omega_{\pm} &= - \frac{i}{2}\lr{\frac{r_{\parallel} \mu_0}{\rho_0} + \frac{\eta_{\perp}}{p_0 + \epsilon_0}}r^2\,,\\
        \text{\eqref{eq:dMHD_Alfven_sol}}:\quad \omega_{+} &= -i \frac{\eta_{\perp}}{p_0 + \epsilon_0} r^2 \,, \qquad \omega_- = - i \frac{r_{\parallel} \mu_0}{\rho_0} r^2\,.
    \end{align}
\end{subequations}
The expansion~\eqref{eq:Alfven_expansion} gives a qualitatively different picture in the $\theta \to \frac{\pi}{2}$ limit compared to the exact solution~\eqref{eq:dMHD_Alfven_sol}. The reason for this is simple: unlike in the boosted sound mode, in the Alfv\'en channel the expansion surfaces $\Sigma_r$ and the critical surfaces $\Sigma_0$ intersect non-trivially. As $\theta \to \frac{\pi}{2}$ the collisions between hydrodynamic modes send the radius of convergence of the small-$r$ expansion to zero.

\begin{figure}
    \centering
        \centering
        \includegraphics[width=0.5\linewidth]{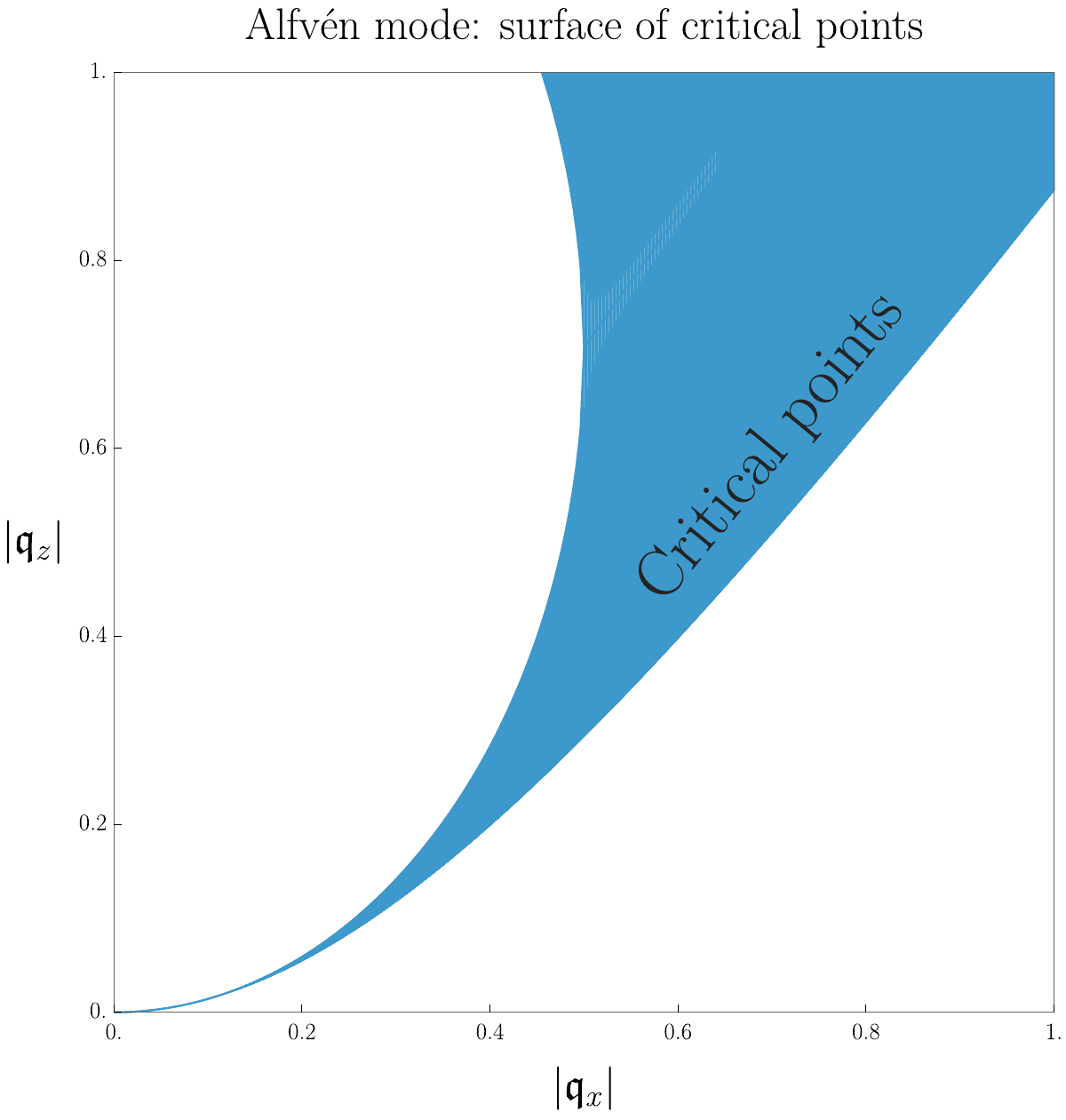}
    \caption{Plot of the critical surfaces for ${\mathfrak b} = 2$, ${\mathfrak c}=1$ and ${\cal V}_A=1/\sqrt{2}$. The critical surfaces are plotted in the space of $|\q_z|$ and $|\q_x|$ as usual. The critical surfaces $\Sigma_0$ reach the origin (as expected), and are ultimately responsible for the non-commutativity of the small-$|\k|$ and the $\theta \to \frac{\pi}{2}$ limits.}
    \label{fig:Alfven_critical_surfaces}
\end{figure}

To see this explicitly, let us identify the critical surfaces $\Sigma_0$ for the Alfv\'en modes. The discriminant in terms of $k_z$ and $k_x^2$ is given in equation~\eqref{eq:dMHD_disc_kzkx}. Written for convenience in terms of the previously-defined unitless parameters $\q_z = k_z/(2 \pi T_0)$, $\q_x=k_x/(2 \pi T_0)$, the critical surfaces $\Sigma_0$ are given by the roots of the discriminant in $\q_z$:
\begin{equation}
\label{eq:dMHD_qz_sols}
\q_z =\begin{cases}
     \pm \sqrt{\frac{2 {\cal V}_A^2 - \mathfrak{b} \mathfrak{c} \q_x^2\pm2\sqrt{{\cal V}_A^2 ( {\cal V}_A^2 - \mathfrak{b} \mathfrak{c} \q_x^2)}}{\mathfrak{c}^2}},& \mathfrak{c} \neq 0\,\,\&\,\,\b \neq 0\\
     \pm \sqrt{ \frac{\mathfrak{b}^2}{4 {\cal V}_A^2} \q_x^4},& \mathfrak{c} = 0\,\,\&\,\, \b \neq 0\\
     \pm \frac{\sqrt{2}}{|\c|} \sqrt{{\cal V}_A^2 \pm {\cal V}_A^2},& \c \neq 0 \,\, \& \,\, \b = 0\\
     0, & \c = 0 \,\,\&\,\, \b = 0
    \end{cases}\,\,\,,
\end{equation}
where $\mathfrak{b} = \lr{\lr{r_{\parallel} \mu_0/\rho_0} - \lr{\eta_{\perp}/(\epsilon_0 + p_0)}}(2 \pi T_0)$ and $\mathfrak{c}=\lr{\lr{r_{\perp} \mu_0/\rho_0} - \lr{\eta_{\parallel}/\lr{\epsilon_0+p_0}}}(2 \pi T_0)$. Note that $\mathfrak{b}=0$ corresponds to the fine-tuned case in which the non-commutativity vanishes, while $\c=0$ is another fine-tuned limit. The $\b\neq 0$, $\c \neq 0$ critical surfaces are plotted in the space of $|\q_x|$, $|\q_z|$ with representative values in Figure~\ref{fig:Alfven_critical_surfaces}. 

The intersections of the critical surfaces~\eqref{eq:dMHD_qz_sols} with the expansion surfaces $\Sigma_r$ may be found by substituting the parametrization~\eqref{eq:Hopf_extended} into the equations~\eqref{eq:dMHD_qz_sols}, and solving for the magnitude $|r|$. If there are no real, positive solutions for the magnitude $|r|$, then there is no intersection with the expansion surfaces. The non-zero solutions (since the expansion surface $\Sigma_r$ and the critical surface $\Sigma_0$ trivially intersect at the origin) for the magnitude $|r|$ are given by
\begin{equation}
\label{eq:dMHD_convR}
    |r| = \begin{cases}
        \pm \frac{2 {\cal V}_A}{\cos(\theta)(\c + \b \tan^2(\theta))}e^{-i \xi}, & \c \neq 0\,\,\&\,\,\b \neq 0\\
        \pm \frac{2}{\b{\cal V}_A \tan(\theta) \sin(\theta) } e^{- i \xi}, & \c = 0\,\,\&\,\,\b \neq 0\\
        \pm \frac{2{\cal V}_A}{\c\cos(\theta)} e^{-i \xi}, & \c \neq 0 \,\,\& \,\, \b = 0
          \end{cases}\,\,\,.
\end{equation}
In all three cases, there are real, positive solutions for $|r|$ when $\xi = 0,\, \pm \pi$ (depending on the sign). The intersection point is given by the real, positive value of $|r|$, which also determines the radius of convergence as a function of $\theta$. In the limit $\theta \to \frac{\pi}{2}$, $|r| \to 0$ in the first two cases, and so the radius of convergence goes to zero. In the third case, $|r|$ diverges\footnote{For large-enough $r$, the first order theory breaks down.} as $\theta \to \frac{\pi}{2}$. We have not included the $\b=\c=0$ case in equation~\eqref{eq:dMHD_convR}, because $|r| = 0$ for $\theta \neq \frac{\pi}{2}$, and then becomes completely unconstrained when $\theta = \frac{\pi}{2}$. 

Note that in the third equation of~\eqref{eq:dMHD_qz_sols}, the critical surfaces collapse to planes of constant $\q_z$ at $\q_z = 0$ and $\q_z = \pm 2 |{\cal V}_A/\mathfrak{c}|$ for all $\q_x$. The limit $\theta \to \frac{\pi}{2}$ corresponds to $\q_z = 0$, and so the expansion in $\q_x=r$ is an expansion entirely within the critical surface. As the discriminant is zero on the critical surface, the solutions~\eqref{eq:dMHD_Alfven_sol} become entire functions in $\q_x$, and the non-commutativity vanishes. This is the reason that the radius of convergence diverges in the limit $\theta \to \frac{\pi}{2}$ for the $\b = 0$, $\c \neq 0$ case of equation~\eqref{eq:dMHD_convR}.

Finally, let us recall that in the case of the boosted sound mode, the non-commutativity of limits in the $\q_x$, $\q_z$ expansions led to a blowup if the limits were taken in the ``wrong" order. This is again the case for the Alfv\'en channel; for $\mathfrak{b} \neq 0$, expanding~\eqref{eq:dMHD_Alfven_sol} to ${\cal O}\lr{r^3}$ (which is, of course, beyond the regime of validity of the first-order theory) yields a term proportional to $\cos^{-1/2}(\theta)$, which diverges in the limit $\theta \to \frac{\pi}{2}$. Each further term of order $r^{2n+1}$ is proportional to $(\cos(\theta))^{\frac{1-2n}{2}}$. We see once again that a blowup comes from the radius of convergence going to zero.

We have now seen three example systems: the boosted shear mode, which had no branch point at the origin; the boosted sound mode, which had a branch point at the origin that did not affect the small-$|\k|$ expansion; and finally, the Alfv\'en mode, which had a branch point at the origin which caused the radius of convergence of the small-$|\k|$ expansion to go to zero when $\theta \to \frac{\pi}{2}$ (except for in a fine-tuned case). In all of the examples, we considered a first-order truncation of the hydrodynamic derivative expansion. We will now demonstrate that these discussions are not merely a product of the first-order theory by considering a hydrodynamic theory where the underlying microscopic behaviour is well understood, namely an ${\cal N}=4$ SYM plasma at strong coupling.

\section{Holographic calculation}
\label{sec:holography}

In this section, we evaluate the quasinormal mode structure of boosted black branes in a gravitational system that is a low-energy limit of the dual to ${\cal N} = 4$ SYM. We first look at the shear channel, and then at the sound channel. In the shear channel, there is no critical point at the origin, and so one can straightforwardly find (for fixed $|\vo|$) the $\theta$-dependent radius of convergence, where $\vo$ is the boost velocity and $\theta = \arccos(\hat{{\mathbf v}}_0{\cdot}\hat{\k})$ as before. On the other hand, for the boosted sound mode, there is a critical point at the origin. Therefore, following the intuition developed in Section~\ref{sec:examples}, we will show evidence that for any $|\qv| = \frac{|\k|}{2\pi T}$ (where $T$ is the Hawking temperature of the black brane), there will \textit{always} be a complex value of $\q_z = \frac{k_z}{2 \pi T}$, $\q_x=\frac{k_x}{2 \pi T}$ at which the sound modes of the boosted black brane collide. 

\subsection{Background}
The dual theory to ${\cal N}=4$ SYM in four dimensions is type IIB string theory compactified on AdS$_5 \times S^5$~\cite{Maldacena:1997re,Aharony:1999ti}. Certain observables in ${\cal N}=4$ SYM are well described in the strong-coupling limit by classical Einstein gravity with a negative cosmological constant. The Einstein-Hilbert action is given by
\begin{equation}\label{eq:EH_action}
    S_{\rm EH} = \frac{1}{16 \pi G_N} \int d^5x \sqrt{-g} \lr{ R - 2 \Lambda}\,,
\end{equation}
where $G_N$ is the five-dimensional Newton's constant, $R$ is the five-dimensional Ricci scalar, and $\Lambda$ is the (negative) cosmological constant, which for AdS$_{D}$ is related to the AdS radius $L$ by
\begin{equation}
    \Lambda = - \frac{\lr{D-1}\lr{D-2}}{2 L^2}\,.
\end{equation}
Let us work in units in which $L=1$; then for AdS$_5$, $\Lambda = -6$. Varying the action along with relevant counterterms yields the vacuum Einstein equations:
\begin{equation}
    R_{MN} - \frac{1}{2} R g_{MN} - 6 g_{MN} = 0\,.
\end{equation}
Among the solutions to this equation is the planar black hole solution, or black brane. Black hole solutions in the gravitational picture (``the bulk") correspond to thermal states in the field theory picture (``the boundary")~\cite{Natsuume2015}. Boosting the black brane generates a larger family of solutions known as the boosted black brane solutions~\cite{Bhattacharyya:2008jc,Banerjee:2008th,Erdmenger:2008rm}, with line elements given by
\begin{equation}
    ds^2 = - r^2 f(r) u_\mu u_\nu dx^\mu dx^\nu + \frac{1}{r^2 f(r)} dr^2 + r^2 \Delta_{\mu\nu} dx^\mu dx^\nu\,,
\end{equation}
where $f(r) = 1 - \frac{r_0^4}{r^4}$. In the above, $r$ denotes the AdS radial coordinate, and $x^\mu$ denotes directions which become the boundary coordinates in the $r\to \infty$ limit. The timelike vector $u^\mu = \gamma(1, v_0^j)$ denotes the boost, and $r_0$ gives the location of the event horizon. Without loss of generality, we align the boost with the $z$-axis, and denote $v_0 \equiv v^z_0$. It will be more convenient to work with coordinates $u = (r_0/r)^2$ instead of $r$; in these units, the event horizon occurs at $u=1$, and the AdS boundary is at $u=0$.

Let us now consider perturbing the black brane. Due to the symmetry of the black brane solution, the perturbations will decompose into three channels: spin 0, spin 1, and spin 2~\cite{Kovtun:2005ev}. We will refer to these respectively as the sound (1), shear (2), and scalar (3) channels. Each of these sets of perturbations will return to equilibrium with characteristic (complex) frequencies $\w = \omega/(2 \pi T)$. These frequencies depend on a wavevector $\qv = \k/(2\pi T)$ due to the planar symmetry of the black brane. 

The structure of the quasinormal modes of the black brane has been known at this point for a long time~\cite{Kovtun:2005ev}, comprising of an infinite set of quasinormal modes of increasingly negative imaginary part; this structure is colloquially known as the ``Christmas tree". Of this infinite set of quasinormal modes, three are gapless; two in the sound channel, and one in the shear channel. The remainder are all gapped. For the ${\cal N}=4$ SYM plasma, the radius of convergence in the isotropic case may be determined by complexifying $\q^2 \equiv \qv{\cdot}\qv$, and increasing $|\q^2|$ until there exists a complex phase of $\q^2$ at which the hydrodynamic modes collide with a gapped mode~\cite{Grozdanov:2019kge,Grozdanov:2019uhi}. We will now investigate the complexification of the wavevector when there is a boost; in particular, we will show that as in Section~\ref{sec:examples} there exists a continuum of collisions between the sound quasinormal modes at complex wavevector due to a critical point at the origin.

\begin{figure}[t]
    \centering
    \begin{subfigure}[t]{0.5\linewidth}
        \centering
        \includegraphics[width=\linewidth]{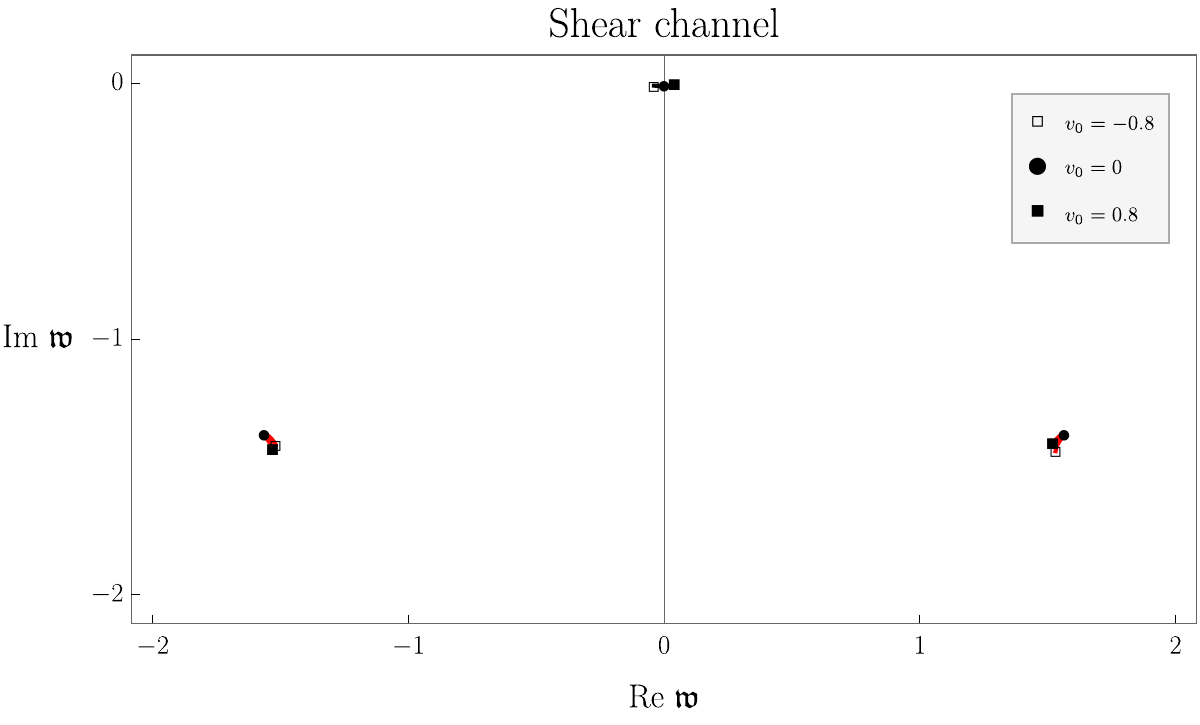}
    \end{subfigure}%
    ~
    \begin{subfigure}[t]{0.5\linewidth}
    \includegraphics[width=\linewidth]{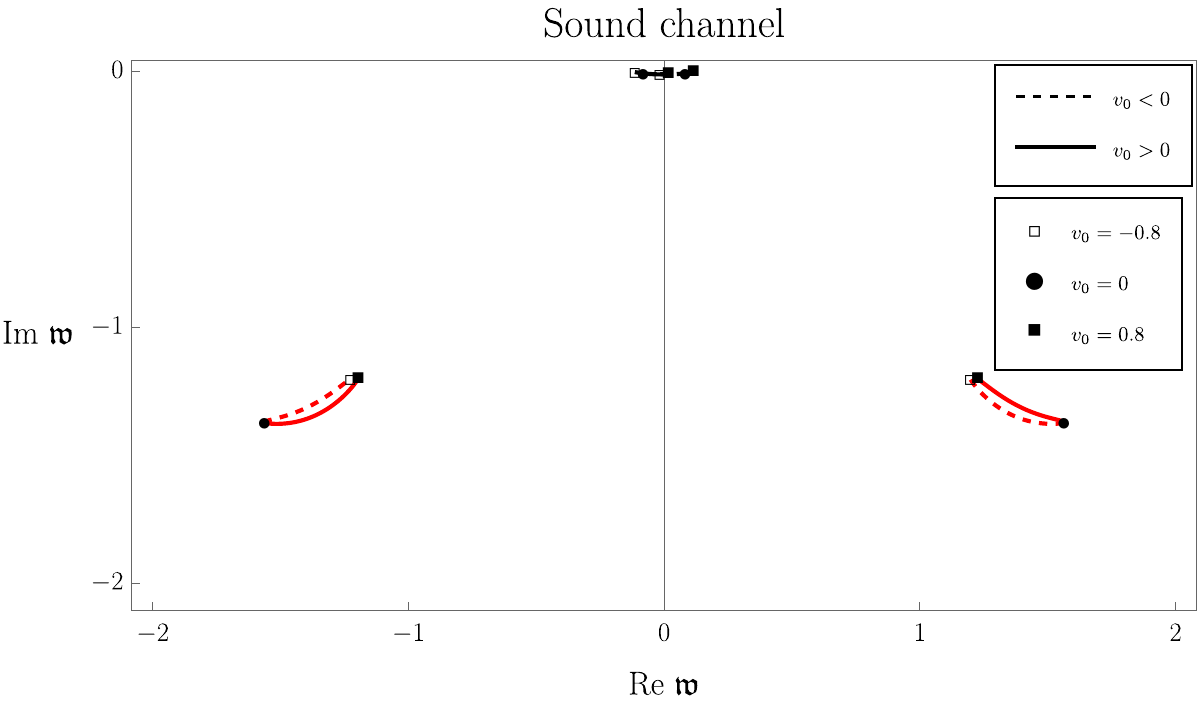}
    \label{fig:boosted_sound_channel_v0}
    \end{subfigure}
    \begin{subfigure}[b]{0.5\linewidth}
        \centering
        \includegraphics[width=\linewidth]{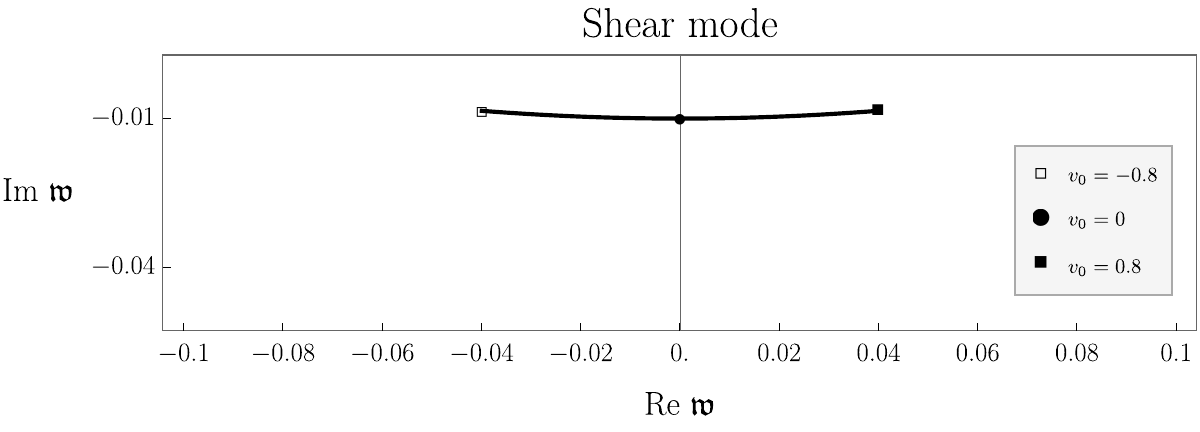}
    \end{subfigure}%
    ~
    \begin{subfigure}[t]{0.5\linewidth}
    \includegraphics[width=\linewidth]{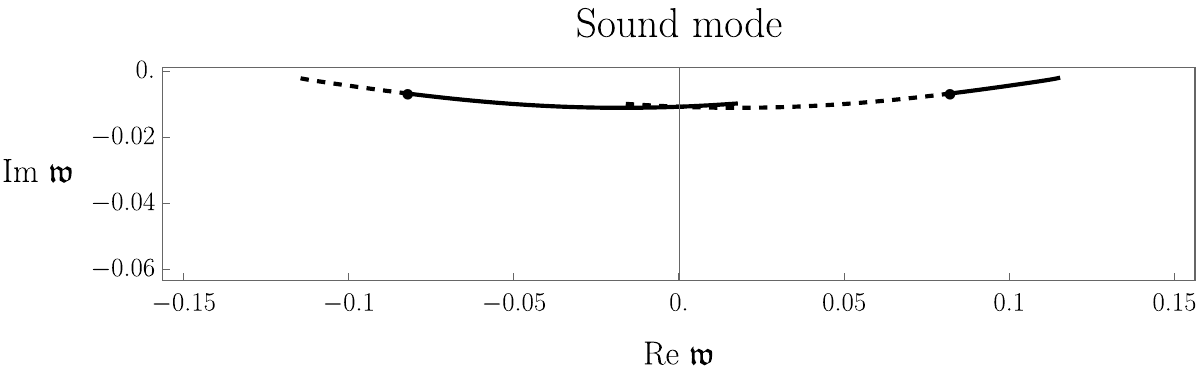}
    \label{fig:boosted_sound_mode_v0}
    \end{subfigure}
    \caption{A survey of the hydrodynamic and lowest-lying non-hydrodynamic quasinormal modes of the shear and sound channels. The top left panel shows the first few modes of the shear channel varying as $v_0$ runs from $-0.8$ to $0.8$. The bottom left panel shows the hydrodynamic shear mode in more detail. The top right panel shows the first few modes of the sound channel varying as $v_0$ runs from $-0.8$ to $0.8$. The bottom right panel shows the hydrodynamic sound modes in more detail. Both plots are at $\q_x = \q_z = 0.1$. 
    }
    \label{fig:mode_surveys}
\end{figure}

The controlling equations for the metric perturbations in each channel for the black brane at rest are of the form~\cite{Kovtun:2005ev}
\begin{equation}
\label{eq:holo_governing_eq}
    {\cal Z}_i''(u) + {\cal A}_i(u,\w,\q) {\cal Z}_i'(u) + {\cal B}_i(u,\w,\q) {\cal Z}_i(u) = 0\,,
\end{equation}
where $i=\{1,2,3\}$ refers respectively to the sound, shear, and scalar channels, and the ${\cal Z}_i(u)$ are gauge-invariant combinations of metric perturbations. In this paper, we are only interested in the channels which exhibit hydrodynamic (gapless) behaviour; therefore, we will only focus on the shear and sound channels. We will obtain the quasinormal modes of the boosted black brane by first computing the spectral curve $F(\w,\q^2)$ describing their position for the black brane at rest, and then boosting $\w \to \gamma(\w'-v_0 \, \q'_z)$, $\q^2 \to \q'^2_x + \gamma^2(\q'_z - v_0\, \w')^2$. For an overview of the sound and shear channels at rest and when boosted, refer to Figure~\ref{fig:mode_surveys}.

 \subsection{Boosted shear channel}
\label{sec:holo_shear}
For the shear channel of the black brane at rest, the controlling equation is given by equation~\eqref{eq:holo_governing_eq} with~\cite{Kovtun:2005ev,Grozdanov:2019kge}
\begin{subequations}
    \begin{align}
        {\cal A}_2(u,\w,\q) &= - \lr{\frac{(\w^2 - \q^2 f(u)) f(u) - u \w^2 f'(u)}{u f(u) ( \w^2 - \q^2 f(u))}}\,,\\
        {\cal B}_2(u,\w,\q) &= \frac{\w^2 - \q^2 f(u)}{u f(u)^2}\,,
    \end{align}
\end{subequations}
where in $u$ coordinates, $f(u)= 1-u^2$. There exist several means of extracting the quasinormal modes from equation~\eqref{eq:holo_governing_eq}; in this paper, we have used the Frobenius method (see Appendix B of \cite{Kovtun:2005ev}). As our interest is primarily in the behaviour of the hydrodynamic mode, we do not need to go to very high order in the expansion to see convergence, and so take $N=35$ for the shear mode.

\begin{figure}
    \centering
    \begin{subfigure}[t]{0.45\linewidth}
        \centering
        \includegraphics[width=\linewidth]{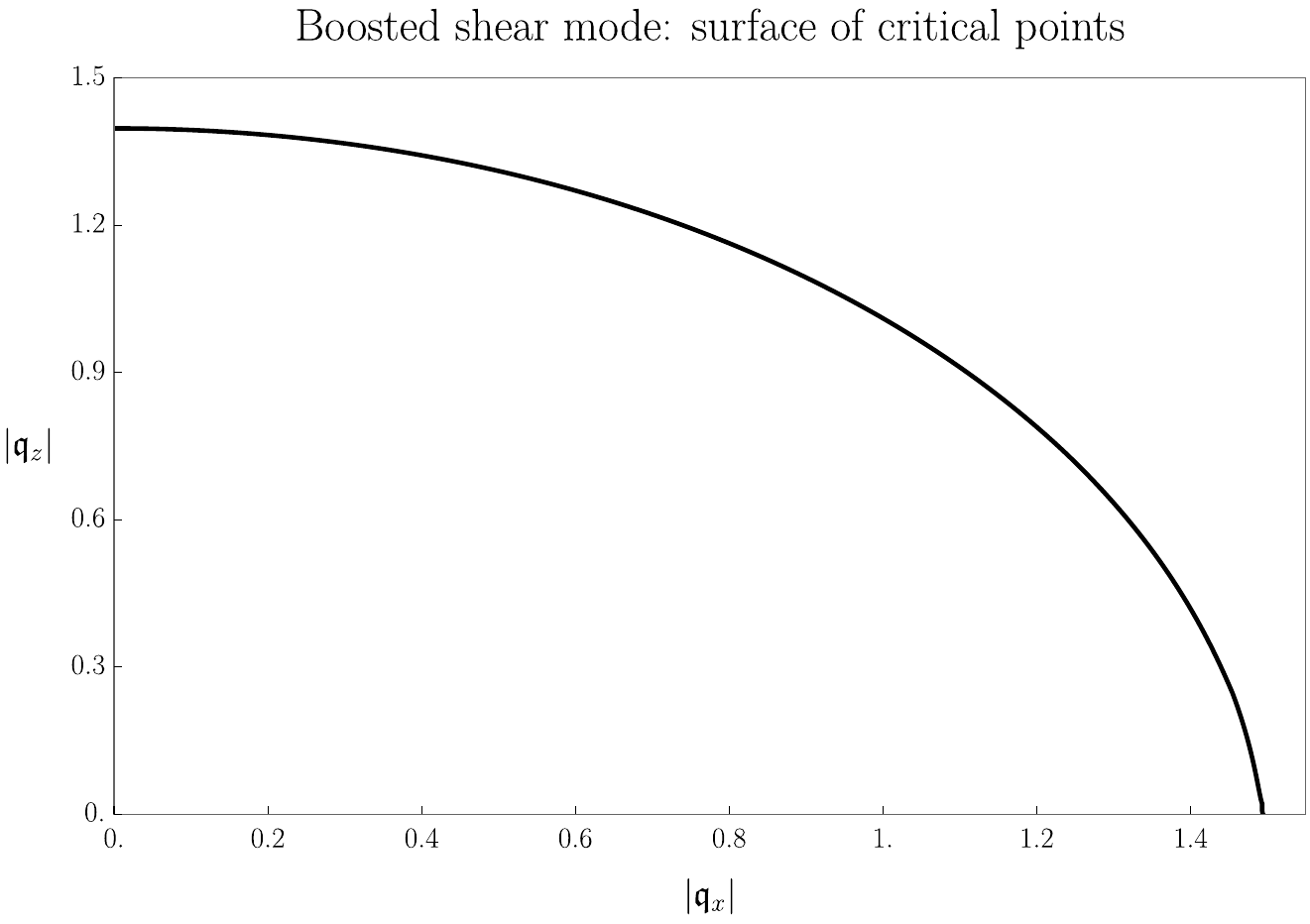}
    \end{subfigure}%
    ~
    \begin{subfigure}[t]{0.45\linewidth}
    \includegraphics[width=\linewidth]{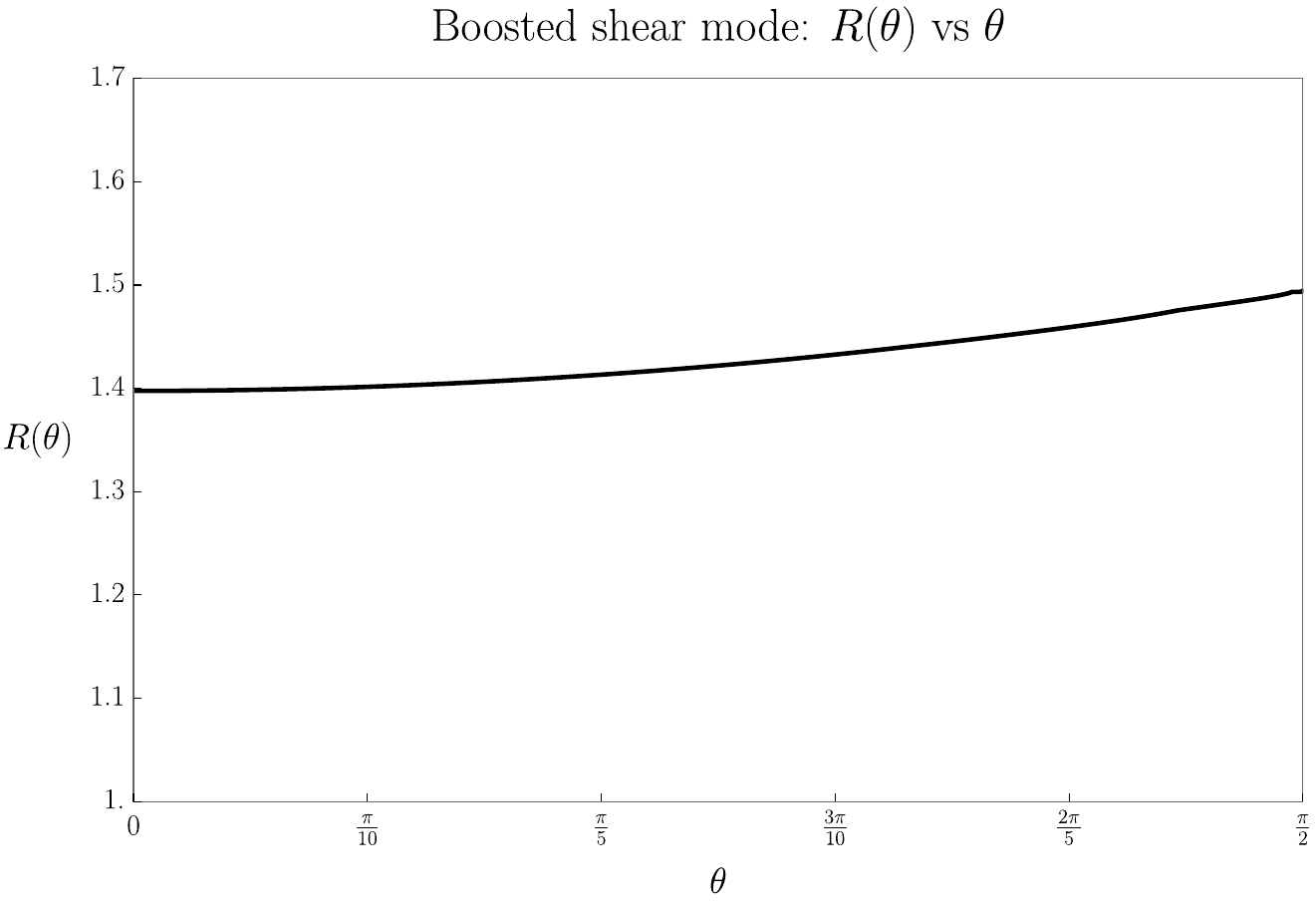}
    \end{subfigure}
    \begin{subfigure}[t]{0.45\linewidth}
    \includegraphics[width=\linewidth]{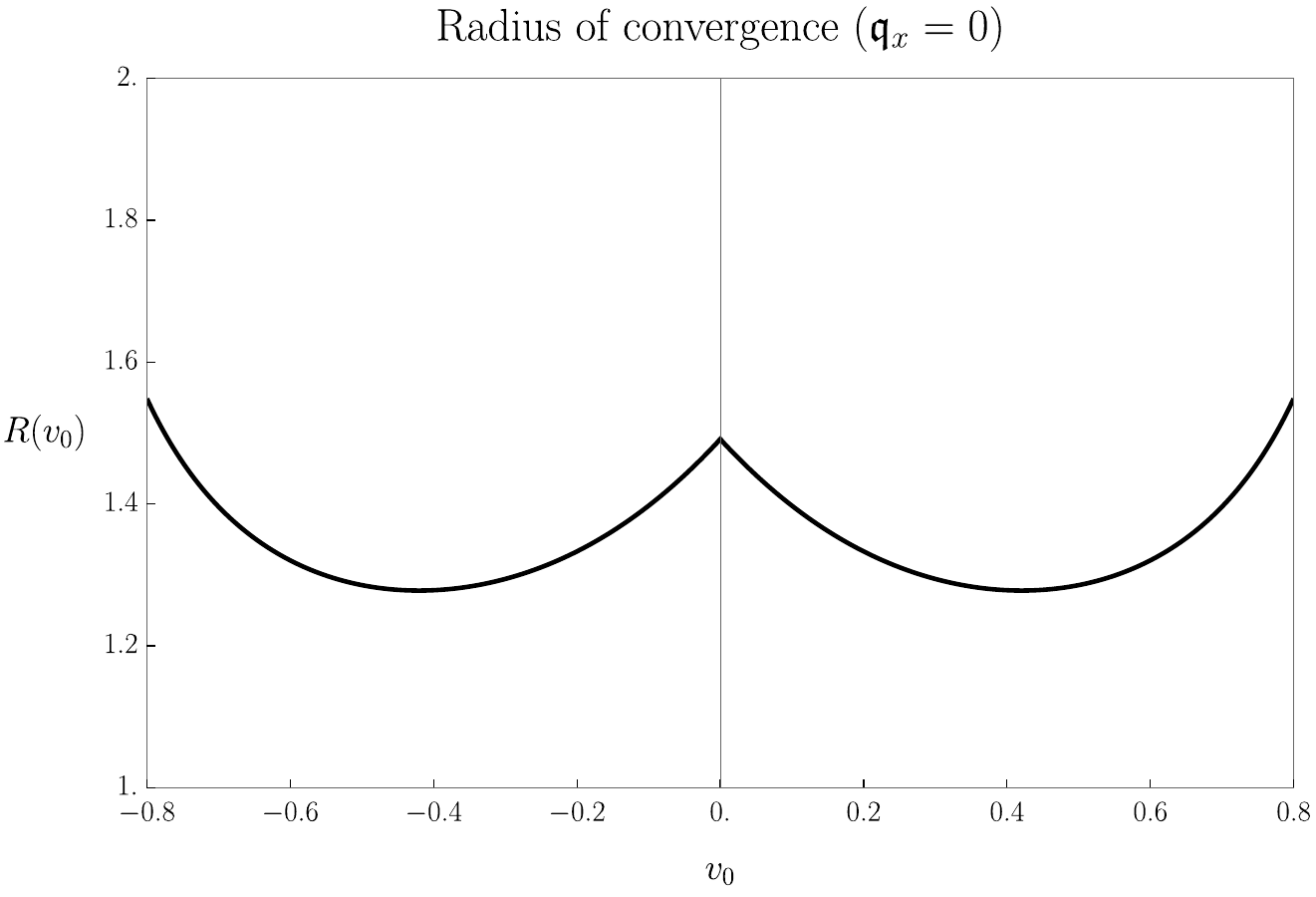}
    \end{subfigure}%
    ~
    \begin{subfigure}[t]{0.45\linewidth}
    \includegraphics[width=\linewidth]{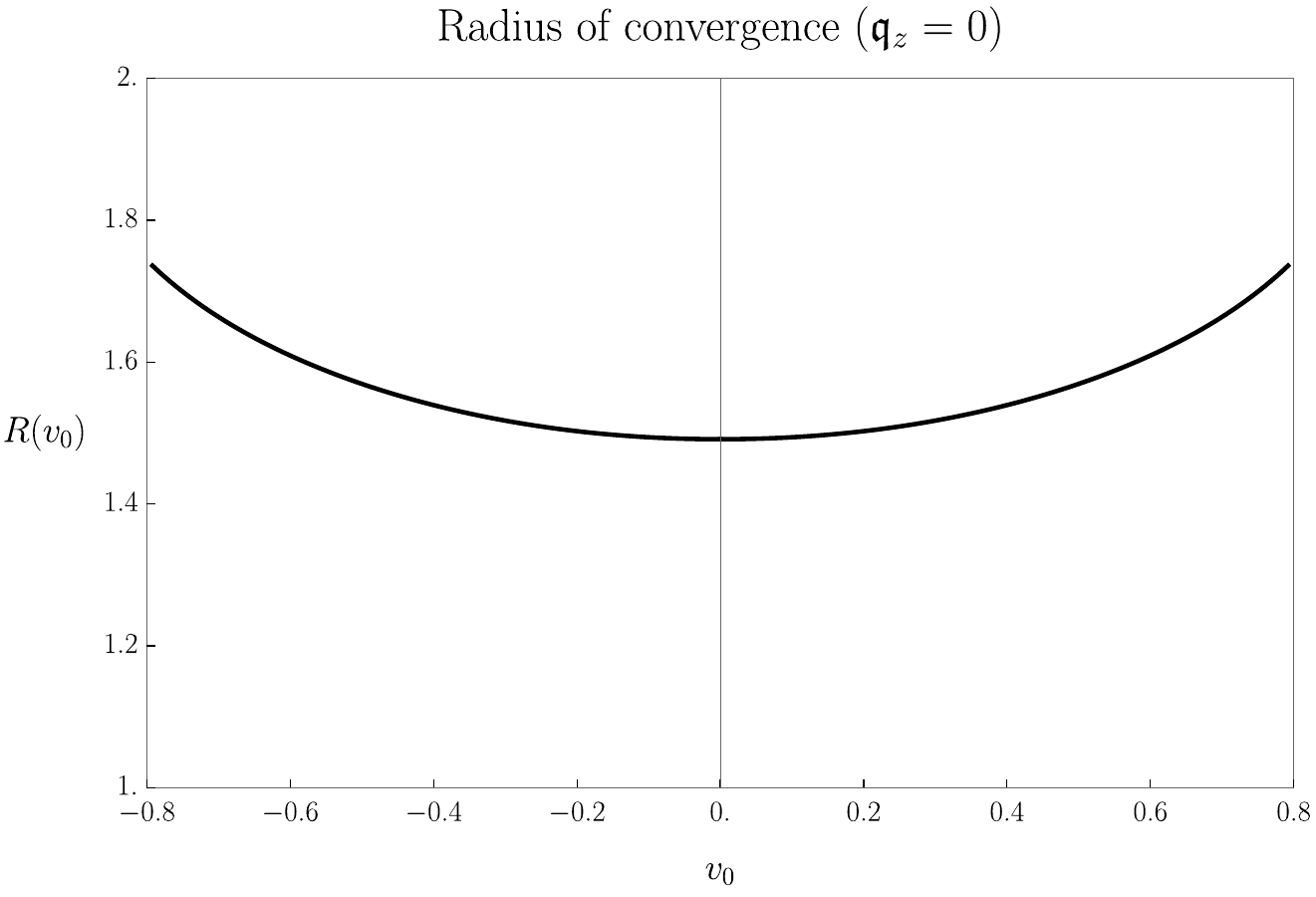}
    \end{subfigure}
    \caption{A set of plots depicting the radii of convergence in the boosted shear mode at certain example values. In the top left, the lower hull of the surface of critical points is plotted against $|\q_z|$ and $|\q_x|$ for $v_0=0.1$. In the top right plot, the radius of convergence $R$ is plotted against $\theta$ for $v_0=0.1$. In the bottom left plot, the radius of convergence is plotted as a function of $v_0$ for $\q_x=0$. In the bottom right plot, the radius of convergence is plotted as a function of $v_0$ for $\q_z=0$. Note that the $R$-axes in the final three plots begin at $1$ to more clearly display detail. The slight kink at the right edge of the first and second plots is a numerical artifact.}
    \label{fig:shear_convergence}
\end{figure}

The radius of convergence of the hydrodynamic expansion is set by collisions of gapless quasinormal modes. Based on the intuition developed in Section~\ref{sec:examples}, we should expect the radius of convergence to depend on the direction the wavevector points relative to the boost velocity. Additionally, there are no collisions between quasinormal modes when $\qv=0$, and so there is no critical surface $\Sigma_0$ going through the origin.

However, there are still collisions for large enough $|\qv|$ described by a critical surface. This critical surface represents a continuum of collisions between the shear quasinormal mode and the lowest-lying non-hydrodynamic quasinormal modes, and sets the radius of convergence of the hydrodynamic expansion. Note that, as there is no mode collision at $\qv=0$, one can define a neighbourhood of the origin $\qv=0$ on which $\w(\q_z,\q_x)$ admits a Taylor series expansion in $\q_z,\q_x$. 

The domain of convergence of the Taylor series may be plotted as a function of $|\q_z|$ and $|\q_x|$, as in Section~\ref{sec:examples}. For some particular examples, refer to Figure~\ref{fig:shear_convergence}. For the first plot, the space of complex $\q_x$, $\q_z$ has been projected onto the subspace spanned by $|\q_x|$, $|\q_z|$ which characterizes the Reinhardt domain. The distance to the black line from the origin for a particular angle $\theta$ gives the radius of convergence in that direction. In the isotropic case, the radius of convergence was found~\cite{Grozdanov:2019uhi} to be at $|\q^2| \approx 2.224$. We see that in the case $\k \perp \vo$, the radius of convergence straightforwardly grows with $|v_0|$. However, when $\k \parallel \vo$, the radius of convergence initially decreases with growing $|v_0|$, before turning around at $|v_0| \approx 0.42$.

\subsection{Boosted sound channel}

\label{sec:holo_sound}
\begin{figure}[t]
    \centering
\begin{subfigure}[t]{0.5\linewidth}
    \includegraphics[width=\linewidth]{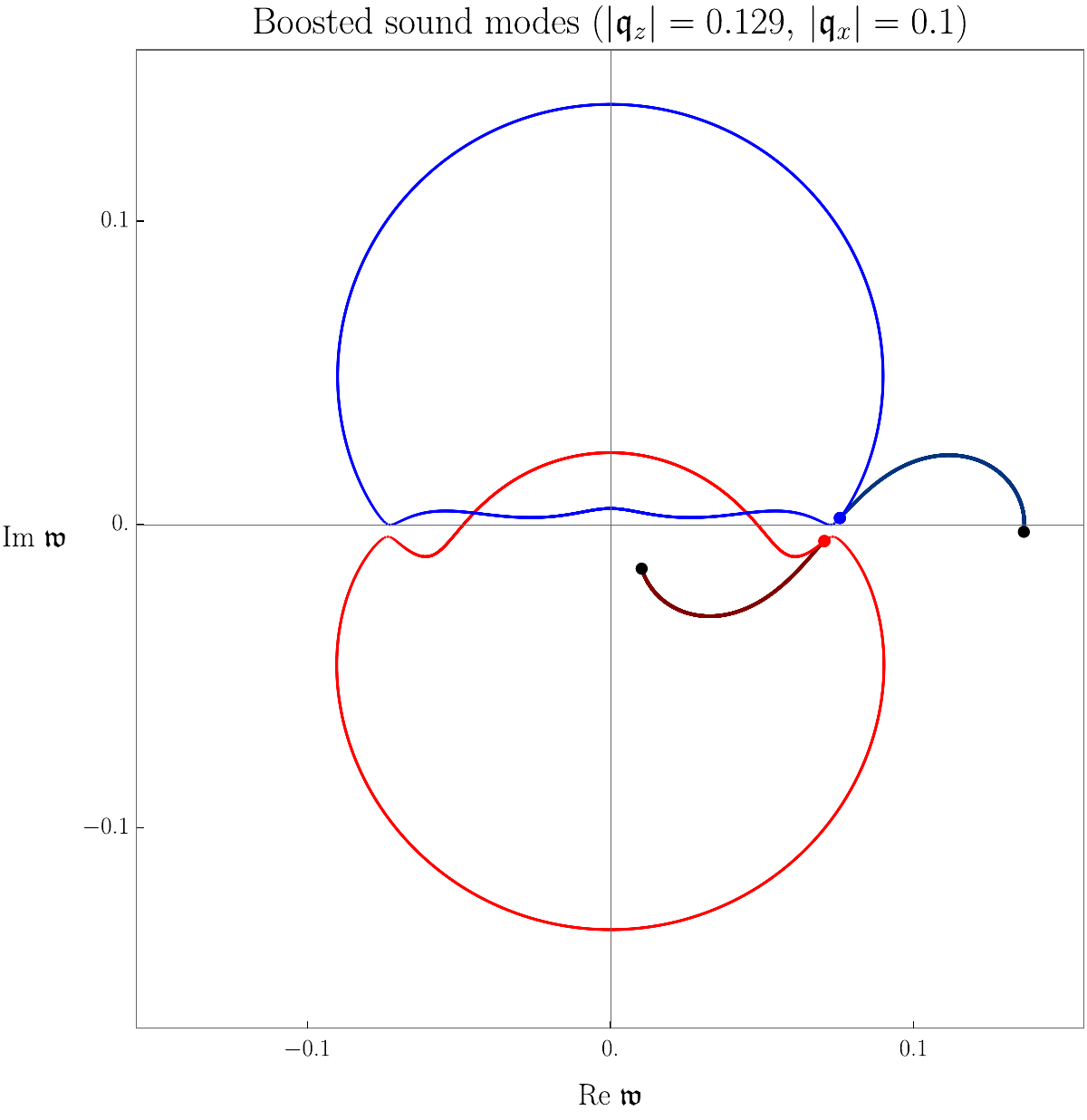}
    \label{fig:boosted_sound_holo_129}
    \end{subfigure}%
    ~
    \begin{subfigure}[t]{0.5\linewidth}
    \includegraphics[width=\linewidth]{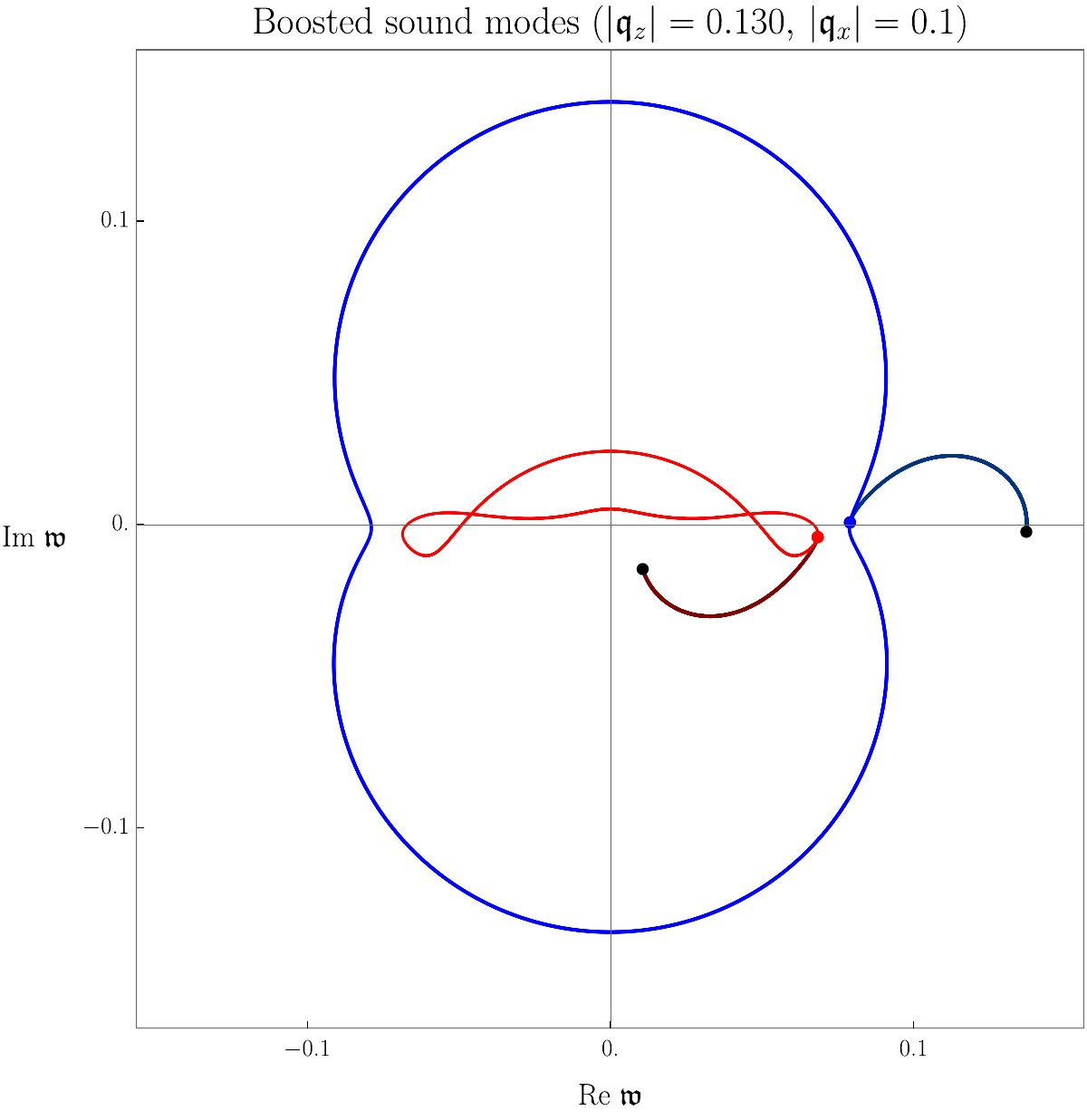}
    \label{fig:boosted_sound_holo_130}
    \end{subfigure}
    \caption{The sound quasinormal modes of the boosted black brane with a boost $v_0=\frac{1}{\sqrt{2}}$. In both plots, the initial positions of the sound modes at real $\q_z$, $\q_x$ are marked with a black dot. The dark red and dark blue lines indicate the path of the modes as $\arg(\q_x)$ is taken from $0$ to $\frac{\pi}{2}$ with $\q_z$ real. The endpoints are marked with red and blue dots. The light red and blue lines denote the complexification of $\q_z$ with $\q_x$ held purely imaginary. The argument of $\q_z$ runs from $0$ to $2\pi$ . The left-hand plot is for $|\q_z|$ slightly below the value of $|\q_z|$ that yields a pole collision for $\q_x = 0.1\, i$ (i.e. $|\q_z| \approx 0.1291$). The right-hand plot is for $|\q_z|$ slightly above the collision value. The behaviour of the sound modes is qualitatively different before and after the collision.}
    \label{fig:boosted_sound_holo_129130}
\end{figure}

\begin{figure}[t]
    \centering
    \begin{subfigure}[t]{0.3\linewidth}
    \centering
    \includegraphics[width=\linewidth]{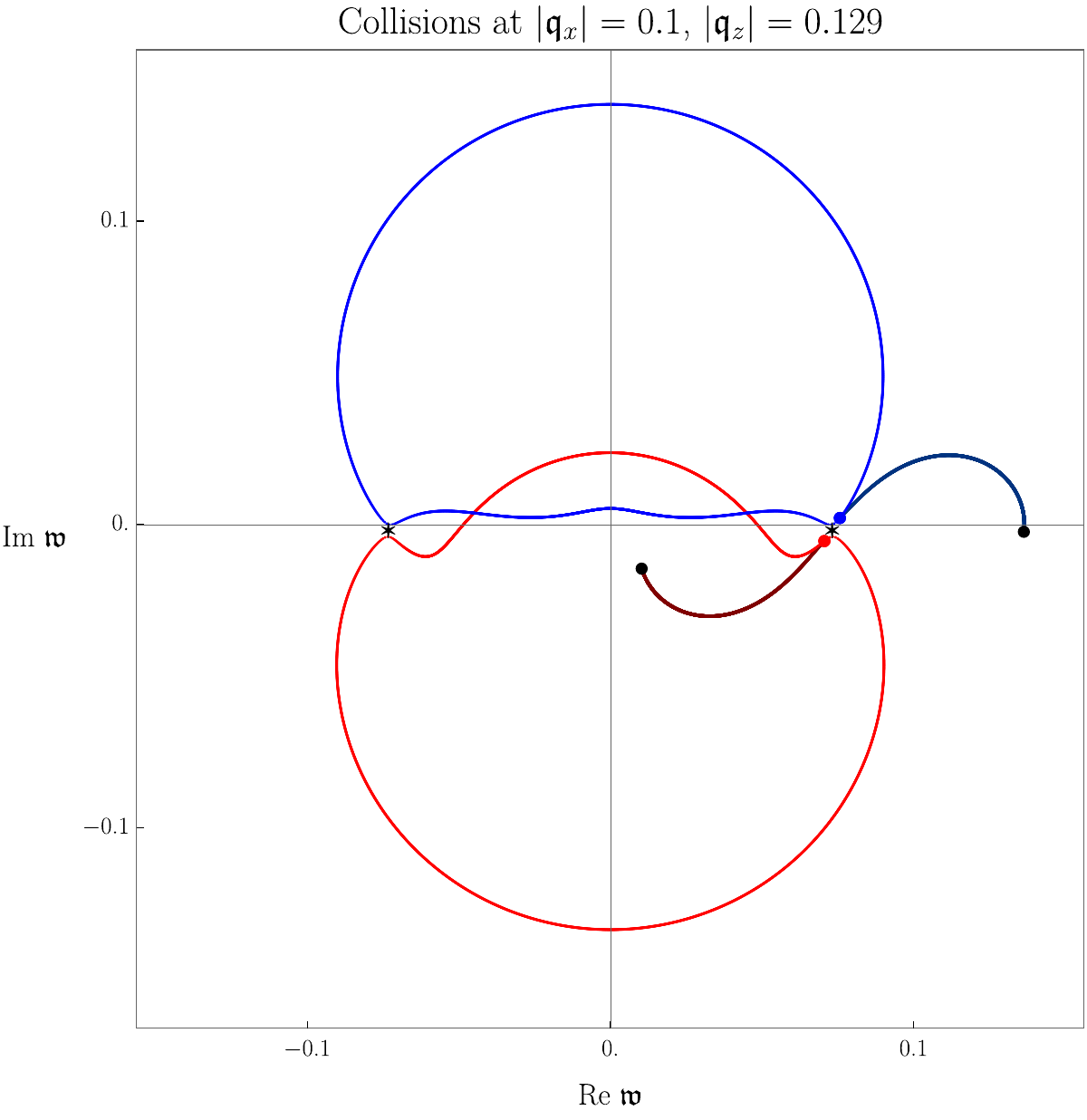}
    \label{fig:boosted_sound_holo_collision_010}
    \end{subfigure}%
    ~
    \begin{subfigure}[t]{0.3\linewidth}
    \centering
    \includegraphics[width=\linewidth]{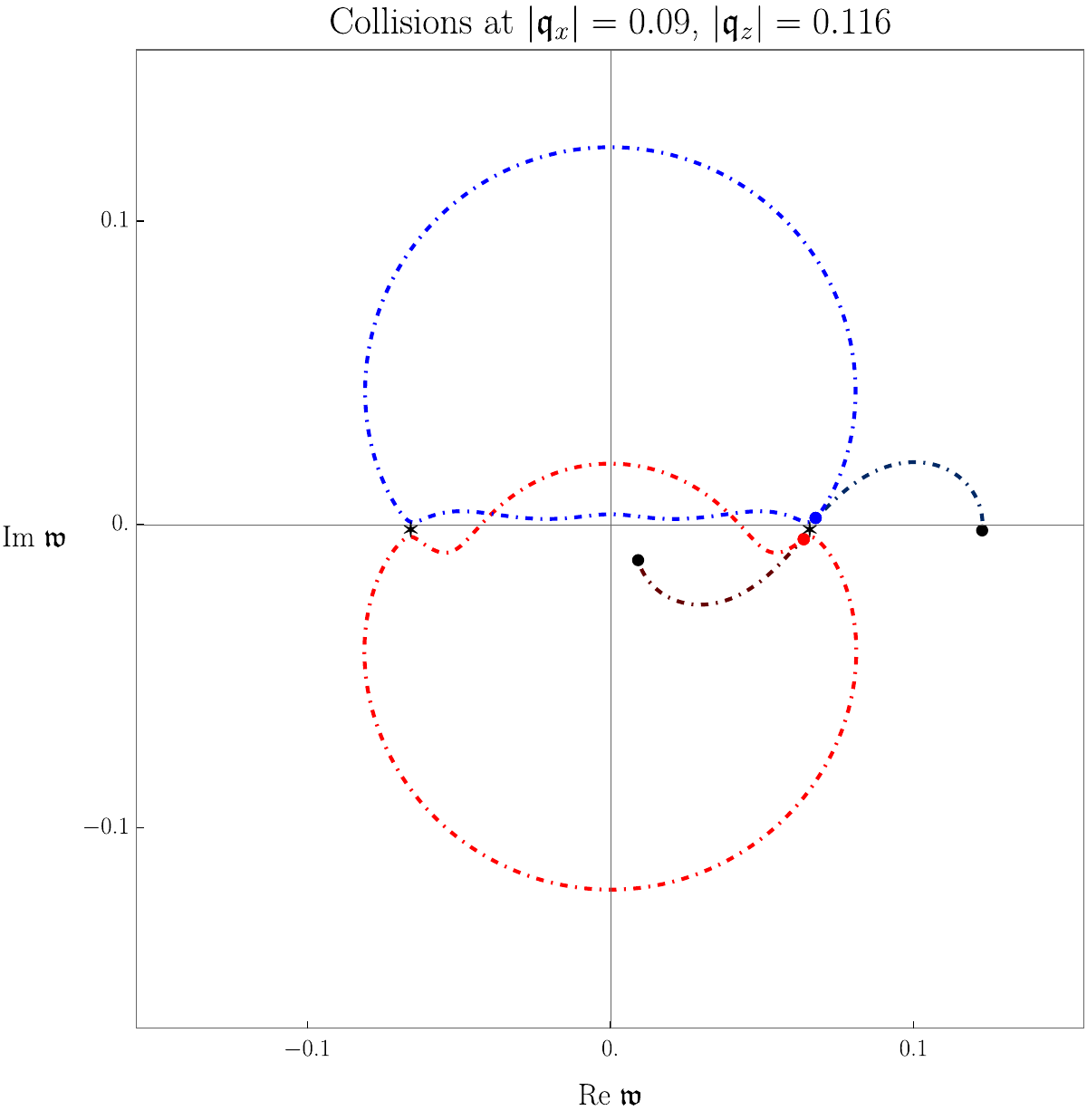}
    \label{fig:boosted_sound_holo_collision_009}
    \end{subfigure}%
    ~
    \begin{subfigure}[t]{0.3\linewidth}
    \centering
    \includegraphics[width=\linewidth]{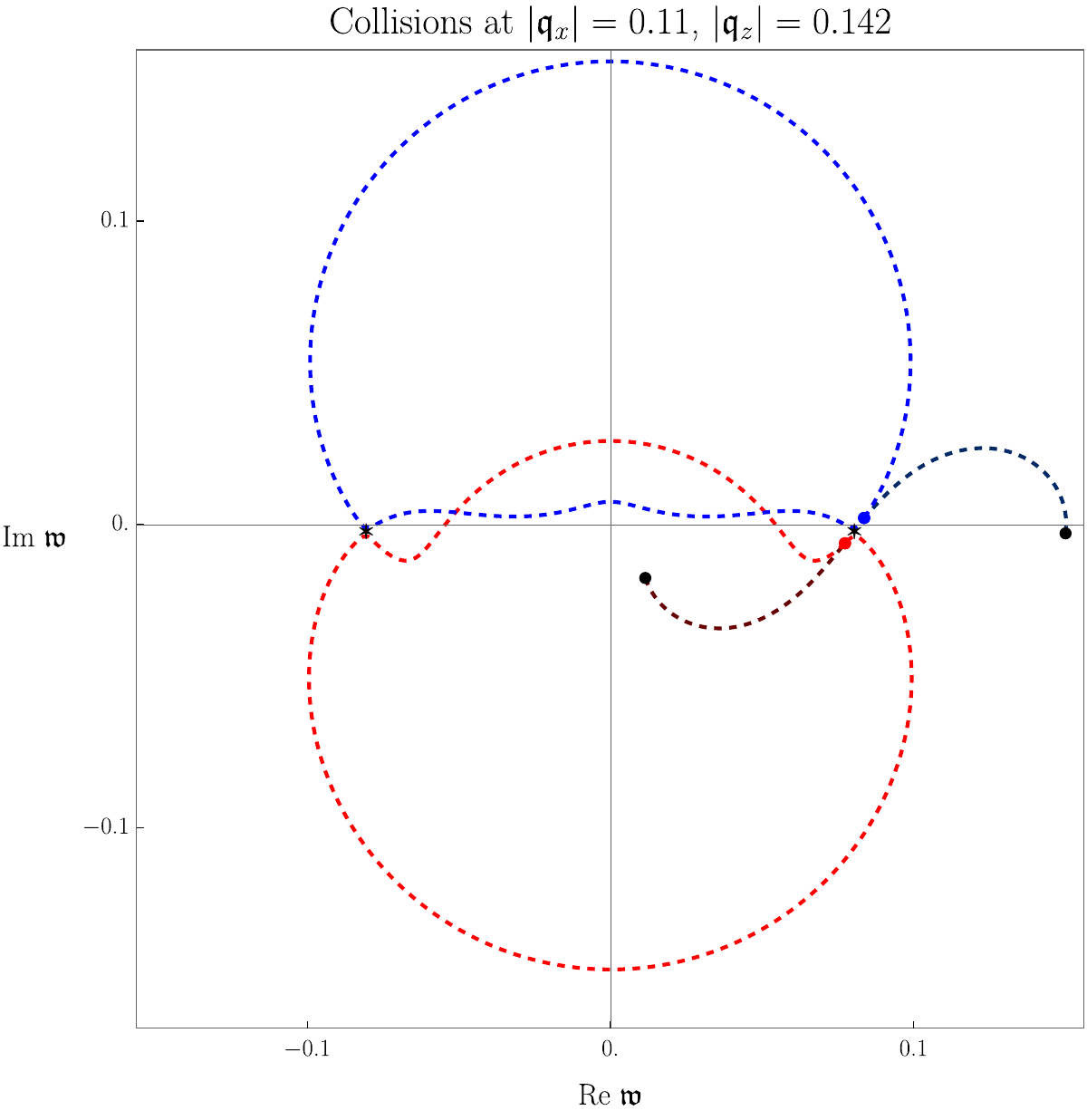}
    \label{fig:boosted_sound_holo_collision_011}
    \end{subfigure}%
    \newline
    \begin{subfigure}[t]{0.3\linewidth}
    \centering
    \includegraphics[width=\linewidth]{Figures/PNG_figs/Holography_Collisions_010.png}
    \label{fig:boosted_sound_holo_collision_010_phase}
    \end{subfigure}%
    ~
    \begin{subfigure}[t]{0.3\linewidth}
    \centering
    \includegraphics[width=\linewidth]{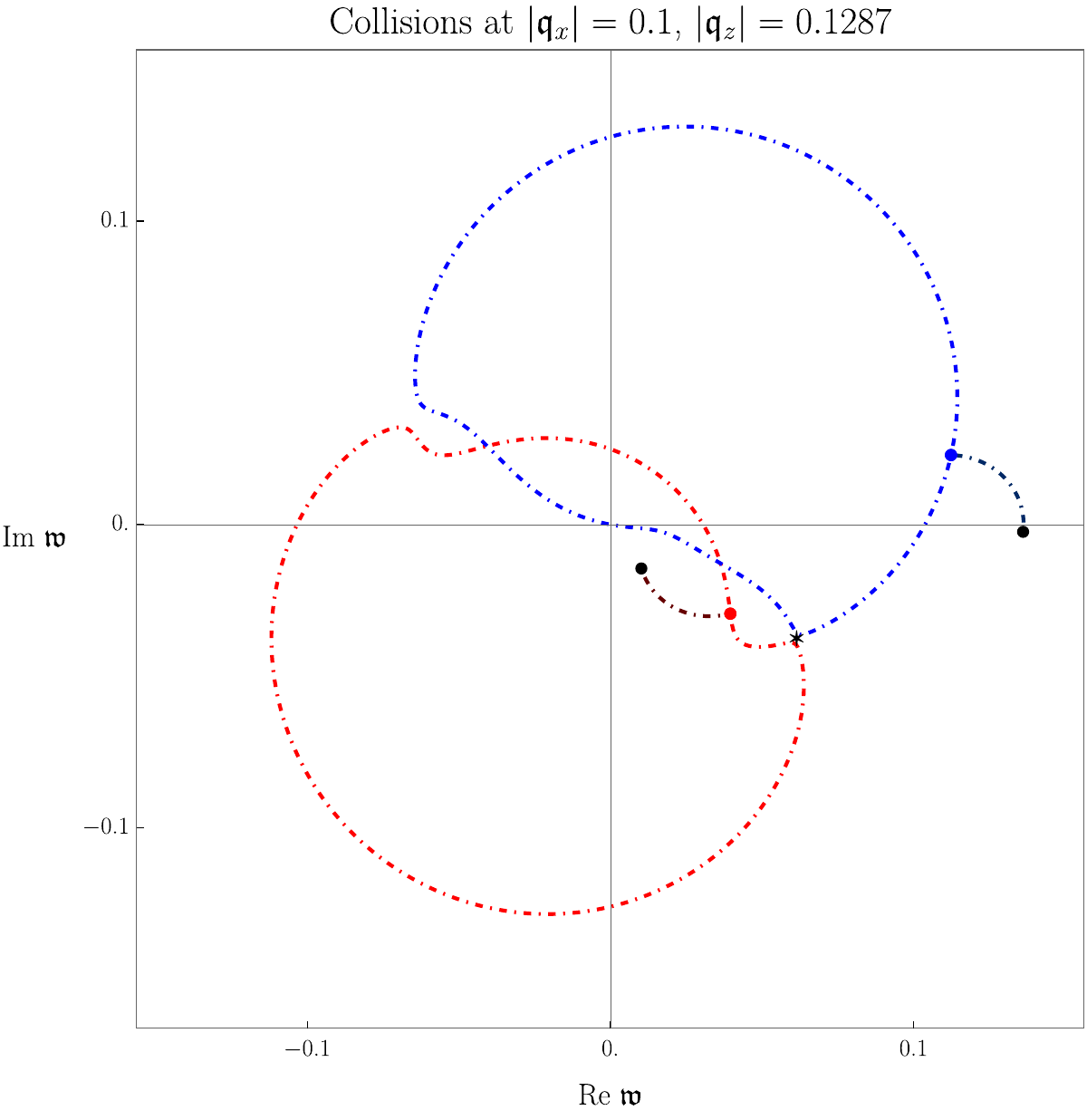}
    \label{fig:boosted_sound_holo_collision_010_pi3}
    \end{subfigure}%
    ~
    \begin{subfigure}[t]{0.3\linewidth}
    \centering
    \includegraphics[width=\linewidth]{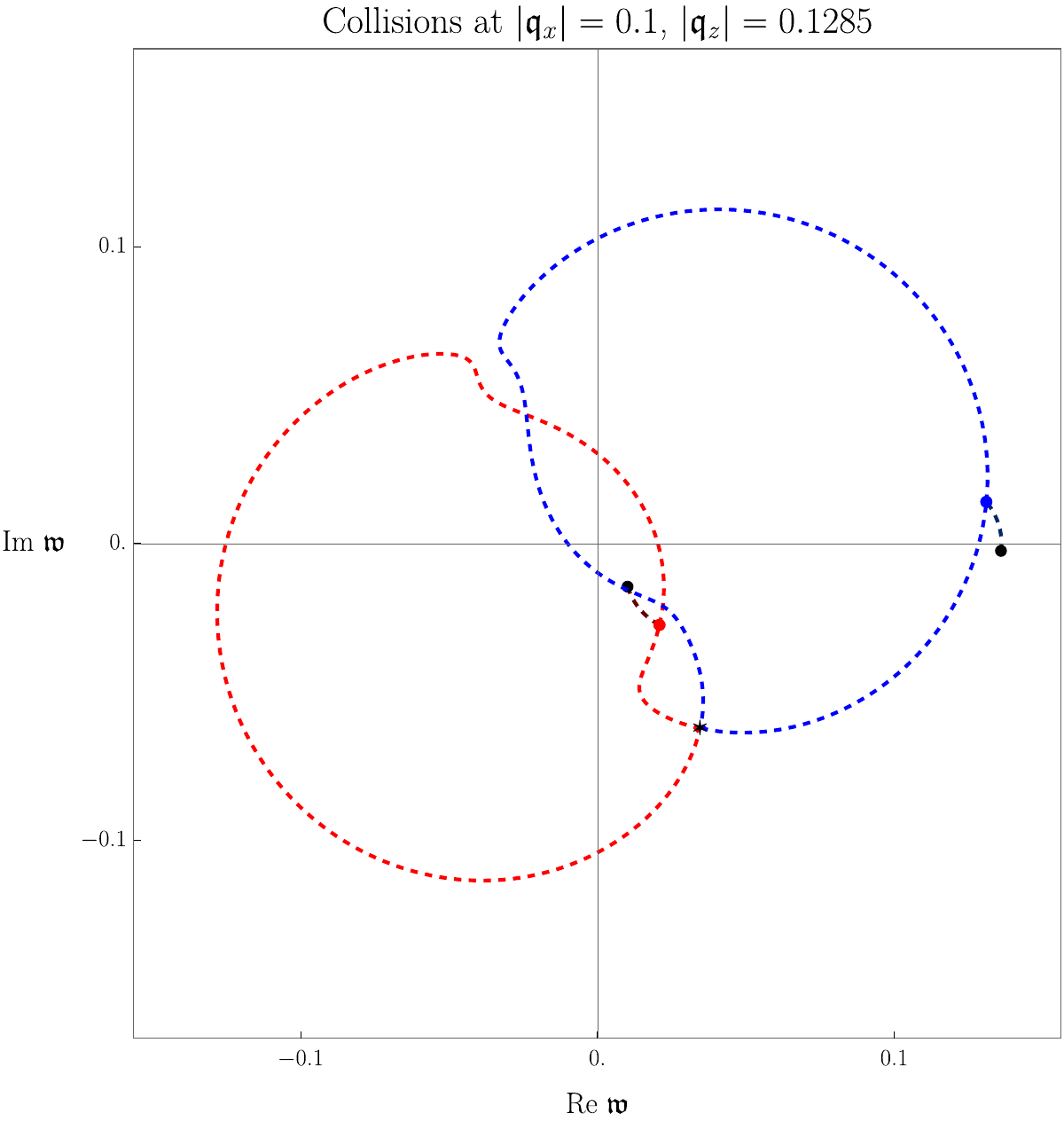}
    \label{fig:boosted_sound_holo_collision_010_pi6}
    \end{subfigure}
    \caption{Collisions between sound modes as the magnitude and phase of $\q_x$ are varied for $v_0 = \frac{1}{\sqrt{2}}$. In the first row of plots, the magnitude of $\q_x$ is adjusted, but the phase is fixed at $\pi/2$. In the second row of plots, the magnitude of $\q_x$ is fixed at $1/10$, but the phase is set to $\pi/2$, $\pi/3$, and $\pi/6$ respectively from left to right. Once again, starting positions are denoted with black dots, and the dark red and blue lines show the movement of the sound modes as the complex phase of $\q_x$ is increased to a set value. The light red and blue lines show the trajectory of the sound modes as the phase of $\q_z$ is then varied from $0$ to $2\pi$. In all cases, there exists a complex value of $\q_z$ at which a collision occurs between the sound modes, showcasing the continuum of pole collisions in the space of complex $\q_x$, $\q_z$ that exists when there is a branch point at the origin.}
    \label{fig:sound_survey_collisions}
\end{figure}

For the sound channel of the black brane at rest, the controlling equation is given by equation~\eqref{eq:holo_governing_eq} with~\cite{Kovtun:2005ev,Grozdanov:2019kge}
\begin{subequations}
    \begin{align}
        {\cal A}_1(u,\w,\q) &= - \lr{\frac{3 \w^2 (1+u^2) + \q^2 (2 u^2 - 3 u^4 - 3)}{u f(u) ( 3 \w^2 + \q^2 (u^2-3))}}\,,\\
        {\cal B}_1(u,\w,\q) &= \frac{3 \w^4 + \q^4 (3-4 u^2 + u^4) + \q^2 (4 u^5 - 4 u^3 + 4 u^2 \w^2 - 6 \w^2)}{u f(u)^2(3 \w^2 + \q^2 (u^2-3))}\,.
    \end{align}
\end{subequations}
We once again use the Frobenius method to determine the quasinormal modes, this time up to $N=25$. 

The quasinormal modes of the sound channel have a more interesting structure than the shear channel. Due to the fact that there are two sound modes, when $\qv\to 0$ there is a mode collision at the origin; therefore, the point $\q_x=\q_z=0$ is a critical point. Based on the intuition developed in Section~\ref{sec:examples}, we should expect that there exists a surface of critical points $\Sigma_0$ extending out of the origin. In other words, for any non-zero fixed values of $|\q_z|$, $|\q_x|$, we should expect that there will be a collision between the two sound quasinormal modes of the boosted black brane for at least one complex value of $\q_z$, $\q_x$. That such a collision occurs, and that there is a crossover in the behaviour of the modes, is illustrated for an example value in Figure~\ref{fig:boosted_sound_holo_129130}. 

We next want to demonstrate that the collisions are continuous. The equations\footnote{Recall that $\de_\w F(\w,\q_z,\q_x)=0$ indicates a breakdown of the implicit function theorem.}
\begin{equation}
    F(\w,\q_z,\q_x)=0, \qquad \de_\w F(\w,\q_z,\q_x)=0\,,
\end{equation} represent two complex equations in three complex variables. This leaves one complex variable free, or equivalently two real parameters. In Figure~\ref{fig:sound_survey_collisions}, we have chosen to vary the magnitude and phase of $\q_x$; for each value of the magnitude and phase, there exists a complex value of $\q_z$ such that there is a collision between the hydrodynamic sound modes, demonstrating the continuum of collisions.

We should note here that the collisions showcased in Figures~\ref{fig:boosted_sound_holo_129130},~\ref{fig:sound_survey_collisions} are of a fundamentally different nature to those that set the radius of convergence in the isotropic case~\cite{Withers:2018srf,Grozdanov:2019uhi}. In the isotropic case, the hydrodynamic sound modes collide with the lowest non-hydrodynamic modes, setting the radius of convergence of the hydrodynamic derivative expansion. In the case considered here, the sound modes are colliding with each other.

The critical question, of course, is whether these collisions between sound modes actually affect the hydrodynamic expansion. As in the equivalent first-order theory in Section~\ref{sec:examples}, the answer remains no. While the self-collisions near the origin imply that one cannot write down a formal power series of $\w$ in $\q_z$, $\q_x$, the hydrodynamic derivative expansion in terms of $|\qv|$ proceeds only on a set of embedded surfaces in the space of complex $\q_x$, $\q_z$, the expansion surfaces $\Sigma_r$. These expansion surfaces $\Sigma_r$ are given by the parametrization in equation~\eqref{eq:Hopf_extended}. The continuum of self-collisions between the hydrodynamic sound modes can only serve to limit the radius of convergence if the critical surfaces $\Sigma_0$ intersect the expansion surfaces $\Sigma_r$. This does not occur in the case of the boosted black brane. 

That the continuum of collisions between the hydrodynamic modes do not appear on the expansion surfaces $\Sigma_r$ is not surprising, and is nothing more than a product of the fact that the ``anisotropy" in this example is generated by a boost. If the critical surfaces $\Sigma_0$ had intersected the expansion surface $\Sigma_r$, the effect would likely have been visible already in the isotropic case; as discussed at the beginning of Section~\ref{sec:examples}, the analysis of dispersion relations in terms of several complex variables holds just as well in isotropic systems.

This avoidance of the expansion surface is not generic, as evidenced by the first-order magnetohydrodynamic theory we investigated in Section ~\ref{sec:examples}. The holographic setup for a thermal system with magnetic fields has been previously investigated~\cite{DHoker:2009ixq,DHoker:2009mmn,Janiszewski:2015ura,Grozdanov:2017kyl,Hofman:2017vwr}, including investigations of quasinormal modes~\cite{Ammon:2017ded} and convergence at complex wavevector~\cite{Cartwright:2024rus} for $\B \parallel \k$. We intend to return to the effects of the critical surfaces $\Sigma_0$ and the expansion surfaces $\Sigma_r$ on quasinormal modes of a magnetic black brane in the near future.

\section{Constraints on anisotropic transport coefficients from causality}
\label{sec:caus_con}
The examples illustrate that convergence properties of anisotropic systems are far richer than in the isotropic case. With the two preceding sections in mind, we will now turn to the second task of the paper: constraining anisotropic transport coefficients from above in terms of radii of convergence, as well as the inverse task of constraining radii of convergence from above in terms of anisotropic transport coefficients. This programme was initiated in~\cite{Heller:2022ejw} in the isotropic case (see also~\cite{Gavassino:2023myj} for a reinterpretation in terms of covariant stability); we now extend these constraints to anisotropic systems. We will consider two cases: dispersion relations which are holomorphic at the origin (such as the boosted shear mode), and dispersion relations with a branch point at the origin (such as the boosted sound mode, and the Alfv\'en channel in magnetohydrodynamics).

\subsection{Review of isotropic case}
We briefly review the isotropic case as described in~\cite{Heller:2022ejw}. For a system to be consistent with microscopic causality, the dispersion relations $\omega = \omega(\k)$ must satisfy the following condition:~\cite{Heller:2022ejw,Gavassino:2023myj}
\begin{equation}
\label{eq:covar_stab_condition}
    \Im(\omega(\k)) \leq |\Im(\k)|\,,
\end{equation}
for complex $\omega$ and complex vector $\k$. We will refer to condition~\eqref{eq:covar_stab_condition} as the covariant stability condition~\cite{Gavassino:2023myj}. The symbol $|\Im(\k)|$ should be read as complexifying the vector $\k$ such that $\k = \Re(\k) + i\, \Im(\k)$, and then taking the magnitude of the (real) vector $\Im(\k)$. Let us now take advantage of isotropy and write $\k = (k_x,0,...,0)$. Let us further write $k_x = r\, e^{i\xi}$, where $r = |k_x|$ and $\xi = \arg(k_x)$. Then the condition~\eqref{eq:covar_stab_condition} becomes
\begin{equation}
    \label{eq:covar_stab_isotropic}
    \Im(\omega(k_x)) \leq r |\sin(\xi)|\,.
\end{equation}
Let us now assume that $\omega(k_x)$ admits a power series solution in $k_x$ with non-zero radius of convergence $R$, given by $\omega = \sum_{n=0}^\infty c_n k_x^n$, where $c_n \in \mathbb{C}$. Then, writing $c_n = \alpha_n + i \beta_n$ ($\alpha_n,\beta_n\in \mathbb{R}$), the expansion can be expressed as
\begin{equation}
    \omega = \sum_{n=0}^\infty r^n  \biggl[(\alpha_n \cos(n \xi) - \beta_n \sin(n \xi)) + i (\beta_n \cos(n\xi) + \alpha_n \sin(n\xi)) \biggr] \equiv U(r,\xi) + i\, V(r,\xi)\,.
\end{equation}
Using elementary Fourier analysis, for $n>0$ one can show
\begin{equation}
    \alpha_n r^n = \frac{1}{\pi} \int_0^{2\pi} V(r,\xi) \sin(n\xi) d\xi, \qquad \beta_n r^n = \frac{1}{\pi} \int_0^{2\pi} V(r,\xi) \cos(n\xi)\,d\xi\,.
\end{equation}
Let us now consider an individual term $|c_n| r^n$. This term may be re-written as
\begin{equation}
    |c_n| r^n = (\alpha_n+i\beta_n)e^{-i\xi_n} r^n = \frac{1}{\pi} \int_{0}^{2\pi} V(r,\xi) \sin(n\xi + \xi_n)\,d\xi\,,
\end{equation}
where $\xi_n = \arg(c_n)$, and we have made use of the fact that $|c_n|r^n$ is real. We would like to make use of condition~\eqref{eq:covar_stab_isotropic}, but $\sin(n\xi+\xi_n)$ is not positive-definite. We may then instead consider
\begin{equation}
\begin{split}
    |c_n| r^n + 2 \beta_0 &= \frac{1}{\pi} \int_0^{2\pi} V(r,\xi) (1 + \sin(n\xi+\xi_n)) \,d\xi\\
    &\leq \frac{r}{\pi} \int_0^{2\pi} |\sin(\xi)|(1 + \sin(n\xi+\xi_n)) \,d\xi\,,
\end{split}
\end{equation}
Evaluating the integral and taking $r\to R$ yields the bound from above~\cite{Heller:2022ejw}
\begin{equation}
\label{eq:isotropic_ineq}
    |c_n| \leq \frac{2 (n^2-1) - (1+(-1)^n)\sin(\xi_n)}{n^2-1} \frac{2}{\pi R^{n-1}} - \frac{2 \beta_0}{R^n}\,.
\end{equation}
Hydrodynamic dispersion relations are gapless, and so in those cases $\beta_0 = 0$. For odd-order terms, the argument $\xi_n$ drops out of the constraint. Of particular note is that, given a sound mode or a shear mode, one can constrain the constant for the diffusive term $-i D k^2$ (which has $\xi_2 = -\pi/2$) from above as
\begin{equation}
\label{eq:isotropic_diffusion_constraint}
    D \leq \frac{16}{3 \pi} \frac{1}{R}\,.
\end{equation}
Something which has perhaps not been previously appreciated is that the inequality~\eqref{eq:isotropic_ineq} for $\beta_0 = 0$ may also be inverted to yield an infinite set of constraints on the radius of convergence from above in terms of (potentially known) transport coefficients:
\begin{equation}
    \label{eq:isotropic_R_constraints}
    R \leq \lr{ \frac{2 (n^2-1) - (1+(-1)^n) \sin(\xi_n)}{n^2-1} \frac{2}{\pi |c_n|}}^{\frac{1}{n-1}}\,.
\end{equation}

\subsection{Anisotropic case: holomorphic at origin}
 We now extend the argument of the isotropic case to constrain anisotropic dispersion relations when $\omega(\k)$ is holomorphic at $\k=0$. Let us consider a system for which there is a single anisotropy in the system. As before, we align this anisotropy with the $z$-axis. There is then a residual spatial $SO(d-1)$ symmetry, which we use to rotate $\k$ into the $xz$-plane. The dispersion relations therefore depend on $k_z$ and $k_x$. 
 
It is useful to parametrize the complex variables $k_x$, $k_z$ in terms of an extension of the so-called Hopf coordinates, which were introduced in equation~\eqref{eq:Hopf}. In these coordinates, the covariant stability condition~\eqref{eq:covar_stab_condition} takes the form
\begin{equation}
\label{eq:covarstab_constraint_nobranch}
    \Im (\omega(\k)) \leq r \sqrt{\cos^2(\theta)\sin^2(\xi_z) + \sin^2(\theta) \sin^2(\xi_x)} \equiv r\,g(\theta,\xi_z,\xi_x)\,,
\end{equation}
where we have implicitly defined the shorthand $g(\theta,\xi_z,\xi_x)$. Since $\omega(\k)$ is holomorphic at the origin, it admits a power-series expansion in $k_x$ and $k_z$, which we will denote by
\begin{equation}
    \omega = \sum_{n=0}^{\infty} \sum_{m=0}^\infty c_{n,m} \,k_z^n \,k_x^{m}\,,
\end{equation}
where $c_{n,m} \in \mathbb{C}$. Rewriting the expansion in the extended Hopf coordinates~\eqref{eq:Hopf} yields
\begin{equation}
    \omega = \sum_{n=0}^\infty \sum_{m=0}^\infty c_{n,m} r^{n+m} \cos^n(\theta) \sin^m(\theta) e^{i (n \xi_z + m \xi_x)} \equiv U(r,\theta,\xi_z,\xi_x) + i V(r,\theta,\xi_z,\xi_x)\,.
\end{equation}
Now, let us write $c_{n,m} = \alpha_{n,m} + i \beta_{n,m}$, where $\alpha_{n,m}$, $\beta_{n,m} \in \mathbb{R}$. Then the real and imaginary parts of $\omega$ can be written
\begin{equation}
\begin{split}
    U(r,\theta,\xi_z,\xi_x) &= \sum_{n=0}^\infty \sum_{m=0}^\infty r^{n+m} \cos^n(\theta) \sin^m(\theta)(\alpha_{n,m} \cos(n \xi_z + m \xi_x) - \beta_{n,m} \sin(n \xi_z + m \xi_x)), \\
    V(r,\theta,\xi_z,\xi_x) &= \sum_{n=0}^\infty \sum_{m=0}^\infty r^{n+m}\cos^n(\theta) \sin^m(\theta)(\beta_{n,m} \cos(n \xi_z + m \xi_x) + \alpha_{n,m} \sin(n \xi_z + m \xi_x))\,.
\end{split}
\end{equation}
Let us now focus on an arbitrary subterm of the expansion of $\omega(\k)$, which we denote by $\omega_{n,m}$. The magnitude of the subterm is given by
\begin{equation}
    |\omega_{n,m}| = |c_{n,m}| r^{n+m} \cos^n(\theta) \sin^m(\theta)\,,
\end{equation}
where the absolute values on the trigonometric functions are dropped due to the restriction of $\theta \in [0,\pi/2]$. This may be re-written as in the isotropic case,
\begin{equation}
    |\omega_{n,m}| = (\alpha_{n,m} + i \beta_{n,m}) e^{-i \varphi_{n,m}} r^{n+m} \cos^n(\theta) \sin^m(\theta)\,,
\end{equation}
where $\varphi_{n,m} = \arg(c_{n,m})$. Let us define the shorthand $s_{n,m} \equiv r^{n+m} \cos^n(\theta) \sin^m(\theta)$ for brevity. We may now note the following identities:
\begin{subequations}
    \begin{align}
        \alpha_{n,m} s_{n,m} &= \begin{cases}
            \frac{1}{\pi^2} \int_{0}^{2\pi} \int_{0}^{2\pi} d\xi_x d\xi_z \, V(r,\theta,\xi_x,\xi_z) \cos(n \xi_z) \sin(m \xi_x) & n>0,\, m> 0\,,\\
            \frac{1}{2 \pi^2} \int_{0}^{2\pi} \int_{0}^{2\pi} d\xi_x d\xi_z \, V(r,\theta,\xi_x,\xi_z) \sin(m \xi_x) & n=0,\, m> 0\,,\\
            \frac{1}{2\pi^2} \int_{0}^{2\pi} \int_{0}^{2\pi} d\xi_x d\xi_z \, V(r,\theta,\xi_x,\xi_z) \sin(n \xi_z)  & n>0,\, m= 0\,,\\
            \frac{1}{4\pi^2} \int_{0}^{2\pi} \int_{0}^{2\pi} d\xi_x d\xi_z \, U(r,\theta,\xi_x,\xi_z)   & n=0,\, m= 0\,,
        \end{cases}\\
        \beta_{n,m}s_{n,m} &= \begin{cases}
            \frac{1}{\pi^2} \int_{0}^{2\pi} \int_{0}^{2\pi} d\xi_x d\xi_z \, V(r,\theta,\xi_x,\xi_z) \cos(n \xi_z) \cos(m \xi_x) & n>0,\, m> 0\,,\\
            \frac{1}{2\pi^2} \int_{0}^{2\pi} \int_{0}^{2\pi} d\xi_x d\xi_z \, V(r,\theta,\xi_x,\xi_z) \cos(n \xi_z)  & n>0,\, m= 0\,,\\
            \frac{1}{2\pi^2} \int_{0}^{2\pi} \int_{0}^{2\pi} d\xi_x d\xi_z \, V(r,\theta,\xi_x,\xi_z)  \cos(m \xi_x) & n=0,\, m> 0\,,\\
            \frac{1}{4\pi^2} \int_{0}^{2\pi} \int_{0}^{2\pi} d\xi_x d\xi_z \, V(r,\theta,\xi_x,\xi_z) & n=0,\, m= 0\,,\\
        \end{cases}
    \end{align}
\end{subequations}
For $m,n>0$, we therefore have
\begin{equation}
    \begin{split}
        |\omega_{n,m}| &= (\alpha_{n,m} + i \beta_{n,m})e^{-i \varphi_{n,m}} s_{n,m}\\
        &= \frac{1}{\pi^2} \int_{0}^{2\pi} \int_{0}^{2\pi} d\xi_z d\xi_x V(r,\theta,\xi_x,\xi_z) \cos(n \xi_z) \sin(m \xi_x + \varphi_{n,m})\,,
    \end{split}
\end{equation}
where we have made use of the fact that $|\omega_{n,m}|$ is real. The inequality~\eqref{eq:covarstab_constraint_nobranch} may then be applied by adding $\beta_{0,0}$:
\begin{align}
    |c_{n,m}| s_{n,m}& + 4 \beta_{0,0} = \frac{1}{\pi^2} \int_0^{2\pi} \int_0^{2\pi} d\xi_x d\xi_z V (1 + \cos(n \xi_z) \sin(m \xi_x + \varphi_{n,m}))\nonumber\\
    &\leq \frac{r}{\pi^2} \int_0^{2\pi} \int_0^{2\pi} d\xi_x d\xi_z\, g(\theta,\xi_z,\xi_x) \,(1 + \cos(n \xi_z) \sin(m \xi_x + \varphi_{n,m}))\,.\label{eq:mn_bound}
\end{align}
If one of $m$ or $n$ is zero, but the other is greater than zero, one can find by essentially the same argument that
\begin{subequations}
\label{eq:morn_bounds}
    \begin{align}
    |c_{n,0}| s_{n,0} + 2 \beta_{0,0} &\leq \frac{r}{2\pi^2} \int_0^{2\pi} \int_0^{2\pi} d\xi_x d\xi_z g(\theta,\xi_z,\xi_x) (1 + \sin(n \xi_z + \varphi_{n,0}))\,,\\
    |c_{0,m}| s_{0,m} + 2 \beta_{0,0} &\leq \frac{r}{2\pi^2} \int_0^{2\pi} \int_0^{2\pi} d\xi_x d\xi_z g(\theta,\xi_z,\xi_x) (1 + \sin(m \xi_x + \varphi_{0,m}))\,.
    \end{align}
\end{subequations}
The integrals~\eqref{eq:mn_bound},~\eqref{eq:morn_bounds} have no closed-form solution for arbitrary $\theta$ as far as we can determine. However, we may still evaluate bounds in particular limits, and the integrals may also be evaluated numerically. In particular, we can note that with the definitions
\begin{subequations}
\begin{align}
    F_{n,m}(\theta,\varphi_{n,m}) &= \begin{cases}
        \frac{1}{\pi^2} \int_0^{2\pi} \int_0^{2\pi} d\xi_x d\xi_z \,g(\theta,\xi_z,\xi_x) (1 + \cos(n \xi_z) \sin(m \xi_x + \varphi_{n,m})), & n>0,\,m>0\\
        \frac{1}{2\pi^2} \int_0^{2\pi} \int_0^{2\pi} d\xi_x d\xi_z \,g(\theta,\xi_z,\xi_x) (1 + \sin(m \xi_x + \varphi_{0,m})), & n = 0,\, m>0\\
        \frac{1}{2\pi^2} \int_0^{2\pi} \int_0^{2\pi} d\xi_x d\xi_z \,g(\theta,\xi_z,\xi_x) (1 + \sin(n \xi_z + \varphi_{n,0})), & n>0,\, m=0
    \end{cases}\,,\nonumber\\
    \ell_{n,m} &= \begin{cases}
        4, & n>0,\, m>0\\
        2, & n=0, \, m>0\\
        2, & n>0,\, m=0
    \end{cases}\,,
\end{align}
\end{subequations}
the bounds may be written concisely as
\begin{equation}
\label{eq:cnm_intermediary}
    |c_{n,m}| \cos^n(\theta) \sin^m(\theta) \leq \frac{F_{n,m}(\theta,\varphi_{n,m})}{r^{n+m-1}} - \frac{\ell_{n,m} \beta_{0,0}}{r^{n+m}}\,.
\end{equation}
Let us now take $r$ up to the (in general, $\theta$-dependent) convergence radius of the expansion, which we denote by $R(\theta)$. Then the condition~\eqref{eq:cnm_intermediary} becomes
\begin{equation}
    |c_{n,m}| \cos^n(\theta) \sin^m(\theta) \leq \frac{F_{n,m}(\theta,\varphi_{n,m})}{R(\theta)^{n+m-1}} - \frac{\ell_{n,m} \beta_{0,0}}{R(\theta)^{n+m}}\,.
\end{equation}
If the dispersion relations under consideration are hydrodynamic, then the dispersion relations must be gapless, and $c_{0,0} = 0$. Then
\begin{equation}
\label{eq:no_branch_bound}
    |c_{n,m}| \leq \frac{F_{n,m}(\theta,\varphi_{n,m})}{R(\theta)^{n+m-1} \cos^n(\theta) \sin^m(\theta)}\,.
\end{equation}
The constraints~\eqref{eq:no_branch_bound} serve as bounds from above on the magnitudes of anisotropic transport coefficients. The bounds may also be inverted, yielding (for $n+m>1$)
\begin{equation}
    R(\theta) \leq \lr{\frac{F_{n,m}(\theta,\varphi_{n,m})}{|c_{n,m}| \cos^n(\theta) \sin^m(\theta)}}^{\frac{1}{n+m-1}}\,.
\end{equation}
Given $R(\theta)$ is not $n,m$-dependent, this represents an (in principle) infinite set of constraints on the radii of convergence, set by each known transport coefficient. Taken together, they can serve as a $\theta$-dependent upper bound on the radii of convergence. 

\paragraph{Behaviour of $F_{n,m}$.} While the integral $F_{n,m}(\theta,\varphi_{n,m})$ cannot be evaluated analytically (except for in particular limits), some broad properties can be determined.

\begin{figure}
    \centering
    \begin{subfigure}[t]{0.25\linewidth}
    \centering
    \includegraphics[width=\linewidth]{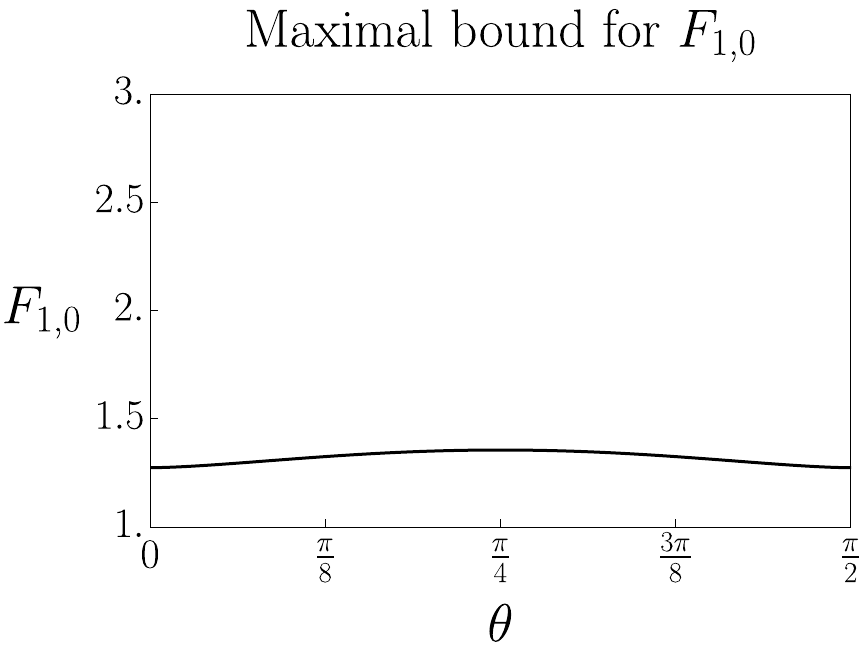}
    \label{fig:F10_plot}
    \end{subfigure}%
    ~
    \begin{subfigure}[t]{0.25\linewidth}
    \centering
    \includegraphics[width=\linewidth]{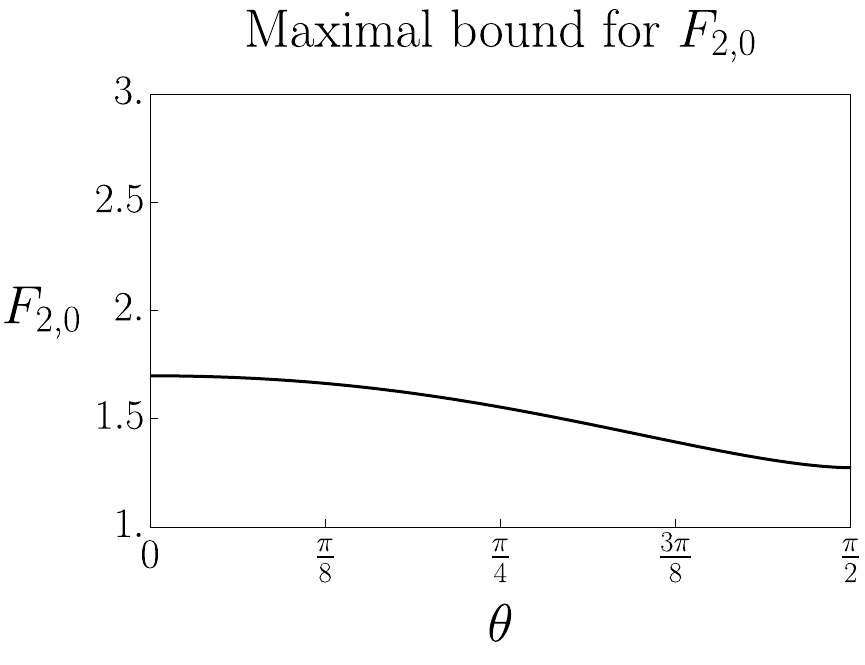}
    \label{fig:F20_plot}
    \end{subfigure}%
    ~
    \begin{subfigure}[t]{0.25\linewidth}
    \centering
    \includegraphics[width=\linewidth]{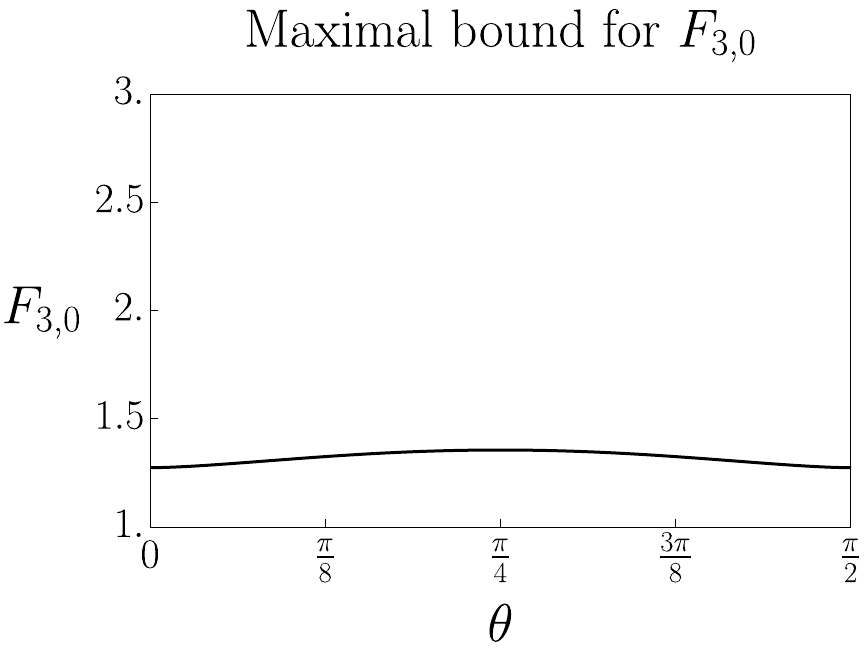}
    \label{fig:F30_plot}
    \end{subfigure}%
    ~
    \begin{subfigure}[t]{0.25\linewidth}
    \centering
    \includegraphics[width=\linewidth]{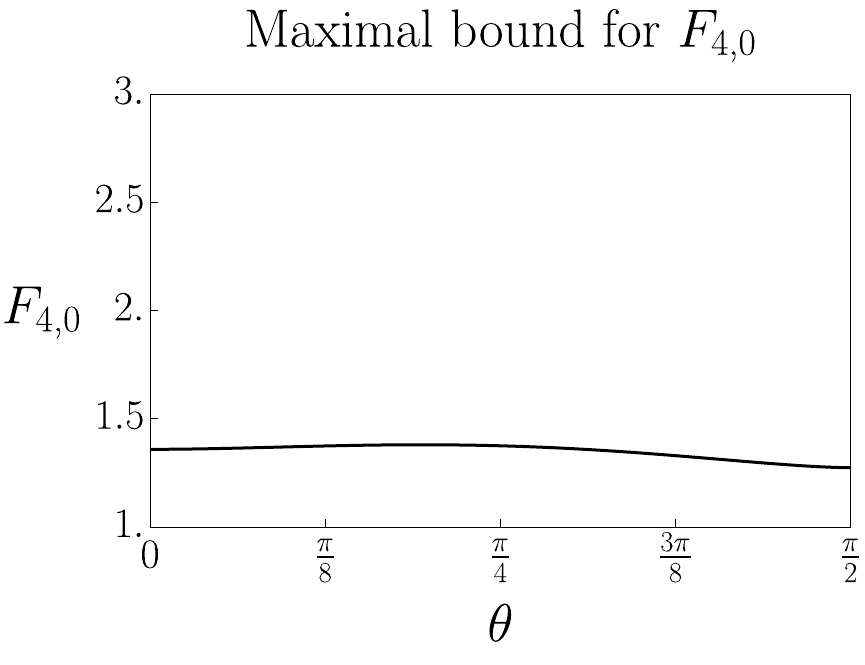}
    \label{fig:F40_plot}
    \end{subfigure}
    \begin{subfigure}[t]{0.25\linewidth}
    \centering
    \includegraphics[width=\linewidth]{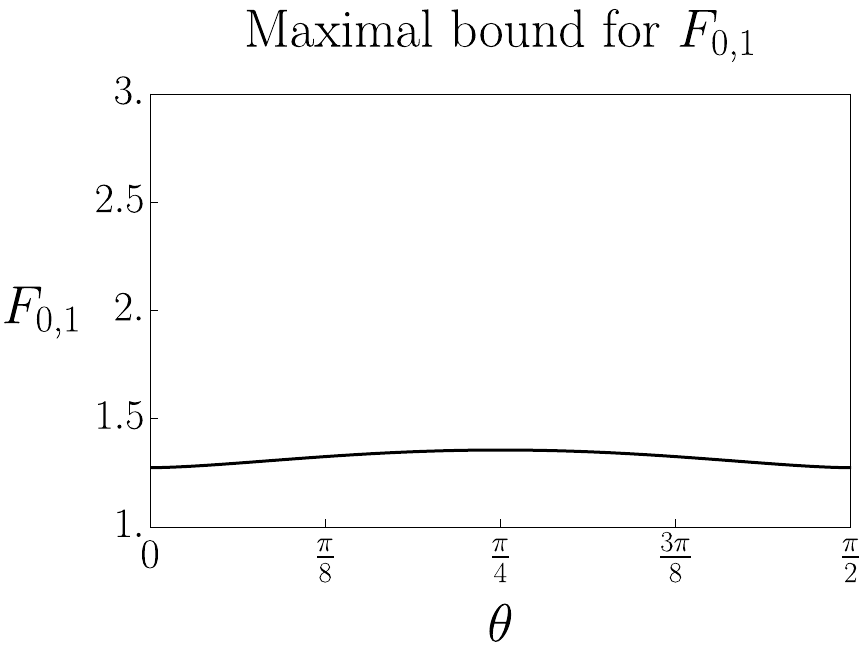}
    \label{fig:F01_plot}
    \end{subfigure}%
    ~
    \begin{subfigure}[t]{0.25\linewidth}
    \centering
    \includegraphics[width=\linewidth]{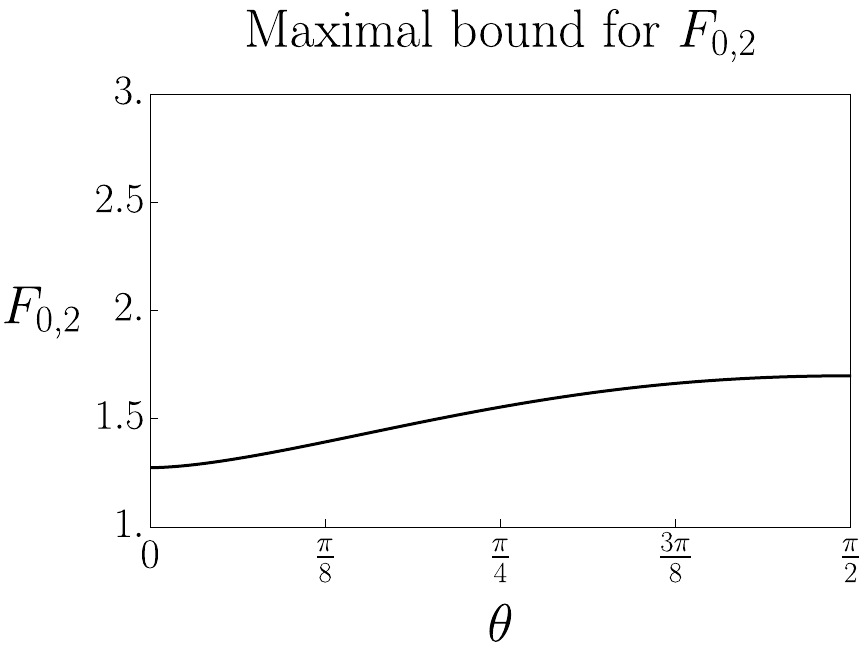}
    \label{fig:F02_plot}
    \end{subfigure}%
    ~
    \begin{subfigure}[t]{0.25\linewidth}
    \centering
    \includegraphics[width=\linewidth]{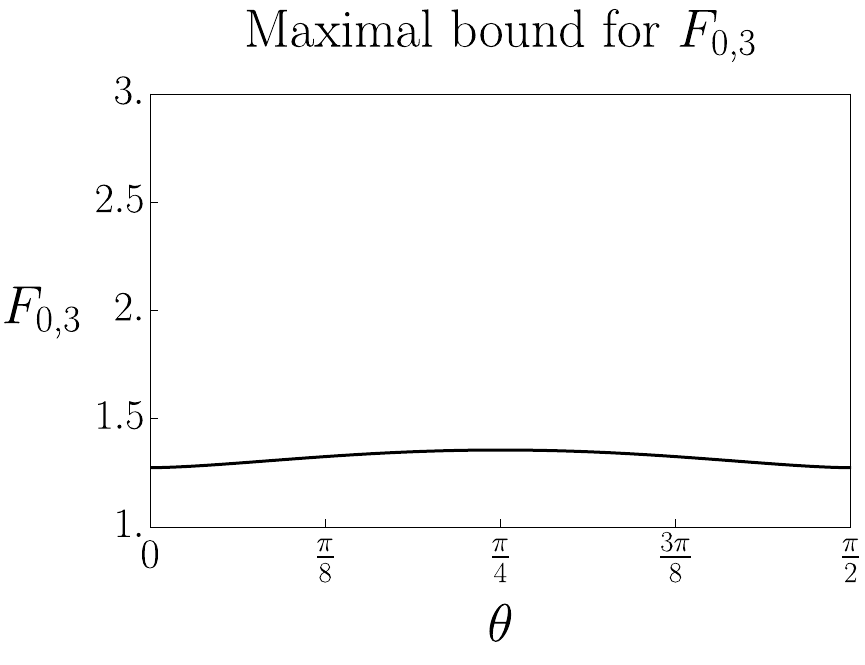}
    \label{fig:F03_plot}
    \end{subfigure}%
    ~
    \begin{subfigure}[t]{0.25\linewidth}
    \centering
    \includegraphics[width=\linewidth]{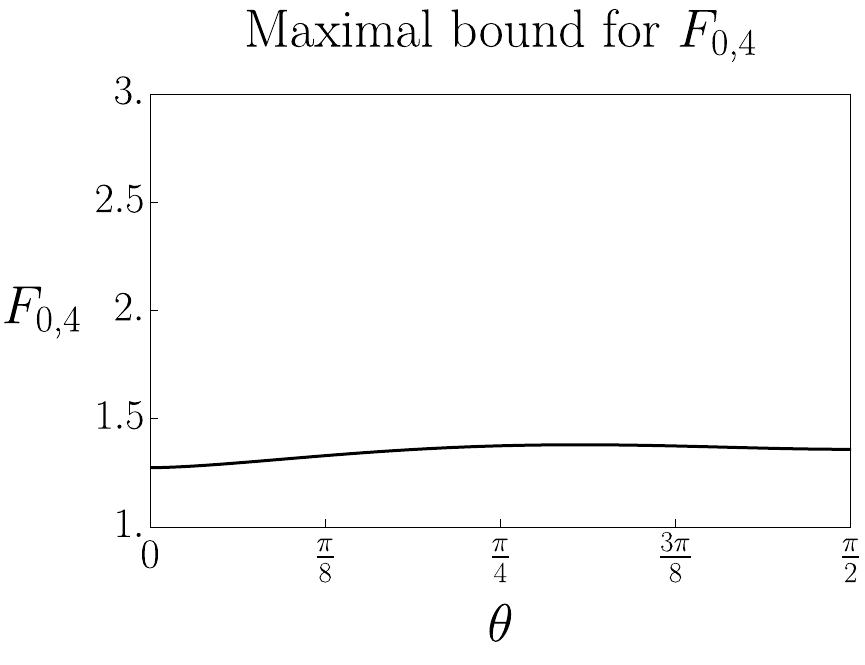}
    \label{fig:F04_plot}
    \end{subfigure}
    \begin{subfigure}[t]{0.25\linewidth}
    \centering
    \includegraphics[width=\linewidth]{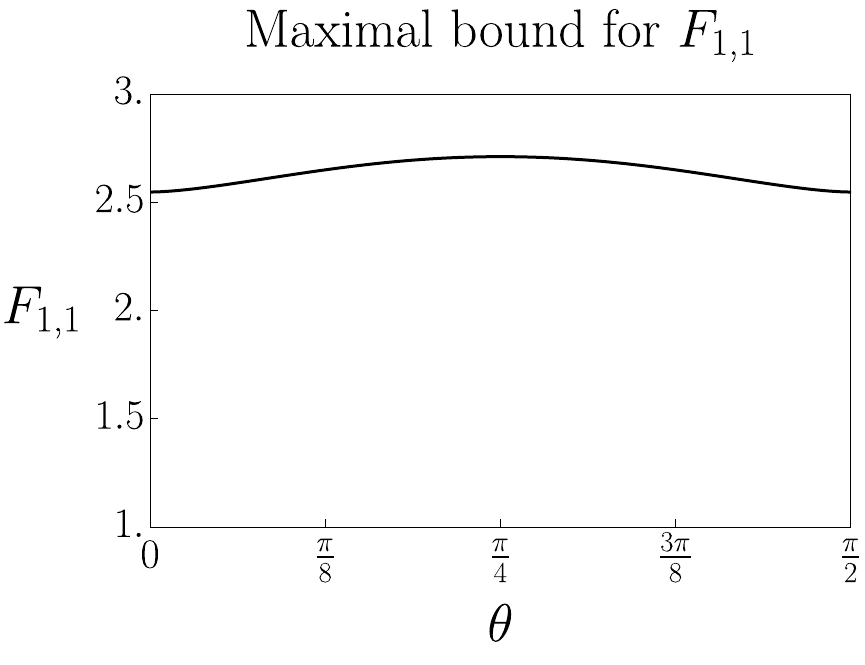}
    \label{fig:F11_plot}
    \end{subfigure}%
    ~
    \begin{subfigure}[t]{0.25\linewidth}
    \centering
    \includegraphics[width=\linewidth]{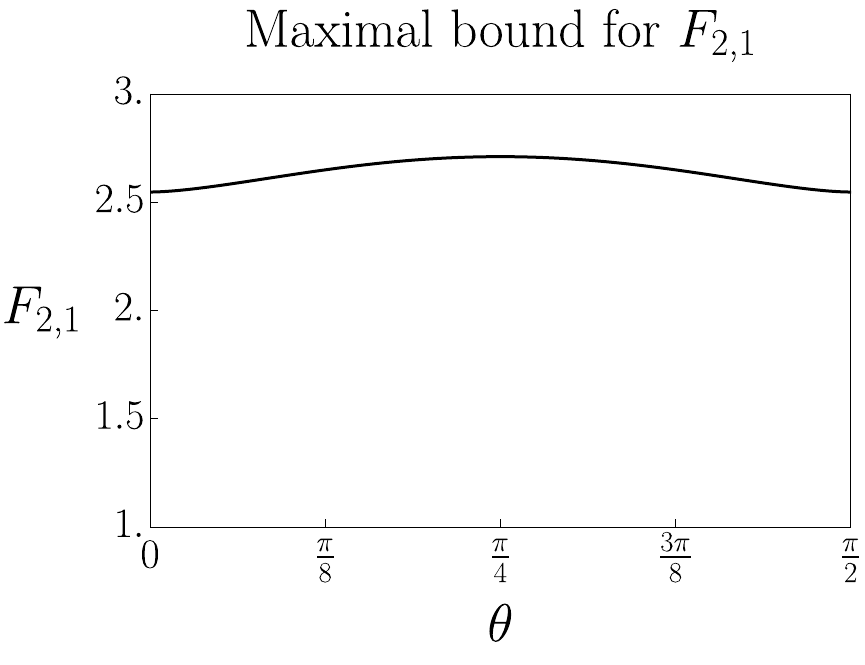}
    \label{fig:F21_plot}
    \end{subfigure}%
    ~
    \begin{subfigure}[t]{0.25\linewidth}
    \centering
    \includegraphics[width=\linewidth]{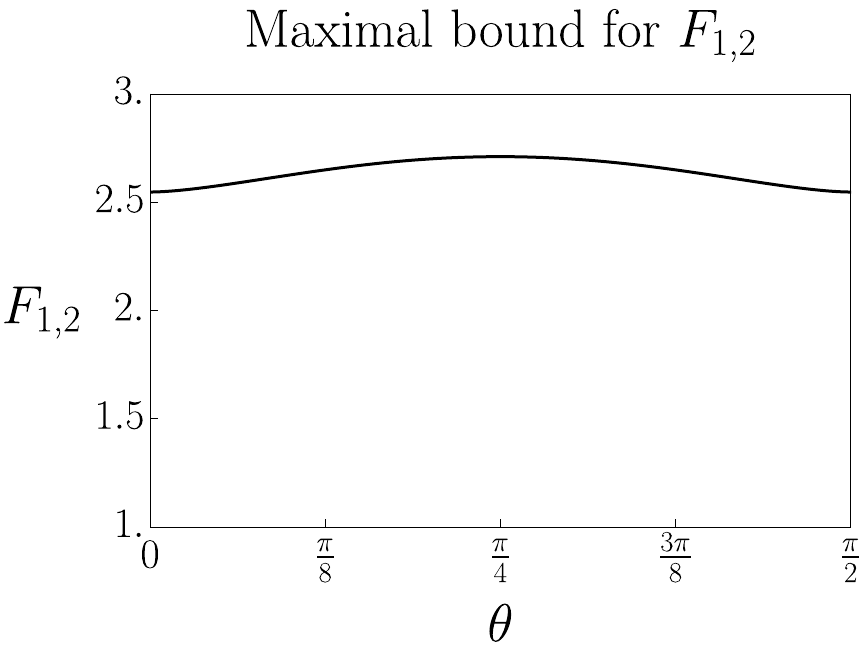}
    \label{fig:F12_plot}
    \end{subfigure}%
    ~
    \begin{subfigure}[t]{0.25\linewidth}
    \centering
    \includegraphics[width=\linewidth]{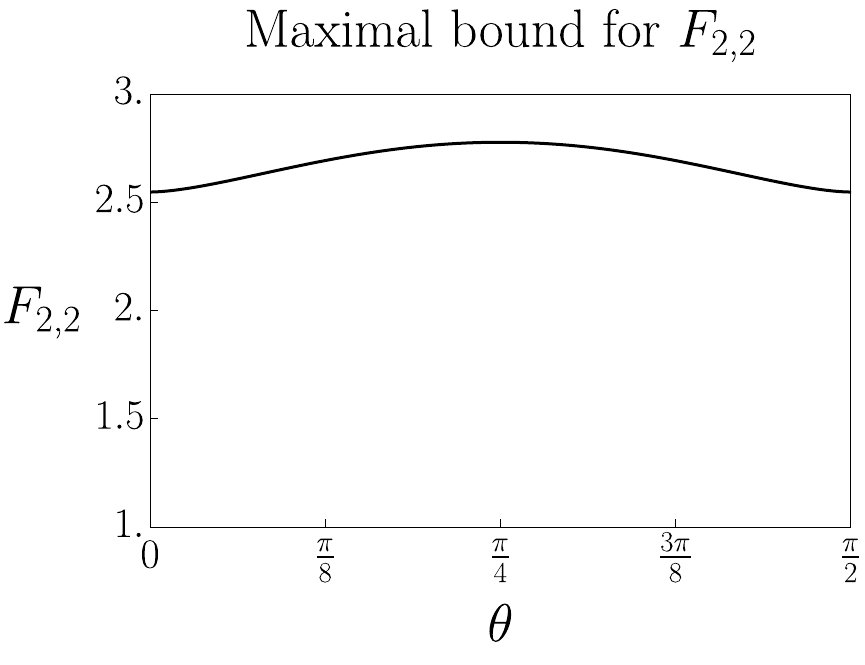}
    \label{fig:F22_plot}
    \end{subfigure}
    \begin{subfigure}[t]{0.25\linewidth}
    \centering
    \includegraphics[width=\linewidth]{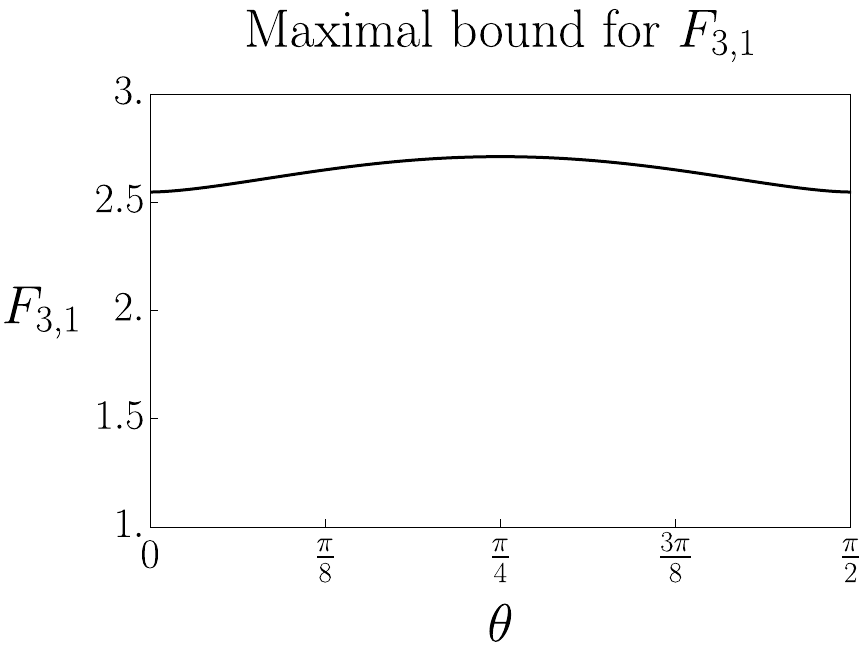}
    \label{fig:F31_plot}
    \end{subfigure}%
    ~
    \begin{subfigure}[t]{0.25\linewidth}
    \centering
    \includegraphics[width=\linewidth]{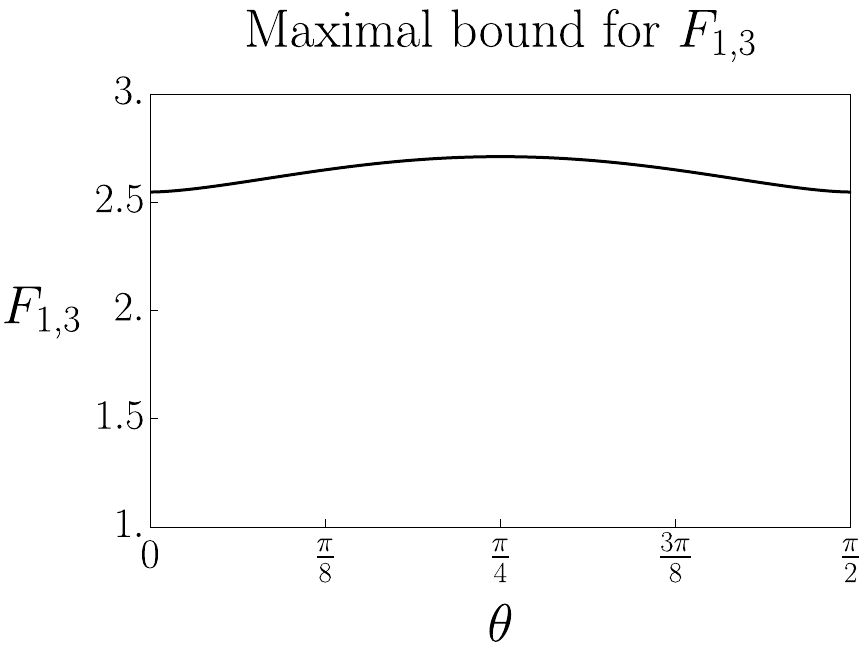}
    \label{fig:F13_plot}
    \end{subfigure}%
    ~
    \caption{A survey of the behaviour of the first fourteen $F_{n,m}$. As expected from the enumerated properties, the dependence on $n,m$ drops out for the terms with odd indices (for $n,m>0$).}
    \label{fig:Fnm_behaviour}
\end{figure}

\begin{itemize}
    \item If either $n$ or $m$ are odd, the dependence of $F_{n,m}$ on $m$, $n$, and $\varphi_{n,m}$ drop out (almost) entirely. This may be straightforwardly determined from the symmetry properties of $g(\theta,\xi_z,\xi_x)$:\begin{subequations}
        \begin{align}
            g(\theta,\xi_z,\xi_x) &= g(\theta,-\xi_z,\xi_x) = g(\theta,\xi_z,-\xi_x)\,,\label{eq:sym_Fourier_one}\\
            g(\theta,\xi_z,\xi_x) &= g(\theta,\xi_z \pm \pi,\xi_x) = g(\theta,\xi_z,\xi_x\pm \pi)\,.\label{eq:sym_Fourier_two}
        \end{align}
    \end{subequations}
    The condition~\eqref{eq:sym_Fourier_one} implies that a Fourier series expansion of $g(\theta,\xi_z,\xi_x)$ in either $\xi_z$ or $\xi_x$ proceeds only in terms of cosines, while the condition~\eqref{eq:sym_Fourier_two} implies that those cosines may be of only even order. Therefore, if either $n$ or $m$ are odd, the term in brackets in $F_{n,m}$ reduces to just $1$, eliminating the dependence on $\varphi_{n,m}$, $n$, and $m$. However, there is a difference by a factor of $2$ between $F_{n,m}$ when one of $n,m$ are odd and both non-zero, and when one of $n$ or $m$ vanishes.
    \item The following six limits hold:
    \begin{subequations}\label{eq:limits_Fnm}
        \begin{alignat}{2}
   & \lim_{\theta \to 0} F_{n>0,m>0}(\theta, \varphi_{n>0,m>0}) &&= \frac{8}{\pi}\,,\qquad \lim_{\theta \to \frac{\pi}{2}} F_{n,m}(\theta,\varphi_{n>0,m>0}) = \frac{8}{\pi}\,,\\
& \qquad\,\lim_{\theta \to 0} F_{0,m>0}(\theta,\varphi_{0,m>0}) &&= \frac{4}{\pi}\,,\qquad  \,\,\,\,\lim_{\theta \to \frac{\pi}{2}} F_{n>0,0}(\theta,\varphi_{n>0,0}) = \frac{4}{\pi}\,,\\
& \qquad \,\,\,\,\lim_{\theta \to 0} F_{n>0,0}(\theta,\varphi_{n>0,0})  &&= \frac{4}{\pi} \lr{ 1 - \frac{1}{2}\frac{(1+(-1)^n) \sin(\varphi_{n,0})}{n^2-1} }\,,\\
&\qquad \lim_{\theta \to \frac{\pi}{2}} F_{0,m>0}(\theta,\varphi_{0,m>0}) &&= \frac{4}{\pi} \lr{ 1 - \frac{1}{2}\frac{(1+(-1)^m) \sin(\varphi_{0,m})}{m^2-1} }\,,
        \end{alignat}
    \end{subequations}
    \item In the following argument, we use $F_{n>0,m>0}$ for concreteness; the same argument carries through analogously when $n=0$ or $m=0$. For $F_{n,m}(\theta,\varphi_{n,m})$ with $n$, $m$ even, the maximum value (and therefore the weakest bound) is obtained when $\varphi_{n,m} = -\pi/2$. While this has not been shown for all $n$, $m$, we have checked it for the lowest 119 terms ($0< n+m \leq 14$). This can be seen by taking the derivative of $F_{n,m}$ with respect to $\varphi_{n,m}$, yielding
    \begin{equation}
    \begin{split}
        \pder{F_{n,m}(\theta,\varphi_{n,m})}{\varphi_{n,m}} &= \frac{1}{\pi^2}\int_0^{2\pi} \int_0^{2\pi} g(\theta,\xi_z,\xi_x) \cos(n \xi_z) \cos(m \xi_x + \varphi_{n,m}) d\xi_x d\xi_z\,,\\
        &=\cos( \varphi_{n,m})\lr{\frac{1}{\pi^2}\int_0^{2\pi} \int_0^{2\pi} g(\theta,\xi_z,\xi_x) \cos(n \xi_z) \cos(m \xi_x) d\xi_x d\xi_z}\,,\\
    \end{split}
    \end{equation}
    where in the last step we have used the symmetry properties of $g$. We see then that $\varphi_{n,m} = \pm \frac{\pi}{2}$ sets the derivative to zero. Taking the second derivative and setting $\varphi_{n,m} = \pm \frac{\pi}{2}$ yields
    \begin{equation}
        \pderr{F_{n,m}(\theta,\varphi_{n,m})}{\varphi_{n,m}}\biggl|_{\varphi_{n,m} = \pm \frac{\pi}{2}} = \mp \frac{1}{\pi^2} \int_0^{2\pi} \int_0^{2\pi}g(\theta,\xi_z,\xi_x) \cos(n \xi_z) \cos(m \xi_x) d\xi_x d\xi_z\,.
    \end{equation}
    The double integral is non-positive at least for the first 119 $F_{n,m}$, and so $\varphi_{n,m} = - \frac{\pi}{2}$ is a maximum for at least those $F_{n,m}(\theta,\varphi_{n,m})$.
\end{itemize}
When $n=0$ or $m=0$, the isotropic case~\eqref{eq:isotropic_ineq} is recovered in~\eqref{eq:limits_Fnm} in the relevant $\theta$-limit, whereas the opposite limit goes to an ($n$-) $m$-independent constant. The first fourteen (i.e. $0<n+m \leq 4$) $F_{n,m}$ are shown in Figure~\ref{fig:Fnm_behaviour}. For those $F_{n,m}$ that depend on $\varphi_{n,m}$, it has been set to $-\pi/2$ to maximize the bound.

\paragraph{Example.} As a simple example, let us take the shear mode in an ${\cal N} = 4$ SYM plasma at strong coupling. The hydrodynamic dispersion relation in the small-$|\qv| = |\k|/(2 \pi T)$ limit has been computed~\cite{Grozdanov:2019uhi} in the isotropic case to ${\cal O}(\q^8)$. We can trivially re-write $\q^2 = \q_z^2 + \q_x^2$ and use the accompanying transport coefficients to try and bound the radius of convergence (which, of course, is already known; refer to Section~\ref{sec:holography}).

\begin{figure}
    \centering
    \includegraphics[width=0.67\linewidth]{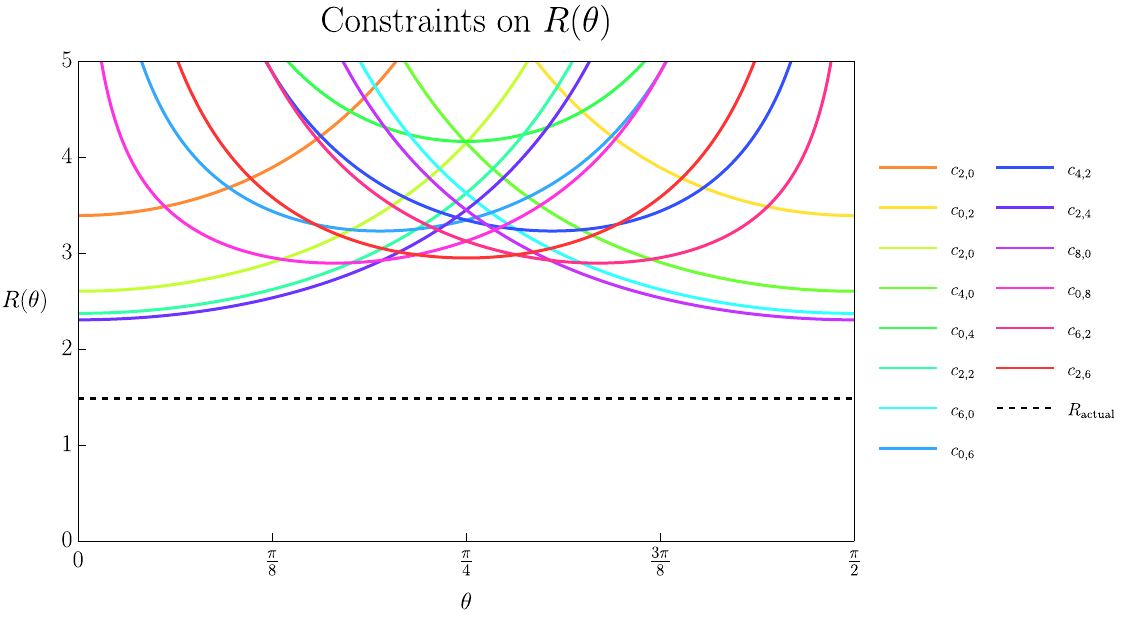}
    \caption{A plot of the fourteen constraints on the radius of convergence associated with the components in equation~\eqref{eq:shear_isotropic_holocomponents}, along with the actual radius of convergence of the isotropic shear mode. All regions above the coloured lines would be excluded by the demands of microcausality.}
    \label{fig:holographic_R_constraints}
\end{figure}

In the language used here, the transport coefficients are given by
\begin{subequations}
\label{eq:shear_isotropic_holocomponents}
    \begin{align}
        c_{2,0} = c_{0,2} &= -\frac{i}{2}\,,\\
        c_{4,0} = c_{0,4} = \frac{1}{2} c_{2,2} &= - \frac{i (1-\ln(2))}{4} \,,\\
        c_{6,0} = c_{0,6} = \frac{1}{3} c_{4,2} = \frac{1}{3} c_{2,4} &= - \frac{i(24 (\ln 2)^2 - \pi^2)}{96}\,,\\
        c_{8,0} = c_{0,8} = \frac{1}{4}c_{2,6} = \frac{1}{4} c_{6,2} = \frac{1}{6} c_{4,4} &= -\frac{i(2 \pi^2(5 \ln 2 - 1) - 21 \zeta(3) - 24 \ln 2(1 + \ln 2(5 \ln 2 - 3)))}{384}\,,
    \end{align}
\end{subequations}
where $\zeta$ is the Riemann zeta function. This yields $14$ constraints from above on the radius of convergence, following equation~\eqref{eq:no_branch_bound}. These constraints have been plotted together in Figure~\ref{fig:holographic_R_constraints}, along with the actual radius of convergence of the isotropic shear mode. Note that some of the strongest bounds are given by the highest-order transport coefficients.

Now let us finally turn to the case of an anisotropic system with a branch point at the origin.

\subsection{Anisotropic case: branch point at origin}
When there is a branch point at the origin, the frequency $\omega(k_z, k_x)$ does not admit a power series solution in $k_x$ and $k_z$. We will instead consider an expansion in the single variable $\mathfrak{r}=\sqrt{k_x^2+k_z^2}$, and then complexify the variable $\mathfrak{r}$ after expanding. The expansion surfaces $\Sigma_r$ in the space of complex $k_x$, $k_z$ were parametrized in equation~\eqref{eq:Hopf_extended}; without the normalization in terms of the temperature, they are given by
\begin{equation}
    k_z = r \cos(\theta) e^{i \xi} \qquad k_x = r \sin(\theta) e^{i \xi}\,,
\end{equation}
where $r=|{\mathfrak r}|$, and $\theta\in [0,\pi)$, $\xi \in [0,2\pi)$ are taken to be real. The covariant stability condition~\eqref{eq:covar_stab_condition} then takes on the same form as in the isotropic case:
\begin{equation}
    \Im \,\omega(r,\theta,\xi) \leq r |\sin(\xi)|\,.
\end{equation}
We will now consider a series expansion in $\mathfrak{r}$, such that $\omega$ is of the form
\begin{equation}
    \omega = \sum_{n=0}^\infty c_n(\theta)\, \mathfrak{r}^n \equiv U(r,\theta,\xi) + i \, V(r,\theta,\xi)\,.
\end{equation}
Let us once again write $c_n(\theta)= \alpha_n(\theta) + i \beta_n(\theta)$. After complexifying ${\mathfrak r}$, the expansion is of the form
\begin{equation}
    \omega = \sum_{n=0}^\infty r^n ((\alpha_n(\theta) \cos(n \xi) - \beta_n(\theta) \sin(n \xi)) + i (\alpha_n(\theta) \sin(n \xi) + \beta_n(\theta) \cos(n \xi))\,.
\end{equation}
We can then consider the magnitude of a given subterm with $n>0$:
\begin{equation}
    |c_n(\theta)| r^n = (\alpha_n(\theta) + i \beta_n(\theta)) e^{-i \xi_n} r^n\,,
\end{equation}
where $\xi_n = \arg(c_n)$. For any given fixed $\theta$, one can use Fourier identities to show that
\begin{equation}
|c_n(\theta)| r^n + 2 \beta_0(\theta)\leq  \frac{r}{\pi} \int_0^{2\pi}\,d\xi\, |\sin(\xi)|(1+\sin(n \xi + \xi_n))\,.
\end{equation}
We then arrive at the analogous statement to the isotropic case by taking $r$ to the (in general, $\theta$-dependent) radius of convergence $R(\theta)$:
\begin{equation}
\label{eq:full_branch}
    |c_n(\theta)| \leq \frac{2 (n^2-1) - (1+(-1)^n) \sin(\xi_n)}{n^2-1} \frac{2}{\pi R^{n-1}(\theta)} - \frac{2 \beta_0(\theta)}{R^n(\theta)}
\end{equation}
For a hydrodynamic mode, $\beta_0 = 0$, and so we arrive at the final expression
\begin{equation}
\label{eq:branch_bound}
    |c_n(\theta)| \leq \frac{2 (n^2-1) - (1+(-1)^n) \sin(\xi_n)}{n^2-1} \frac{2}{\pi R^{n-1}(\theta)}\,.
\end{equation}
Note that the inequality~\eqref{eq:full_branch} is simply the isotropic inequality~\eqref{eq:isotropic_ineq} with $c_n$, $R$ dependent on $\theta$. The inequalities~\eqref{eq:branch_bound} may be inverted to give an infinite set of inequalities that constrain the radius of convergence from above:
\begin{equation}
    R(\theta) \leq \lr{\frac{2 (n^2-1) - (1+(-1)^n) \sin(\xi_n)}{n^2-1} \frac{2}{\pi |c_n(\theta)|}}^{\frac{1}{n-1}}\,,
\end{equation}
for all $n>1$. One can also straightforwardly repeat the analysis of the isotropic case to show that for a sound-type mode $\omega = \pm v(\theta) r - i \Gamma(\theta) r^2 + ...$, the same inequalities must hold, modified for a dependence on $\theta$:
\begin{equation}
    |v(\theta)| \leq 1, \qquad 0 \leq \Gamma(\theta) \leq \frac{16}{3 \pi} \frac{1}{R(\theta)}\,.
\end{equation}
Note that for $n=1$, the bound~\eqref{eq:branch_bound} gives a weaker condition on $v(\theta)$ than simply inserting $\omega = \pm v(\theta) r + \oser{r^2}$ into the covariant stability condition~\eqref{eq:covar_stab_condition} and taking the limit as $r \to 0$.

The constraints~\eqref{eq:branch_bound} generalize the isotropic constraints found in~\cite{Heller:2022ejw} to anisotropic dispersion relations $\omega(\k)$ with a branch point at the origin, while the constraints~\eqref{eq:no_branch_bound} generalize to anisotropic dispersion relations $\omega(\k)$ that are holomorphic at the origin. Most anisotropic cases of interest will have a branch point at the origin corresponding to propagation; to name a few, the Alfv\'en modes and magnetosonic modes in MHD~\cite{Anile}, and the second sound mode in the theory of superfluidity~\cite{Carter:1992,Bhattacharya:2011eea,Bhattacharya:2011tra,Herzog:2011ec} all have branch points at the origin. 

In the event that the surfaces of critical points and the expansion surfaces intersect in such a way as to reduce the radius of convergence to zero at some $\theta$, the bound~\eqref{eq:branch_bound} will become useless at that angle, as the right-hand side will diverge. However, unless the radius of convergence is zero for all angles, the constraint may still be applied for any non-divergent values of $\theta$.

\section{Conclusion}
\label{sec:conc}
In this paper, we have explored two main topics. Firstly, we have investigated the effect of anisotropy on convergence properties of hydrodynamic dispersion relations. Using the analysis of several complex variables, we have seen that for any set of hydrodynamic dispersion relations $\omega(\k)$ which have a branch point at the origin, there is a continuum of collisions between hydrodynamic modes which occur at any value of $|\k|$. These collisions between hydrodynamic modes at complex $\k$ only serve to obstruct the convergence of the hydrodynamic derivative expansion if the surfaces of branch points (the ``critical surfaces", $\Sigma_0$) intersects with the surfaces in the space of complex $\k$ (the ``expansion surfaces" $\Sigma_r$) upon which the small-$|\k|$ expansion actually proceeds. 

A boosted sound mode and boosted shear mode were investigated in the context of an ${\cal N}=4$ SYM plasma. The continuum of pole collisions was confirmed to indeed be present in the UV-complete model. While there was no intersection of the expansion surfaces $\Sigma_r$ and the critical surfaces $\Sigma_0$ in the boosted sound mode (except at the origin itself), this is not particularly surprising; the ${\cal N} =4$ SYM plasma is known to have a finite (i.e. non-zero) radius of convergence for an observer at rest, and this situation should not change after a boost.

Secondly, by demanding consistency of dispersion relation with microcausality, we obtained bounds on radii of convergence of hydrodynamic dispersion relations $\omega(\k)$ from above in terms of anisotropic transport coefficients, and vice-versa. In particular, an infinite set of constraints were found which bound the radius of convergence from above at each angle $\theta$, where $\theta$ is the angle between $\k$ and the anisotropic direction. This extends the work of~\cite{Heller:2022ejw} to anisotropic transport.

There are a number of remaining directions one could explore. One immediately pressing direction is that of the holographic dual to the Alfv\'en waves discussed in Section~\ref{sec:examples}. In the first-order theory, the radius of convergence of the small-$|\k|$ expansion goes to zero as the wavevector becomes perpendicular to the magnetic field. This occurs because the expansion surfaces and the critical surfaces intersect, and the intersection moves into the origin as $\theta \to \pi/2$. This is the source of the previously-known non-commutativity between the $\theta \to \pi/2$ and the small-$|\k|$ limits (see e.g.~\cite{Hernandez:2017mch,Grozdanov:2017kyl,Fang:2024skm,Fang:2024hxa,Hoult:2025exx}). Whether such a non-commutativity persists in the holographic case is a subject which we will address in work to come.

Another remaining direction comes from extending our analysis to other anisotropic systems. An immediate example is that of a relativistic superfluid~\cite{Carter:1992,Bhattacharya:2011eea,Herzog:2011ec,Bhattacharya:2011tra}. It may be the case that in relativistic superfluids, there exists an intersection between the expansion surfaces $\Sigma_r$ and the critical surfaces $\Sigma_0$, potentially leading to a zero radius of convergence at specific angles.

Finally, we found conditions on radii of convergence and anisotropic transport coefficients in terms of one another in Section~\ref{sec:caus_con}. In the isotropic case, one can combine transport coefficients and radii of convergence into sets of unitless parameters which may be constrained in terms of one another. This programme was initiated in~\cite{Heller:2023jtd}, generating an infinite-dimensional polyhedron that bound the space of allowed transport coefficients and radii of convergence. This ``hydrohedron" can almost certainly be analogously developed in the anisotropic case, with an additional dependence on an angle $\theta$.

In conclusion, the presence of anisotropies both enriches and complexifies the question of convergence for the hydrodynamic derivative expansion. It is not only collisions between hydrodynamic and non-hydrodynamic modes that can restrict the regime of validity of the small-$|\k|$ expansion, but also collisions between hydrodynamic modes themselves. Care should be taken when considering convergence and perturbations of systems in anisotropic states.

\paragraph{Acknowledgements.} I would like to thank P. Kovtun for extensive and insightful discussions, as well as B. Aitken, J. Coffey, P. Dhivakar, W. Harvey, D. I. Hoult, K. Jensen, and A. Omar for helpful conversations. I would also like to thank P. Kovtun and P. Dhivakar for feedback on a draft of this manuscript. This work was supported by the NSERC of Canada.

\appendix

\section{Analysis in several complex variables}
\label{appendix:scv}
Many of the results in the field of complex analysis for a single variable generalize straightforwardly to functions of several complex variables; however, there are a number of key results which are often taken for granted in physical applications which do not generalize. The most striking of these is the (non-)existence of isolated singularities. Here, we present a (very) brief discussion on the subject of analysis in several complex variables, with a particular focus on domains of convergence. References include~\cite{Shabat:1992}, and the classical~\cite{HormanderBook}. See also~\cite{Lebl:2025} for a gentle introduction. In the following, for concreteness we restrict to $\mathbb{C}^2$; however, results straightforwardly generalize to $\mathbb{C}^n$.

\paragraph{Single complex variable.} The following are some standard results in one complex variable. A function in one complex variable $f(z)$ is called ``holomorphic" in the neighbourhood of a point $z_0$ if $\pder{f(z)}{\bar{z}}\vert_{z=z_0} = 0$, where $\bar{z}$ is the complex conjugate of $z$. This is equivalent to the statement that $f(z)$ is analytic in the neighbourhood of $z_0$. If $f(z)$ is holomorphic in the neighbourhood of $z_0$, then it admits a Taylor expansion about $z_0$
\begin{equation}\label{app:ocv_Taylor}
    f(z) = \sum_{n=0}^\infty c_n (z-z_0)^n\,.
\end{equation}
The domain $D$ on which $f(z)$ is holomorphic is called the ``domain of holomorphy" of $f(z)$. In the context of the Taylor series~\eqref{app:ocv_Taylor}, this domain $D$ is also called the ``domain of convergence" of \eqref{app:ocv_Taylor} about $z_0$. In $\mathbb{C}$, this domain $D$ is always a disc characterized by a single number $R$ such that $D = \{ z: |z-z_0| < R\}$. The number $R$ is called the ``radius of convergence", and is set by the distance from $z_0$ to the nearest point in $\mathbb{C}$ at which $f(z)$ is not holomorphic. This could be due to a branch point, or a pole, or other types of non-removable singularities. 

If the point $z_0$ is itself a singular point, then it is still possible to define a series expansion on a disc around $z_0$; however, this expansion will not, in general, be a Taylor series. Possibilities include a Laurent series (negative powers of $z$) and a Puiseux series (fractional powers of $z$). Puiseux series in particular are very familiar in hydrodynamics; the hydrodynamic sound mode is a Puiseux series in $\xi \equiv k^2$, $\omega = \pm v_s \xi^{1/2} - i \frac{\Gamma}{2} \xi \pm\,...$, where the $...$ indicate higher-order terms in $\xi$.

\paragraph{Several complex variables}
We move now to several complex variables. Let us define $\mathbf{z} = (z_1, z_2) \in \mathbb{C}^2$. A function $f(\z)$ is called holomorphic in the neighbourhood of a point $\z_0$ if it is holomorphic individually in each of the complex components of $\z$ (Hartogs' fundamental theorem)\footnote{Note that this is actually a somewhat remarkable result. The corresponding statement is not true for real functions of several variables.}. The domain in ${\mathbb C}^2$ on which $f(\z)$ is holomorphic is once again called the ``domain of holomorphy". 

There are a number of extensions to the notion of a disc used in the single variable case, and they are not equivalent to one another. One natural extension is the ball $\mathbb{B}^4$ in $\mathbb{R}^{4}\simeq \mathbb{C}^2$, which is given by the condition $\z {\cdot} \overline{\z} = R^2$, where `` ${\cdot}$ " is the Euclidean inner product (i.e. dot product), and $\overline{\z} = (\overline{z_1}, \overline{z_2})$. Another natural extension is the polydisc\footnote{The polydisc is often denoted with a $\Delta$; given the content of this paper, this notation has been changed here to a ${\cal D}$ to avoid any possibility of confusion with the discriminant.} ${\cal D}$ about a point $\z_0$, which is defined by ${\cal D} = \{ z: |z^j - z_0^j| < r^j\}$ for some collection of positive real numbers $\mathbf{r} = ( r^1, r^2)$. The ball and the polydisc are not equivalent to one another.

\begin{figure}
    \centering
    \begin{subfigure}[t]{0.5\linewidth}
            \centering
            \includegraphics[width=\linewidth]{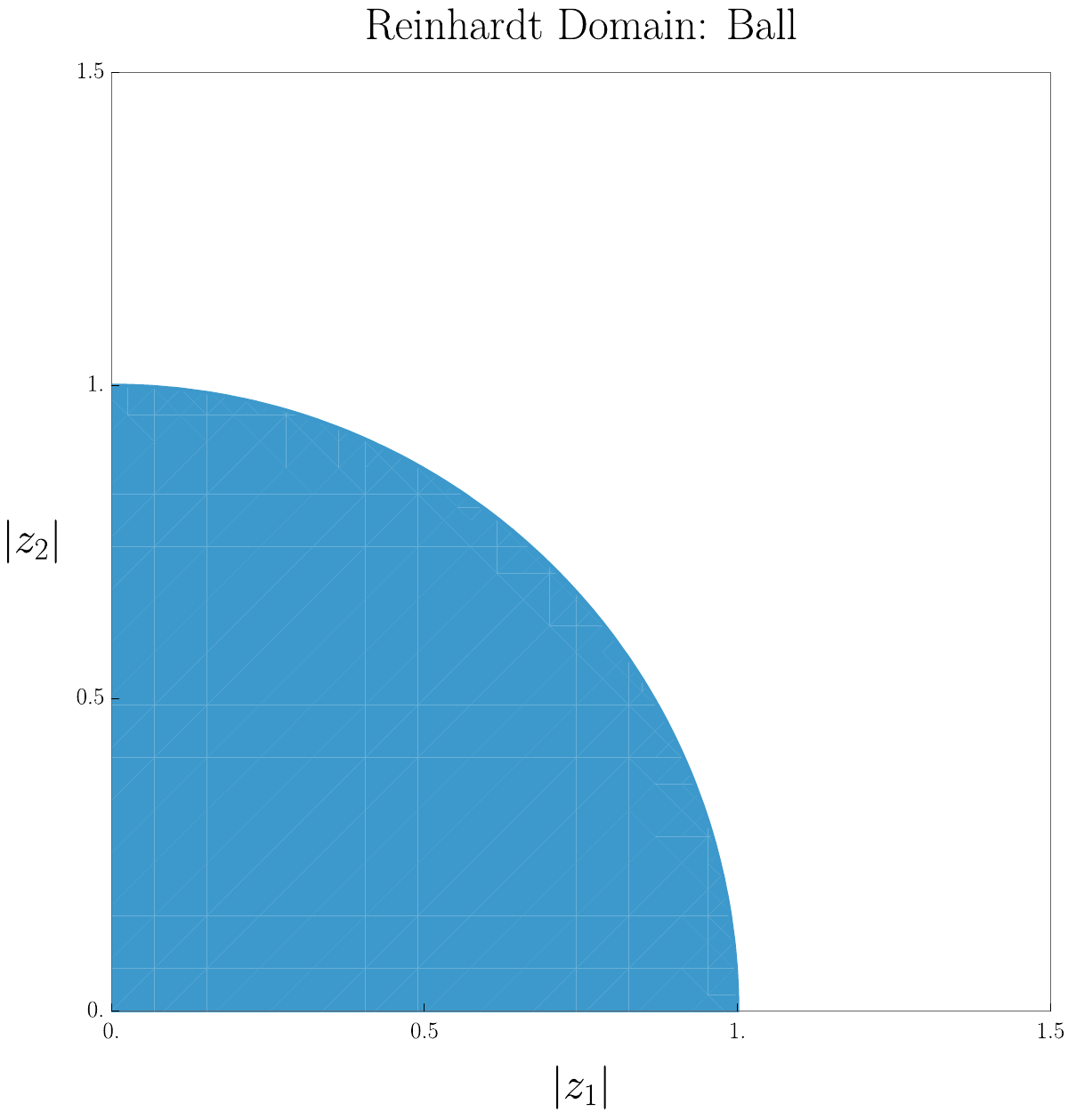}
    \end{subfigure}%
    ~
    \begin{subfigure}[t]{0.5\linewidth}
            \centering
            \includegraphics[width=\linewidth]{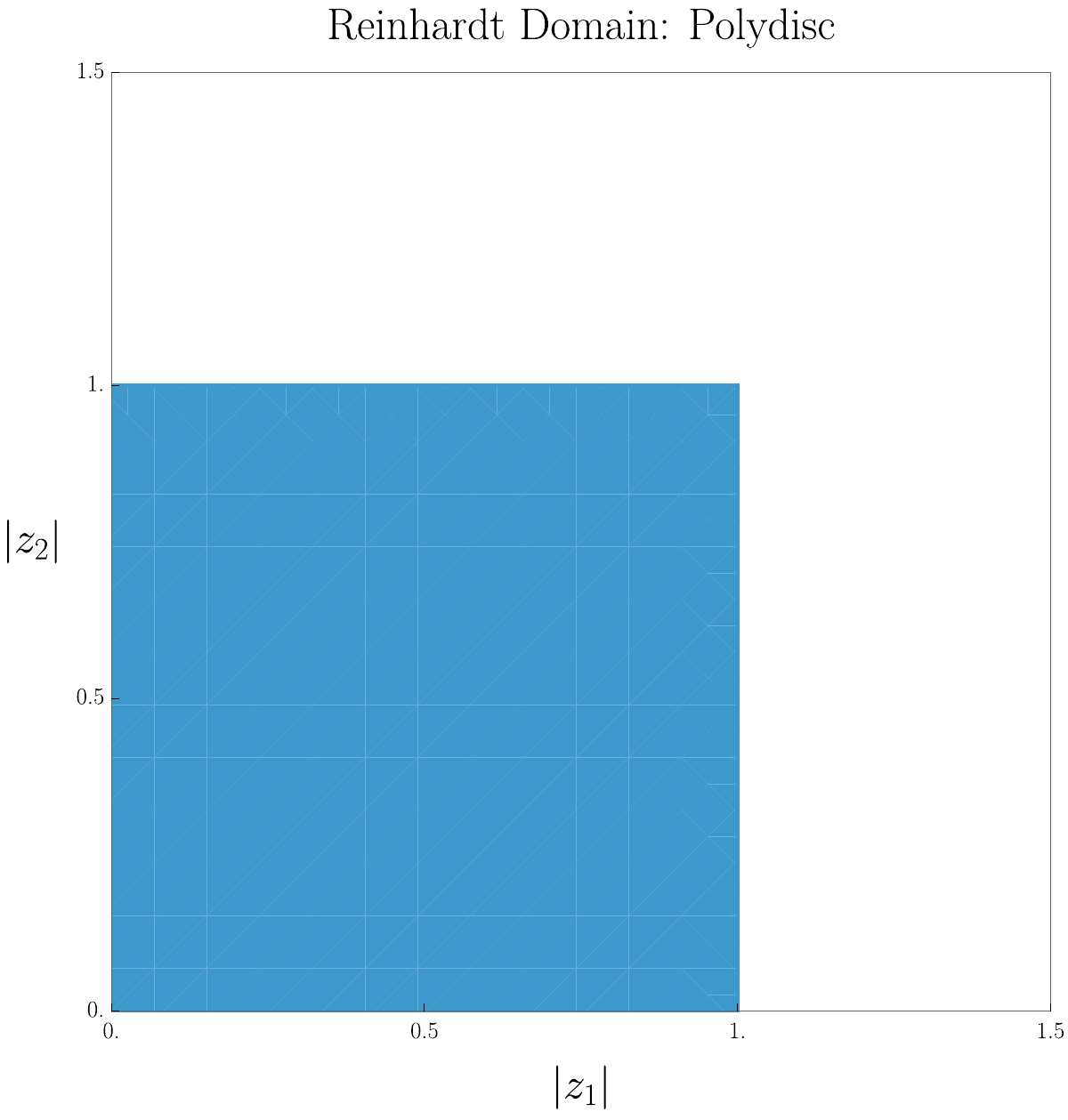}
    \end{subfigure}
    \caption{On the left, a plot of a ball as a Reinhardt domain. On the right, a plot of a polydisc as a Reinhardt domain.}
    \label{fig:RD}
\end{figure}

If a function $f(\z)$ is holomorphic in a neighbourhood of a point $\z_0$, then it may be written in the Taylor series
\begin{equation}\label{app:scv_Taylor}
    f(\z) = \sum_{n=0}^\infty\sum_{m=0}^\infty c_{n,m} z_1^n z_2^m\,.
\end{equation}
The domain of convergence of such series as~\eqref{app:scv_Taylor} is given by a class of domains known as ``Reinhardt domains", or ``circular domains", which are ultimately the correct generalization of the disc to use. About a point $\z_0$, a Reinhardt domain $D$ is defined by the condition that if a point $\z$ is in the Reinhardt domain, then all points $\z'$ with components given by $z'^j = z^j_0 + (z^j - z_0^j)e^{i\theta_j}$ for $\theta_j \in (0, 2\pi)$ are \textit{also} in the Reinhardt domain\footnote{This generalizes the condition in the main body of the text to Reinhardt domains centred on a point besides the origin.}.  Polydiscs and balls are both examples of Reinhardt domains. The fact that Reinhardt domains are closed under rotations by complex phases leads to the convenient property that Reinhardt domains can be characterized entirely by the magnitudes of the variables $z_1$, $z_2$. As an example, the unit ball and the unit polydisc are plotted in Figure~\ref{fig:RD} solely in terms of $|z_1|$, $|z_2|$. Reinhardt domains were also plotted in the main body of the text in Figures~\ref{fig:boosted_shear} and~\ref{fig:shear_convergence}.

A Reinhardt domain is called ``complete" if, for any point $\z$ in the Reinhardt domain, any second point $\z'$ for which $|\z' - \z_0| \leq |\z - \z_0|$ is \textit{also} in the Reinhardt domain. It is called ``logarithmically convex" if, given the mapping $\alpha(D) = (\ln|z_1|, \ln|z_2|)$ (excluding the origin), the set $\alpha(D)$ is a convex set. Domains of convergence of Taylor series such as~\eqref{app:scv_Taylor} in several complex variables are given by logarithmically convex complete Reinhardt domains.

Finally, we come to the question of singularities -- and this is where the analyses of single and several complex variables differ dramatically. One of the most fundamental results in the analysis of several complex variables is Hartogs' extension theorem. The precise statement, quoting Theorem 2.3.2 from~\cite{HormanderBook}, reads
\begin{center}
    \textbf{Hartogs' extension theorem: } \\
    \fbox{
    \parbox{0.8\linewidth}{
    \centering
    ``Let $\Omega$ be an open set in $\mathbb{C}^n$, $n>1$, and let $K$ be a compact subset of $\Omega$ such that $\Omega\backslash K$ is connected. For every $u \in A(\Omega \backslash K)$ one can then find $U \in A(\Omega)$ such that $u = U$ in $\Omega \backslash K$."
    }
    }
\end{center}
where $A(D)$ is the set of all holomorphic functions on the domain $D$. Loosely, the theorem may be re-stated as follows:
\begin{center}
    \textbf{Hartogs' extension theorem (paraphrased): } \\
    \boxed{\text{Any compact set of singular points in $\mathbb{C}^n$ for $n\geq 2$ is removable.}}
\end{center}
In other words, any set of singular points over which a function $f(\z)$ cannot be analytically continued must necessarily extend to the boundary of a domain such that the complement is not connected; for $\mathbb{C}^2$, this means any set of singular points must extend to infinity, and cut $\mathbb{C}^2$ in two (or more). This is a drastic difference from the $\mathbb{C}$ case, where we are used to having isolated singularities. The consequences of this difference may be readily seen in the body of the paper.

\bibliographystyle{JHEP}
\bibliography{hydro-general-bib}

\end{document}